\pdfoutput=1
\documentclass[sigplan,10pt]{acmart}
\renewcommand\footnotetextcopyrightpermission[1]{}
\AtBeginDocument{%
  }

\setcopyright{none}
\acmDOI{}
\acmISBN{}

\usepackage{graphicx}
\graphicspath{{fig/}}

\usepackage{amsmath}

\usepackage{amssymb}
\usepackage{amsthm}

\usepackage{relsize}
\usepackage{xspace}
\usepackage{textcomp}

\usepackage{xcolor}

\usepackage{listings}
\usepackage[most]{tcolorbox}

\usepackage{booktabs}
\usepackage{multirow}
\usepackage{threeparttable}
\usepackage{array}

\usepackage{longtable}

\usepackage{subcaption}

\usepackage{float}
\usepackage{placeins}

\usepackage[ruled,vlined]{algorithm2e}

\usepackage{enumitem}

\usepackage{tikz}

\usepackage{comment}
\usepackage{todonotes}

\usepackage{xurl}

\newcommand{\ccbf}[1]{\mbox{\smaller[0.5]\textbf{\texttt{#1}}}}
\newcommand{\cc}[1]{\mbox{\smaller[0.5]\texttt{#1}}}

\newcommand{\tool}{\textsc{Origin}\xspace}
\newcommand\cldg{\ccbf{CLDG}\xspace}
\newcommand\agent{\cc{GPT-Researcher}\xspace}
\newcommand\kb{function-to-information-source mapping table\xspace}

\newtheorem{definition}{Definition}[section]

\begin{document}
\title[Backtracking Agent Applications Security Incidents]{Backtracking Agent Applications Security Incidents}

\author{Yicheng Cai}
\affiliation{%
  \institution{Pennsylvania State University}
  \city{University Park}
  \state{PA}
  \country{USA}
}
\email{yccai@psu.edu}

\author{Taegyu Kim}
\affiliation{%
  \institution{Pennsylvania State University}
  \city{University Park}
  \state{PA}
  \country{USA}
}
\email{tgkim@psu.edu}

\author{Peng Liu}
\authornote{Corresponding author.}
\affiliation{%
  \institution{Pennsylvania State University}
  \city{University Park}
  \state{PA}
  \country{USA}
}
\email{pxl20@psu.edu}

\renewcommand{\shortauthors}{Cai et al.}

\begin{abstract}

Agent applications are increasingly deployed, yet agent-specific attacks pose real threats to them.
Upon a security incident, organizations backtrack to identify the CEP (compromise entry point), remediate the damage, and prevent its recurrence.
The main artifacts available for backtracking are trajectory logs, which record agent functionalities, and source code; app(lication)-specific function call logs are often fairly limited.
However, trajectory logs are at the agent functionality abstraction layer, while the CEP candidates are function calls at the app code execution layer.
Due to this semantic gap, the analyst needs to manually inspect the relevant code for each CEP candidate to infer its corresponding agent functionality.
In this work, we present \tool, a backtracking tool for the analyst to submit a small number of queries to identify CEPs that existing tools cannot provide.
Our approach poses two challenges: (1) define the intermediate abstraction that captures the mapping relationships between the two semantic layers; (2) determine the region-to-variable mapping for constructing the cross-layer graph defined in the intermediate abstraction.
Each region is a specific group of agent functionalities.
\tool addresses the challenges by (1) the cross-layer dependency graph (\cldg) that connects region-inferred functions at the semantic level to the variables in the low-level taint propagation subgraph, plus the associated query language, and (2) the region inference algorithm.
Case studies on five popular agent applications across eight attack incidents
show that an analyst using \tool correctly identifies all the CEPs.

\end{abstract}




\settopmatter{printfolios=true}
\settopmatter{printacmref=false}
\maketitle
\thispagestyle{plain}
\pagestyle{plain}

\section{Introduction}
\label{sec:introduction}
Agent applications powered by large language models~(LLMs) have been widely deployed.
For example, an open-source agent application, \agent,  accumulated over 24{,}000 GitHub stars after release \cite{GPTResearcher2024}.
Moreover, enterprise agent applications may increase from under 5\% in 2025 to about 40\% by the end of 2026, according to a recent Gartner forecast~\cite{GartnerPredicts402025}.

Meanwhile, such agent applications are under threat from emerging agent-specific attacks, among which indirect prompt injection (IPI)~\cite{greshakeNotWhatYouve2023,zhanInjecAgentBenchmarkingIndirect2024,debenedettiAgentDojoDynamicEnvironment2024,liuFormalizingBenchmarkingPrompt2024} and memory \& context poisoning (MC-Poison)~\cite{chenAgentPoisonRedteamingLLM2024, wangBadAgentInsertingActivating2024,zouPoisonedRAGKnowledgeCorruption2025} are the most prevalent two of the OWASP Top 10 threats for agent applications~\cite{editorOWASPTop10}.
For instance, a number of IPI CVEs~\cite{CursorCve202554130, 5ireCve202558357,NeuronCve202567510, ClaudeCodeIPI} and MC-Poison-based red-teaming~\cite{ChatGPTMemory,LyonsAtlasVuln2025,GoodinGeminiMemory2025} have been reported across multiple commercial agent applications.

When a security incident occurs in deployed agent applications, organizations backtrack the incident in order to identify the CEP (compromise entry point), remediate the damage, and prevent its recurrence.
The main artifacts available for backtracking are: (1) trajectory logs, which record a sequence of agent observations and actions~\cite{zhuWhereLLMAgents2025} mapping to agent functionalities, such as
planning, acting by invoking tools, and recalling information from memory~\cite{liuAdvancesChallengesFoundation2025}; (2) source code of the compromised applications.
It is worth noting that application-specific function call logs are often fairly limited: developers may not anticipate which calls are diagnostically relevant to emerging agent-specific attacks and consequently omit the corresponding log printing statements~\cite{yuanBeConservativeEnhancing2012}, or complete logging is disabled due to overhead thresholds and privacy considerations~\cite{zhaoLog20FullyAutomated2017,yuanImprovingSoftwareDiagnosability2012}.

{\bf Motivating example.}
Consider an MC-Poison security incident on \agent, where an analyst received an incident report from a user/victim of the application (Section~\ref{subsec:motivating example}).
There are three stakeholders: the analyst, the remedy administrator (usually the user), and the developers.
The analyst aimed to backtrack the incident to identify the CEP, i.e., the function call that first introduced the malicious content to the application (Section \ref{subsec:cep-def}).
When the corresponding input arguments of the CEP function call are already logged, the analyst can (1) provide the remedy admin with a concrete, actionable remedy, e.g., the exact poisoned file to remove from a knowledge base that may contain hundreds of documents in various formats; when the corresponding function call log entry is missing, the analyst can (2) deliver to the developers the log printing statement placement around the CEP function call, enabling more efficient logging that makes the application diagnosable for future incidents.

{\bf Identifying the CEPs requires significant manual effort from the analyst.}
The analyst first triaged and examined the trajectory log and narrowed down the suspect cause to a particular agent functionality (i.e., knowledge-base retrieval).
However, it remains difficult to identify the CEPs based only on the agent functionality since one agent functionality typically corresponds to many function calls, and no trajectory log records concrete function calls (Listing~\ref{lst:triaged-log}).
Fundamentally speaking, trajectory logs are at the agent functionality abstraction layer,  while function calls are at a lower layer (i.e., the app code execution layer).
Static analysis on the source code of the application can complement trajectory-log-based agent-functionality-level diagnosis by pinpointing the CEPs through variable-level taint analysis.
However, the analysis is hindered by a large set of CEP candidates: only a small subset is associated with the suspected initial compromise.
After
log-based pruning, there are still 1,518 candidates left in \agent source code; to infer if
a candidate belongs to the suspected initial compromise, the analyst spent hours on manual code inspection, detailed in Section~\ref{subsec:motivating example}.
The main reason is that the code around the candidates carries little agent functionality semantics, and the analyst needs to gather enough evidence along the corresponding code paths before making a deduction.

{\bf None of the existing approaches can provide stakeholders with the two ``can provide'' benefits mentioned in the motivating example}, since trajectory-log-based failure diagnosis~\cite{zhuWhereLLMAgents2025,maDiagnosingFailureRoot2025,zhangWhichAgentCauses2025}, detection-based failure attribution~\cite{kimPromptFlowIntegrity2025,zhangAgentSentryMitigatingIndirect2026,chocronWhoOwnsThis2026}, and preventive information flow control~\cite{debenedettiDefeatingPromptInjections2025} \textbf{do not provide the analyst with the CEPs} but faulty agent execution state~\cite{zhuWhereLLMAgents2025,maDiagnosingFailureRoot2025,zhangAgentSentryMitigatingIndirect2026}, faulty agent~\cite{zhangWhichAgentCauses2025} or agent-owner~\cite{chocronWhoOwnsThis2026}, and LLM tool calls (\textbf{not} function calls at the source code level)~\cite{kimPromptFlowIntegrity2025,debenedettiDefeatingPromptInjections2025} instead.

{\bf Our approach can provide stakeholders with the two benefits because it backtracks the security incident.}
As mentioned in the motivating example, narrowing the CEPs during backtracking requires significant manual effort since the analyst needs to do code inspection to recover the agent functionality semantics from the program source code.
\textbf{Our insight is to introduce a new intermediate abstraction} (Section~\ref{subsec:problem-scope}) that enables the analyst to write a few queries at the agent-functionality-level, and the query processing outputs can exclude irrelevant CEP candidates directly.
However, identifying which CEP candidates semantically correspond to specific agent functionalities is hard since program structures and implementations vary, making it difficult to generalize how to identify the mapping relationships.
Fortunately, we observe that agent functionalities can be grouped into a small number of \emph{regions} depending on the type of information source (Section~\ref{sec:background}) with which they interact.
Each \emph{region} is a set of program variables that implement agent functionalities of the same group.
The variables that receive the return values of the CEP candidates are also included.
{\bf Based on this insight, we build a tool named \tool that realizes this abstraction as a graph across two layers.}
\tool
takes as input the source code of an agent application and a small set of developer-provided labels, and produces the cross-layer graph that can be queried by analysts using a query
language during security incident investigations:
(i) narrow down candidates at the region level/layer, (ii) identify the \emph{CEPs} at the variable level.

{\bf Challenges.}
Our approach poses two challenges.
First, define the intermediate abstraction that captures the mapping relationships between two semantic layers: the agent functionality layer vs. the application code execution layer.
Since the implementation of the same agent functionality by different agent applications varies (e.g., the functions/variables naming, the used agent frameworks), we believe such an intermediate abstraction can provide an analyst with the same interface when backtracking security incidents of different applications.
Second, constructing the cross-layer graph (defined in the intermediate abstraction) requires determining which region each variable belongs to.
This is because a variable's region is determined by which information source its data came from or flows into, but a variable's name, type, or enclosing function does not carry these semantics.

{\bf Solutions.}
For the first challenge, we design the Cross-Layer Dependency Graph (\cldg) (Section~\ref{sec:cldg}) that connects region-inferred functions to the underlying variables in the taint graph through cross-plane edges.
We also design the associated query language.
As in BackTracker~\cite{kingBacktrackingIntrusions2003}, the essence of backtracking is dependence analysis.
Since the agent application CEPs are specific function calls, static taint analysis is suitable to identify the two types of dependencies a backtracking tool can rely on: variable-level data dependencies and function-level data dependencies.
For the second challenge, we design region inference (Section~\ref{subsec:region-inference}) that mimics the analyst's code inspection process by rules,
which requires only a small set of developer-provided labels and automatically infers the regions for the variables.

{\bf Contributions.}
We make the following contributions.
\begin{itemize}[leftmargin=*,parsep=1.0pt,topsep=1.0pt, partopsep=1.0pt]
\item \textbf{We design the \cldg plus the associated query language}, a new intermediate abstraction that bridges the semantic gap between agent-functionality-level reasoning and program-variable-level application code execution.
\item \textbf{We design region inference}, a technique that maps program variables into the relevant regions by rule-guided propagation of label information to the relevant functions along tainted program paths.
\item \textbf{We build and evaluate \tool} on five real-world
  agent applications across eight attack incidents.
  \tool correctly identifies every \textit{CEP} with a single
  candidate per incident, reducing the analyst's inspection
  scope by 64--94\% compared to the full taint graph and
  replacing exhaustive source code inspection with
  region-guided diagnosis.
\end{itemize}

To foster research on this topic, we will release the source code of \tool to GitHub.
\section{Background}
\label{sec:background}

{\bf Agent Applications.}
An LLM-powered \emph{agent application} interacts with diverse information sources for implementing various \emph{agent functionalities} (e.g., perceiving the user instructions, planning, acting by invoking tools, and recalling information from memory)~\cite{liuAdvancesChallengesFoundation2025}.
Such applications produce \emph{trajectory logs} to record the invocations of agent functionalities during a concrete execution~\cite{zhuWhereLLMAgents2025,maDiagnosingFailureRoot2025}.
\textit{Two types of information sources} are: (1)~\emph{external information sources} outside the applications, including but not limited to user inputs, interactions with humans or agents, sensor readings, network-accessible content (e.g., web pages and emails), and LLM-generated content;
(2)~\emph{internal information sources} within the applications such as knowledge bases and memory.
The memory stores multi-round interactions between the user and the agent, persisted on disk in either original or summarized form, so that the agent can recall prior context across sessions~\cite{liuAdvancesChallengesFoundation2025}.

\noindent{\bf Attacks on Agent Applications.}
Agent-specific attacks such as IPI~\cite{greshakeNotWhatYouve2023,zhanInjecAgentBenchmarkingIndirect2024,liuFormalizingBenchmarkingPrompt2024}
and MC-Poison~\cite{editorOWASPTop10,chenAgentPoisonRedteamingLLM2024,zouPoisonedRAGKnowledgeCorruption2025}
compromise the agent applications by injecting manipulated content into legitimate information sources of the applications.
MC-Poison attacks usually involve two phases that occur
in separate executions: a \emph{poisoning phase}, in which malicious data is written to an internal information source, and an \emph{exploitation phase}, in which the stored data is retrieved and causes harm. In some incidents, poisoning occurs outside the application (e.g., an attacker directly modifies a knowledge base document), and only the exploitation phase is observable.

\noindent{\bf Static Taint Analyzers.}
Existing taint-analysis frameworks such as Joern~\cite{JoernBugHunters2025}, Pysa~\cite{PysaStaticAnalysis}, and CodeQL~\cite{CodeQL2025} can already compute propagation paths over program variables and expressions.
These capabilities can be repurposed for backtracking agent application security incidents.
However, they expose only program code-level semantics and do not provide abstractions aligned with agent functionalities.

\section{Motivation}
\label{sec:motivation}
\subsection{Motivating Example}\label{subsec:motivating example}
{\bf Security Incident.}
Consider \agent, an agent application that conducts research for users and generates synthesized content~\cite{GPTResearcher2024}.
A security analyst receives an incident report from a user of the application.
The report states that the application produced a research synthesis containing incorrect information for one particular research task, while other tasks were completed correctly.
The report includes: a description of how the application is used, a PDF of the research synthesis,
a trajectory log~\cite{zhuWhereLLMAgents2025} of the application (Listing~\ref{lst:triaged-log}), and the application's source code.
According to the user description, the application used both a knowledge base and web data for this failed task.
The analyst listed three likely accident causes: the LLM was hallucinating, the scraped webpages were manipulated, or the knowledge base was poisoned, where the latter two are security-oriented.

\noindent{\bf Three Stakeholders.}
There are three stakeholders in the incident: the analyst, the remedy administrator (usually the user), and the developers.
The remedy admin needs a concrete, actionable remedy for the damage, e.g., removing the exact poisoned file from the knowledge base
with hundreds of documents;
the developers need the log printing statement placement suggestion that makes the application more diagnosable for future incidents, e.g., log the parameter values of the function call corresponding to the initial compromise immediately before that call (Listing~\ref{lst:CEP-function-call}, Line 83).
To provide the two stakeholders with the two benefits, the analyst aimed to identify the CEP, i.e., the function call that first introduced the malicious content to the application (Section~\ref{subsec:cep-def}).

\noindent{\bf Log Triage.}
To identify the CEP of the incident, the analyst first triaged the trajectory log (Listing~\ref{lst:triaged-log}), which records the user task, the application's research plan, links to scraped webpages, and HTTP requests to an LLM provider.
These coarse-grained traces were sufficient to rule out LLM hallucination and manipulated webpages, leaving knowledge-base poisoning as the remaining suspect cause.
However, the log could not help identify which function call is responsible for the initial compromise (i.e., retrieving poisoned content from the knowledge base), because the trajectory log did not record concrete function calls.
Fundamentally speaking, trajectory logs are at the agent functionality abstraction layer, while function calls are at the app code execution layer.

\begin{lstlisting}[
    numbers=none,
    language=,
    % frame=tb,
    caption={Triaged trajectory log.},
    label={lst:triaged-log},
    captionpos=b,
    % abovecaptionskip=2pt,
    aboveskip=4pt,
    belowskip=0pt
]
2025-09-07 20:43:26,657-research-INFO- <@\textcolor{red}{Starting~research~for}@> query: Is Nvidia stock price going up?
...
2025-09-07 20:43:36,913-httpx-INFO- HTTP Request: POST https://api.openai.com/v1/chat/completions "200 OK"
2025-09-07 20:43:36,915- research - INFO - <@\textcolor{red}{Research~outline~planned}@>: ['NVIDIA stock price forecast September 2025 analysts consensus', 'NVIDIA real...]
...
2025-09-07 20:43:46,662-gpt_researcher.scraper.scraper-INFO- URL: https://www.cnn.com/markets/stocks/AMD
\end{lstlisting}

\noindent{\bf Static Taint Analysis.}
To identify the CEP, the analyst complemented the trajectory-log-based agent-functionality-level diagnosis by variable-level static taint analysis on the application's source code.
The analysis extracted 1,895 API calls as taint source candidates, i.e., function calls that could potentially introduce untrusted content.
Using distinctive log patterns (e.g., ``Starting research for~\ldots'' colored red in Listing~\ref{lst:triaged-log}), the analyst located the corresponding log printing statements in the source code, identified their callers, and pruned program paths that cannot reach the callers, filtering out candidates on irrelevant paths.
This log-guided pruning reduced the candidates from 1,895 to 1,518, but the analyst still needs to determine which of these candidates are related to knowledge-base retrieval functionality.

\begin{lstlisting}[
numbers=none,
language=Python,
escapechar=|,
captionpos=b,
caption={Code snippet around the CEP function call.},
label={lst:CEP-function-call},
% abovecaptionskip=2pt,
aboveskip=4pt,
belowskip=0pt
]
|4| from langchain_community.document_loaders import (
|5|     PDFLoader, TextLoader, ..., HTMLLoader )
|63|   async def _load_document(self, file_path, file_ext):
|64|     ret_data = []
|65|     |$\color{red}{\texttt{loader\_dict}}$| = {"pdf": PDFLoader(file_path),
|66|       "txt": TextLoader(file_path),
...
|72|       "html": HTMLLoader(file_path)}
|80|     loader = |$\color{red}{\texttt{loader\_dict}}$|.get(file_extension, None)
|83|     ret_data = loader.|$\color{red}{\texttt{load()}}$|
|84|     return ret_data
\end{lstlisting}

\noindent{\bf Manual Code Inspection Burden.}
Without tool support to filter candidates by their functionality, the analyst's only option is to inspect each candidate manually.
Yet even determining whether a candidate is related to knowledge-base retrieval requires multi-step code inspection.
For instance, to determine that the candidate \texttt{loader.load()} in Listing~\ref{lst:CEP-function-call} retrieves facts from a knowledge base, the analyst had to (a)~trace the variable \texttt{loader} back to its definition, (b)~examine each class in \texttt{loader\_dict} and its \texttt{load} function, and (c)~combine these observations with the caller name \texttt{\_load\_document} to infer the developer's intent.
Repeating this process across 1,518 candidates would take hours.

\subsection{Limitations of Existing Approaches}\label{subsec:limitation}

Existing approaches for analyzing and defending agent applications \textbf{cannot} provide stakeholders with the two benefits illustrated above.
The main reason is that trajectory-log-based failure diagnosis~\cite{zhuWhereLLMAgents2025,maDiagnosingFailureRoot2025,zhangWhichAgentCauses2025}, detection-based failure attribution~\cite{kimPromptFlowIntegrity2025,zhangAgentSentryMitigatingIndirect2026,chocronWhoOwnsThis2026}, and preventive information flow control~\cite{debenedettiDefeatingPromptInjections2025} \textbf{do not provide the analyst with the CEPs}.
They instead provide the analyst with the faulty agent execution state~\cite{zhuWhereLLMAgents2025,maDiagnosingFailureRoot2025,zhangAgentSentryMitigatingIndirect2026} (e.g., an agent forming an incorrect plan), faulty agent~\cite{zhangWhichAgentCauses2025} or agent-owner~\cite{chocronWhoOwnsThis2026}, and LLM tool calls (\textbf{not} function calls at the source code level)~\cite{kimPromptFlowIntegrity2025,debenedettiDefeatingPromptInjections2025}.
\section{Approach Overview}

\subsection{Scope}
\label{subsec:scope}
Existing agent applications can be classified into two categories: predefined workflow (C1),
where application source code rather than LLM decides control flow among agent functionalities;
OpenClaw-like agents~\cite{OpenClaw2026,Hermes2026,AutoGPT2025} (C2), where the LLM decides at runtime which agent functionality to invoke next.
The C2 agents have recently emerged; however, most existing applications belong to C1.
Our approach is directly motivated by C1, not by C2.
Hence, the scope of this work is C1; C2 and agent library frameworks~\cite{LangChain2025,wuAutoGenEnablingNextgen2024,Smolagents} are out of scope and remain an area for future research.

\vspace{-1mm}
\subsection{Threat Model}
\label{subsec:threat-model}
We assume that adversaries can launch two categories of agent-specific attacks by methods proposed in existing works: IPI~\cite{greshakeNotWhatYouve2023} and MC-Poison~\cite{dongPracticalMemoryInjection2025,chenAgentPoisonRedteamingLLM2024}.
The two attack classes are the most prevalent among the OWASP Top 10 threats for agent applications and can cover all the  risk categories in the report that target agent functionalities of the applications~\cite{editorOWASPTop10}.
The remaining categories are either risks
inherited from traditional software security (i.e., remote code execution, privilege escalation, and supply chain integrity), or
those
reside outside the application (i.e., inter-agent communication protocol, social engineering, and insider threats).
Note: the non-protocol part of the insecure inter-agent communication risk category can be caused by IPI and MC-Poison.
Such an attack incident is investigated in Section~\ref{subsec:case-study-unique}.

\vspace{-1mm}
\subsection{Compromise Entry Point}
\label{subsec:cep-def}

The agent-specific attack classes
in Section~\ref{subsec:threat-model} share a common characteristic: malicious data enters the application through the return value of an API call and propagates through data flow to cause harm.
We define the compromise entry point (CEP) based on the characteristic and standard static taint tracking~\cite{yinPanoramaCapturingSystemwide2007}.
Given
a security incident in an agent application, the \textbf{\emph{CEP} is identified by: }
(i) the name of the API call whose return value contains the malicious data,
(ii) the name of the corresponding source code program file, and
(iii) the line number of the call site.
We also provide
the analyst with a set of taint propagation paths, where each path starts from a variable receiving the return value of the CEP function call and ends at a variable capturing the attack effect.
Each variable in the paths is contained in exactly one function.
Mapping each variable to its enclosing function gives a function-level path: the sequence of functions through which the data passes from entry to effect.

\vspace{-2mm}
\subsection{Problem Statement}
\label{subsec:problem-scope}

{\bf Our insight: the need for an intermediate abstraction.}
As the motivating example illustrates, the analyst reasons about executed agent functionalities observed from the trajectory log (e.g., ``which code retrieves from the knowledge base?''), while static taint analysis operates at the level of program variables.
However, no existing static analysis technique can determine which program variables semantically correspond to which specific agent functionalities, because program structures and naming conventions vary across applications, making it hard to generalize such a mapping.

Fortunately, we observe that agent functionalities can be grouped into a small number of groups based on the type of information source (Section~\ref{sec:background}) with which they interact.
For instance, retrieving from a knowledge base and recalling memory are distinct functionalities, but both interact with internal information sources; hence, they belong to the same group.
We call such a group a \emph{region}: a set of program variables that collectively implement agent functionalities of the same group.
An analyst can query at the region level (e.g.,
``which region is for retrieving content from the knowledge base?''); the query processing outputs can exclude irrelevant CEP candidates directly, without manual code inspection.

Concretely, the intermediate abstraction is composed of four elements:
(1) a \emph{region set} that partitions agent functionalities into finite groups;
(2) a \emph{region-to-variable mapping} that associates each region with the program variables that implement it;
(3) a \emph{region-level query interface (including the query language)} through which analysts express their diagnostic queries in terms of regions rather than  variables;
(4) a \emph{region-to-variable query mapping} that translates region-level queries into variable-level queries, supported by existing static analysis infrastructure.
The goal of this work is to design a tool that realizes the new abstraction.

\vspace{-2mm}
\subsection{Challenges}
\label{subsec:challenges}

{\bf Challenge 1: Define the intermediate abstraction that captures the mapping relationships between two semantic layers: agent functionality layer vs. the application code execution layer.}
The implementation of the same agent functionality by different agent applications varies (e.g., the functions/variables naming, the used agent frameworks).
We believe such an intermediate abstraction is needed so that an analyst can use the same interface for backtracking security incidents of different applications.

\noindent{\bf Challenge 2: Constructing the region-to-variable mapping.}
The abstraction requires a concrete mapping from regions to program variables (element~2).
No existing static analysis technique can infer agent functionalities from program source code.
As illustrated in Section~\ref{subsec:motivating example}, variable names, such as \texttt{ret\_data}, \texttt{loader\_dict}, and \texttt{loader} in the function \texttt{\_load\_document}, do not carry the hints that they are used for retrieving from a knowledge base.
Therefore, it is hard to determine which region they belong to.

\subsection{\tool Overview}
\label{sec:overview}
\begin{figure}[htbp]
    \centering
    \vspace{-2mm}
    \includegraphics[width=0.483\textwidth]{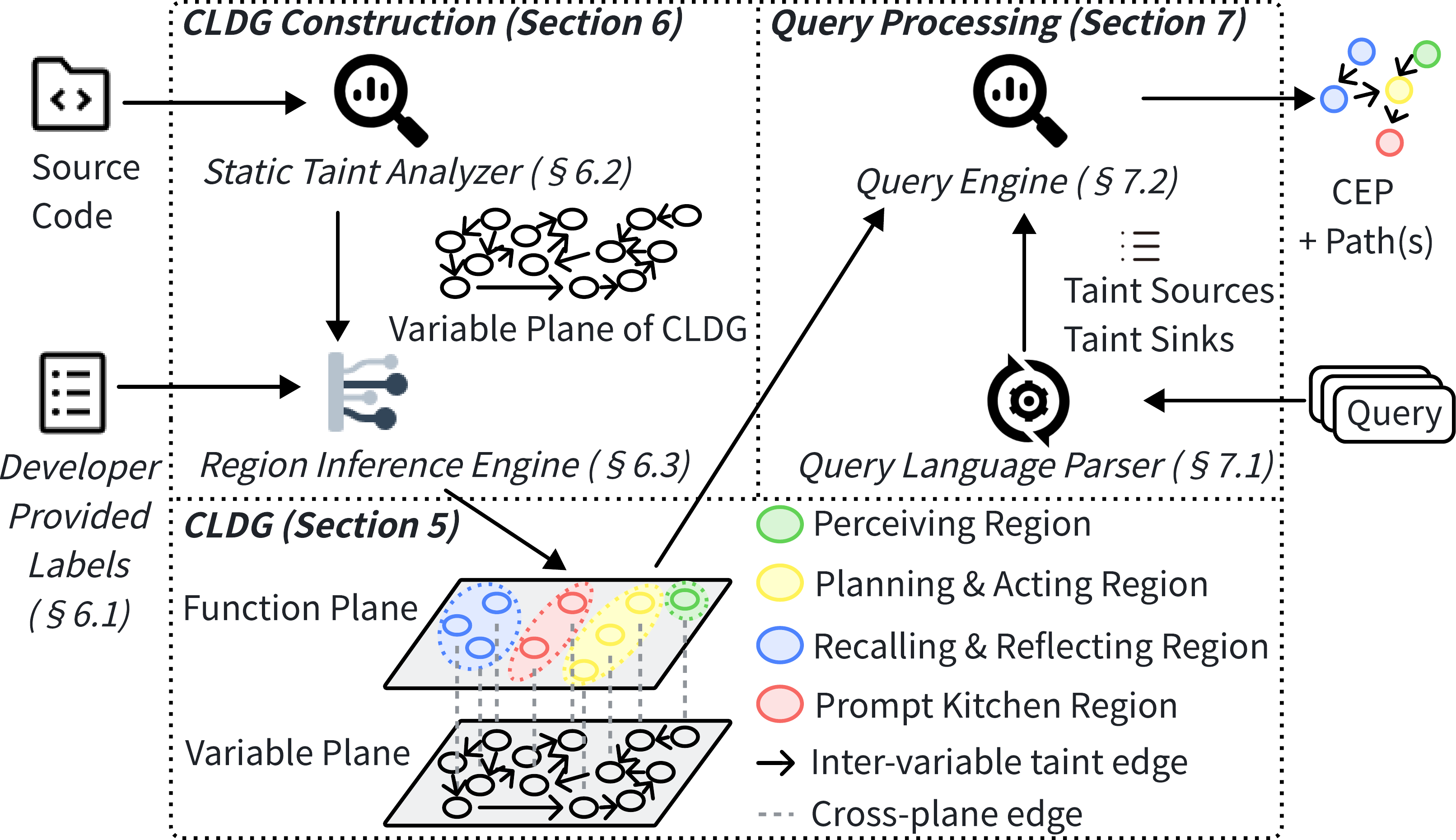}
    \vspace{-7mm}
    \caption{\tool overview.}\label{fig:tool}
    \vspace{-3mm}
\end{figure}

\tool takes as input (1)~the source code of an agent application and (2)~developer-provided labels.
It outputs a \cldg that analysts query to identify the CEPs of security incidents on the agent application.
Figure~\ref{fig:tool} illustrates \tool.

{\bf \cldg (Section~\ref{sec:cldg}).}
The \cldg is a two-plane directed graph:
The \emph{function plane} is a set of region-inferred function nodes (abstraction element~1 in Section~\ref{subsec:problem-scope}); the \emph{cross-plane edges} connect each function to the variables in the \emph{variable plane} (element~2).
The \cldg, together with the associated query language (elements~3 and 4, Section~\ref{sec:query-stage}), addresses Challenge~1.

{\bf \cldg Construction (Section~\ref{sec:construction}).}
\tool (i) performs forward taint propagation from the return values of every API call in the source code to build the variable plane of the \cldg (Section~\ref{subsec:taint-construction}); (ii) uses the developer-provided labels (Section~\ref{subsec:dev-kb}) as input for region inference (Section~\ref{subsec:region-inference}), which classifies each function to a region
 (addressing Challenge~2).

{\bf Query Processing (Section~\ref{sec:query-stage}).}
The analyst first submits a region-level query against the \cldg; then inspects the returned graph to identify the source/sink variables for a variable-level query (Section~\ref{sec:query-stage}).
The output is the CEP and the corresponding paths in Section~\ref{subsec:cep-def}.

\section{Cross-Layer Dependency Graph}
\label{sec:cldg}

The \cldg realizes the intermediate abstraction as a directed graph of two planes: the \emph{function plane} and the \emph{variable plane}.

\begin{definition}[Cross-Layer Dependency Graph]
Let $P$ be an agent application with the function set $\mathcal{F}$. The \emph{Cross-Layer Dependency Graph} $G = (\mathcal{R},\, V_{\mathrm{fn}},\, \lambda,\, E_{\mathrm{cross}},\, V_{\mathrm{var}},\, E_{\mathrm{var}})$ of $P$ is a two-plane directed graph,
where:
\begin{itemize}[leftmargin=*,parsep=1.0pt,topsep=1.0pt, partopsep=1.0pt]
    \item \textbf{Region set.} $\mathcal{R}$ is a mapping from each region name to a set of agent functionalities.
    Each agent functionality is a mapping from the functionality name to a set of functions in the agent application program.
    Five representative regions (detailed in Appendix~\ref{apdx:region-set}) are: Per (Perceiving), PA (Planning \& Acting), RR (Recalling \& Reflecting), PK (Prompt Kitchen), and UNK (Unknown).

    \item \textbf{Function plane.} $V_{\mathrm{fn}} \subseteq \mathcal{F}$ is the set of functions reachable from all API calls in $P$ via taint propagation. $\lambda\colon V_{\mathrm{fn}} \to \mathcal{R} \cup \{\mathrm{UNK}\}$ assigns each function node its inferred region.

    \item \textbf{Variable plane.} $V_{\mathrm{var}}$ is the set of program variables reachable from all API calls in $P$ via taint propagation, and $E_{\mathrm{var}} \subseteq V_{\mathrm{var}} \times V_{\mathrm{var}}$ is the set of taint edges among them.
    The same variable in different code statements is treated as different nodes (with different identifiers) in the variable plane.

    \item \textbf{Cross-plane edges.} $E_{\mathrm{cross}} \subseteq V_{\mathrm{fn}} \times V_{\mathrm{var}}$ connects each function node $f$ to every program variable defined/used in $f$.
\end{itemize}
\vspace{-1.6mm}
\end{definition}

The region set $\mathcal{R}$ can be determined by the security analyst as long as the mapping relation is provided.
Our design is applicable to any analyst-defined region set.
Once a particular region set is adopted, \tool requires a corresponding \kb from developers of the agent application (Section~\ref{subsec:dev-kb}).
In addition, every cross-function taint edge in the variable plane is naturally mapped to an edge connecting two functions.
Such edges appear on the function-level paths mentioned in Section~\ref{subsec:cep-def}.

\section{\cldg Construction}
\label{sec:construction}

Constructing the \cldg involves two steps: (1) taint propagation to build its variable plane (Section~\ref{subsec:taint-construction}) and (2) infer regions for nodes in the planes (Section~\ref{subsec:region-inference}).
Step (2) requires developer knowledge as input (Section~\ref{subsec:dev-kb}).

\subsection{Developer Knowledge Collection}
\label{subsec:dev-kb}

\tool runs with a \kb showing which functions in an application interact with external (EXT) or internal (INT) information sources.
\textbf{This information is collected once from the agent application developer} and reused across all
incident investigations.

Well-trained developers know only functions that directly call built-in or library APIs need examination, but they may still find it time-consuming to label this subset of functions, e.g., 353 out of 607 functions in GPT-Researcher.
As in Listing~\ref{lst:knowledge-source}, to determine whether an API call \texttt{execute()} accesses an internal information source (INT), the developer needs to (1) trace variable \texttt{commons} back to \texttt{common\_dependencies} function definition to identify the Supabase client, (2) check the preceding calls before \texttt{execute()} against the API documentation to determine that \texttt{execute()} retrieves content from the \texttt{chat\_history} table, and (3) confirm that the \texttt{chat\_\\history} table is used as agent memory.
The developer needs to examine \textbf{all remaining API calls} in the function to ensure they are \textbf{not} interacting with external (\texttt{EXT}) information sources (e.g., a user, an LLM, or a web browser) before assigning a label \texttt{INT} to the function.

\begin{lstlisting}[
language=Python,
deletekeywords={[2]filter},
basicstyle=\ttfamily\scriptsize,
numbers=none,
escapechar=|,
xleftmargin=1.5em,
captionpos=b,
breaklines=true,
caption={An example function for labeling, simplified.},
label={lst:knowledge-source},
% abovecaptionskip=2pt,
aboveskip=4pt,
belowskip=4pt
]
|1| def get_chat_history(chat_id):
|2|     commons = common_dependencies()
|3|     history = (commons["supabase"]
|4|         .from_("chat_history").select("*")
...
|7|         .execute()).data
|8|     return [ChatHistory(m) for m in history or []]
\end{lstlisting}

To address this, we use an LLM with a prompt containing the function's API calls, relevant source code, and rules for identifying information sources.
The LLM produces a structured report:
a decision for each call, its explanation, supporting code locations, and any missing information for labeling a function (Appendix~\ref{apdx:knowledge-prompt}).
Although human validation is still needed, this reduces the developer's work from labeling to reviewing a report. The experiment results show that the knowledge collection speed increases by 15.4–75.0\% and the accuracy increases by 8.3–50.4\% (Section~\ref{subsec:rq6}).

\subsection{Taint Graph Construction}
\label{subsec:taint-construction}

With the developer labels in place, \tool constructs the graph.
The variable plane ($V_{\mathrm{var}}$, $E_{\mathrm{var}}$) is
produced by performing context-, flow-, field-sensitive forward taint propagation from the return values of every API call (implementation described in Section~\ref{sec:implementation}).
Because propagation starts from all API calls, any CEP function call that matches the definition in Section~\ref{subsec:cep-def} lies on a path in the variable plane and is locatable by queries against the \cldg.
$V_{\mathrm{fn}}$ and $E_{\mathrm{cross}}$ are derived from the variable plane: $f \in V_{\mathrm{fn}}$ if any of its variables appears in $V_{\mathrm{var}}$; $E_{\mathrm{cross}}$ connects each function to its variables.

\subsection{Region Inference}
\label{subsec:region-inference}

The region inference algorithm implements the \textit{region-to-variable mapping} element in the abstraction (Section~\ref{subsec:problem-scope}).
The goal is to classify the functions in the function plane of \cldg into the five regions detailed in Appendix~\ref{apdx:region-set}.

We observe that a naive feature-based clustering approach, i.e., clustering by function names, API names, or program code syntactic features, is \textbf{not} suitable for this task:
(1) functions/APIs of the same name could implement completely different agent functionality in different agent applications, and thus belong to different regions.
For example, two functions named \texttt{load} may read a webpage (fall into the {\tt PA} region) or read a file from a knowledge base (fall into the {\tt RR} region).
Similarly, (2) syntactically similar string operations, e.g., \texttt{x.format(y)}, may be used for prompt assembly or
for
non-agent functionalities, e.g., composing an error message.

Fortunately, we find that an analyst, fully aware of the semantics of the five regions, usually gathers evidence during code inspection to answer four questions, and the fusion of the answers can determine which region a function belongs to: (1) does the function interact with information source(s)? (2) does the function read from or write to the information source(s), or do both? (3) if the function interacts with multiple information source(s), which type of information source dominates (Section~\ref{sec:background})? and (4) is the function mainly assembling prompts to appear in the LLM context window?
\textbf{\tool uses two categories of rules to mimic the aforementioned analyst's reasoning process during code inspection for region inference:}

\emph{Label propagation rules} (3 in total) mimic the process of gathering evidence to answer the four questions about a function.
Two rules propagate (along the tainted paths) the labels of the API calls that interact with information sources.
Concretely, the two rules are:
(1) if an API call reads an information source, propagate the label forward;
(2) if an API call writes an information source, propagate the label backward.
We also have a rule for propagating a special label of function calls that assemble prompts rather than interact with information sources.
Prompt-assembly-related function calls are
assigned with such a label (e.g., text-composition functions like
\texttt{format}; function names or parameters with special names consisting of ``prompt'').
This rule propagates labels only within a function because we observe that prompt assembly is an agent functionality local to discrete function(s).

\emph{Condition-action rules} (9 in total, Table~\ref{tab:region-rules}) mimic the analyst's reasoning process of region determination by combining answers to the four questions against the gathered evidence.
For example, if a function receives a user request, it gathers the label of an external information source and no label of prompt assembly according to the label propagation rules; consequently, it is classified into the {\tt Per} region, which only passively reads external information sources.

\begin{table}[htbp]
\centering
\setlength{\tabcolsep}{9pt}
\renewcommand{\arraystretch}{0.85}
\footnotesize
\vspace{-2mm}
\caption{Condition-action rules for region inference.}
\label{tab:region-rules}
\vspace{-3mm}
\begin{tabular}{c c c c c c}
\hline
\textbf{Rule} & \textbf{IS?} & \textbf{R/W} & \textbf{Main IS} & \textbf{Asm.?} & \textbf{Region} \\
\hline
1 & No  & --  & --  & No  & UNK \\
2 & Yes & R   & Ext & No  & Per \\
3 & Yes & W   & Ext & No  & PA  \\
4 & Yes & R+W & Ext & No  & PA  \\
5 & Yes & R   & Int & No  & RR  \\
6 & Yes & W   & Int & No  & RR  \\
7 & Yes & R+W & Int & No  & RR  \\
8 & Yes & R+W & Ext & Yes & PK  \\
9 & Yes & R+W & Int & Yes & PK  \\
\hline
\end{tabular}
\par\vspace{0.2em}
\noindent
\begin{minipage}{0.96\linewidth}
\footnotesize
IS: information source; R: read from an information source;
W: write to an information source; Ext/Int: external/internal information source;
Asm.: whether the function mainly assembles the prompt.
\end{minipage}
\vspace{-4mm}
\end{table}

Our evaluation shows that the two categories of rules work well in practice (Section~\ref{subsec:rq2}).
Meanwhile, our condition-action rules are intended to be practical rather than exhaustive.
Initially, we observed that agents usually prepare a prompt shortly before writing to the LLM context window.
Therefore, we only have Rule~8 for the PK region, since LLMs are treated as external information sources in \tool.
However, our case studies later showed that prompt assembly can also begin shortly after reading any information source, without immediately passing the partial prompt to the LLM context window. We added Rule~9 to cover this case.

\section{Query Processing}
\label{sec:query-stage}

Given a constructed \cldg, the analyst investigates a security incident by issuing queries against it using a query language.

\subsection{Query Language}\label{subsec:query-language}

The query language implements the \textit{region-level query interface} element in the abstraction (Section~\ref{subsec:problem-scope}).
The full grammar is provided in Appendix~\ref{apdx:spec-grammar}.

Each query begins with the keyword \texttt{CHECK}, followed by one or more atomic formulas joined by logical connectives.
An atomic formula consists of the predicate \texttt{DEPENDS\_ON}, which asserts a taint data flow between two terms.
A term can be either a region name or a program variable identifier, allowing queries at the region level or the variable level.
Logical connectives include \texttt{AND} and \texttt{OR}.
For example, \texttt{CHECK (PK DEPENDS\_ON PA) AND (PA DEPENDS\_ON PK)} is a commonly used region-level query for backtracking IPI incidents by an analyst (Table~\ref{tab:rq4}).
The atom \texttt{PK DEPENDS\_ON PA} asks whether data from the {\tt PA} region (e.g., the crawled webpage), can flow into the {\tt PK} region (e.g., to assemble a prompt). Similarly, the atom \texttt{PA DEPENDS\_ON PK} asks whether data in the PK region (e.g., the assembled prompt) can flow back to the PA region (e.g., the LLM takes the prompt as input and the generated content flows to write a document).

The language also provides a chaining operator $\rightarrow$: the formula after $\rightarrow$ is evaluated against the subgraph returned by the preceding formula, allowing the analyst to progressively narrow down the variable-level results.
For example, after writing the aforementioned region-level query, an analyst can write $\rightarrow$, followed by a second region/variable-level query.
\tool first evaluates the first query; if the result is not empty, it continues to evaluate the second query against the result of the first query (Section~\ref{subsec:query-evaluation}).

\subsection{Query Evaluation}
\label{subsec:query-evaluation}
This subsection implements the \textit{region-to-variable query mapping} element of abstraction (Section~\ref{subsec:problem-scope}).

\noindent\textbf{Region-Level Query Evaluation.}
A region-level query, such as \texttt{CHECK PA DEPENDS\_ON RR}, returns a data dependency graph over function nodes through two steps:
(1) identify the variables in $R_{\mathrm{src}}$ functions and the variables in $R_{\mathrm{sink}}$ functions by using $\lambda$ and $E_{\mathrm{cross}}$;
(2) compute forward reachability from the $R_{\mathrm{src}}$ variables and backward reachability from the $R_{\mathrm{sink}}$ variables in the variable plane, producing two sets of reachable variable nodes.
The intersection of these two sets contains all variables that lie on some taint path from $R_{\mathrm{src}}$ to $R_{\mathrm{sink}}$.
An edge $(f_1, f_2)$ appears in the result if a taint edge in $E_{\mathrm{var}}$ connects a variable in $f_1$ to a variable in $f_2$ within this intersection.
When a query contains logical connectives, \texttt{AND} and \texttt{OR} combine result graphs by intersection and union of their node and edge sets, respectively.

\noindent\textbf{Variable-Level Query Evaluation.}
In a variable-level query, the source and sink are specified by their node identifiers in the variable plane (visible when the analyst inspects a function's variables in the region-level query result).
For example, \texttt{CHECK node.5110 DEPENDS\_ON node.4216} queries whether there is a taint path from variable \texttt{4216} to variable \texttt{5110}.
To evaluate the variable-level query, a naive graph slicing of the \cldg variable plane may include paths that are not feasible in any real execution.
\tool uses the application's call graph and control flow graphs constructed during \cldg construction to enumerate only the control-flow feasible taint paths between the source and sink variables.
The returned paths are variable-level taint paths.
As mentioned in Section~\ref{subsec:cep-def}, the taint paths can be naturally mapped to the corresponding function-level paths.

\vspace{-1mm}
\section{Implementation}
\label{sec:implementation}
 We implement \tool as a Python program ($\sim$16 KLoC) on top of Joern~(v4.0.360)~\cite{CodePropertyGraph}, with type inference enhancement by Pyrefly~(v0.45.0)~\cite{FacebookPyrefly2026}.
We address three limitations of Joern that cause false negatives in taint propagation.
First, Joern does not parse generator expressions; we rewrite \texttt{yield} and \texttt{yield from} to \texttt{return} before analysis.
Second, taint stored in class instance variables (e.g., \texttt{self.x}) is not tracked across functions;
we propagate it explicitly via control-flow graph reachability.
Third, Joern's call graph is incomplete under dynamic dispatch. We first incorporate type-informed indirect call resolution by patching Pyrefly~\cite{FacebookPyrefly2026}; then we infer the targets of indirect calls by following Xia et al.~\cite{xiaDEEPTYPERefiningIndirect2024} and Liu et al.~\cite{liuImprovingIndirectcallAnalysis2024} with two enhancements: context-sensitive type chain tracking in place of a global type-chain map, and value set analysis in place of full data-flow analysis.

\section{Case Studies}
\label{sec:case-studies}
We now discuss our experience applying \tool to backtracking security incidents on real-world agent applications.

\noindent{\bf Selected Applications.}
We have chosen five actively maintained open-source agent applications (Table~\ref{tab:subjects}): two single-agent applications (SAAs) and three multi-agent applications (MAAs) with substantial real-world adoption, from four major domains, i.e., deep research (DR), conversational assistant (CA), coding copilot (CC), and software development (SD).

\begin{table}[htbp]\footnotesize
\centering
\renewcommand{\arraystretch}{0.9}
\vspace{-2mm}
\caption{Selected agent applications.}
\vspace{-3mm}
\label{tab:subjects}
\begin{tabular}{l@{\hspace{5pt}}c@{\hspace{10pt}}rrrl}
\hline
\textbf{Application} & \textbf{Domain} &
\textbf{Stars} & \textbf{LoC} &
\textbf{Version} & \textbf{Type} \\
\hline
GPT-Researcher~\cite{GPTResearcher2024} & DR & 24,027 & 9,317   & v3.3.3 & MAA \\
Quivr~\cite{Quivr2024}          & CA & 38,575 & 2,615   & v0.0.33 & SAA \\
aider~\cite{Aider2026}          & CC & 38,237 & 25,945  & v0.86.0 & SAA \\
MetaGPT~\cite{MetaGPT2026}        & SD & 59,192 & 43,284  & v0.8.2  & MAA \\
SWE-agent~\cite{SWEagent2026}      & SD & 17,716 & 8,995   & v1.1.0  & MAA \\
\hline
\end{tabular}
\end{table}

\begin{table*}[!t]\footnotesize
\centering
\renewcommand{\arraystretch}{0.9}
\caption{CEP backtracking results.
UI means whether the attack is launched through user interface or not.
\#Queries denotes the number of queries submitted per diagnosis phase.
\#Paths is the number of function-level paths returned by the variable-level query.
\#Func is the number of functions on the path matching the actual attack execution.
\#Cand.\ is the number of CEP candidates identified by the analyst from the returned paths.
Succ.\ indicates whether the diagnosis satisfies the success criteria.}
\label{tab:case-study}
\vspace{-3mm}
\begin{tabular}{ll c | l rrr | r c l}
\hline
\multicolumn{3}{c|}{\textbf{Attack Description}} &
\multicolumn{4}{c|}{\textbf{Diagnosis Detail}} &
\multicolumn{3}{c}{\textbf{Diagnosis Result}} \\
\cmidrule(lr){1-3} \cmidrule(lr){4-7} \cmidrule(lr){8-10}
\textbf{Application} & \textbf{Attack Class} & \textbf{UI} &
\textbf{Phase} & \textbf{\#Queries} & \textbf{\#Paths} & \textbf{\#Func} &
\textbf{\#Cand.} & \textbf{Succ.} & \textbf{CEP Description} \\
\hline
GPT-Researcher
  & IPI            & No & A & 2 &  3  & 13 & 1 & \checkmark & Obtaining web search result \\
  & MC-Poison (KB) & No & A & 2 & 4 & 19 & 1 & \checkmark & Loading knowledge document \\
\hline
\multirow{6}{*}{Quivr}
  & \multirow{2}{*}{MC-Poison (M)}
                   & \multirow{2}{*}{Yes}
                        & A & 2 & 1 & 6
                                    & \multirow{2}{*}{1} & \multirow{2}{*}{\checkmark} & \multirow{2}{*}{Receiving chat message} \\
  &                &     & B & 2 & 2 & 8  &   &            & \\
\cline{2-10}
  & \multirow{2}{*}{MC-Poison (KB)}
                   & \multirow{2}{*}{Yes}
                        & A & 2 &  2  & 9
                                    & \multirow{2}{*}{1} & \multirow{2}{*}{\checkmark} & \multirow{2}{*}{Receiving file upload} \\
  &                &     & B & 2 &  1 &  13  &   &            & \\
\cline{2-10}
  & \multirow{2}{*}{IPI + MC-Poison (KB)}
                   & \multirow{2}{*}{Yes}
                        & A & 2 & 1 & 9
                                    & \multirow{2}{*}{1} & \multirow{2}{*}{\checkmark} & \multirow{2}{*}{Crawling URL content} \\
  &                &     & B & 2 & 1 & 10 &   &            & \\
\hline
aider
  & IPI            & No  & A & 2 & 4  &  13  & 1 & \checkmark & Reading code file \\
\hline
MetaGPT
  & IPI            & No  & A & 2 & 2  &  14  & 1 & \checkmark & Reading agent planned spec file \\
\hline
SWE-agent
  & IPI            & No  & A & 2 & 1 & 21 & 1 & \checkmark & Fetching GitHub issue \\
\hline
\noalign{\vskip 3pt}
\multicolumn{10}{l}{\scriptsize An MC-Poison attack is not applicable to aider, MetaGPT, and SWE-agent; IPI attack alone is not applicable to Quivr because any external content (e.g., crawled webpages) is} \\
\multicolumn{10}{l}{\scriptsize first transformed into vectors, saved to the vector database, and only reaches the LLM after retrieval, making it MC-Poison.} \\
\vspace{-6mm}
\end{tabular}
\end{table*}

\noindent{\bf Investigated Security Incidents.}
Our eight investigated incidents are diverse in attack entry points and attack classes.
As shown in Table~\ref{tab:case-study}, we cover both IPI~\cite{greshakeNotWhatYouve2023} and MC-Poison \cite{dongPracticalMemoryInjection2025,chenAgentPoisonRedteamingLLM2024} attack classes, where MC-Poison is further divided into the subclasses of memory poisoning (M) and knowledge-base poisoning (KB); one incident is on a compound attack that combines IPI and MC-Poison~(KB).
Beyond directly modifying files on the file system (e.g., the GPT-Researcher MC-Poison~(KB) incident in the motivating example), we also consider attack scenarios where the malicious data enters through the application's user interface: for instance, in the Quivr MC-Poison~(M) incident, the attacker sent a crafted chat message that poisoned the memory of the agent.

\subsection{Shared Backtracking Workflows}

\textit{The analyst's backtracking workflows are similar across the investigated incidents, despite the diversity in attack classes and entry points.}
Specifically, the analyst
(i)~identifies the suspect agent functionality from the trajectory log,
(ii)~submits a region-level query,
(iii)~inspects the resulting graph to identify variable-level taint source/sink candidates,
(iv)~selects the source/sink to submit a variable-level query,
and (v) identifies the CEP and the corresponding paths.
Typically, a CEP identification requires two queries (Table~\ref{tab:case-study}): one region-level query to narrow down the scope and one variable-level query to pinpoint the CEP.
We show in Section~\ref{subsec:graph-inspect} that inspecting the resulting graph of a region-level query usually narrows down the source/sink candidates for a variable-level query to one or two, respectively.
With the knowledge obtained from the trajectory logs, the analyst can have a suspect ranking for source and sink candidates (detailed in Appendix~\ref{apdx:case-studies}). In our case studies, these help the analyst submit the correct variable-level query with minimal trials.
Besides, for attack incidents that span multiple phases in separate executions (e.g., Quivr MC-Poison (M)), the analyst repeats steps (i)--(iv) for each phase before identifying the CEP.

\textit{We observe that the variable-level query result, presented as a set of function-level paths, is typically sufficient for the analyst to identify and understand the CEP without inspecting the underlying variable-level taint propagation paths.}
In the Quivr MC-Poison~(M) incident, the attack spans two phases (Section~\ref{sec:background}). The analyst submitted one variable-level query for each phase and obtained two valid function-level paths.
Although the two paths correspond to two executions within different time periods, the analyst identified the shared function call arguments (i.e., the \texttt{chat\_history} table of Supabase) in the source node (blue, {\tt RR} region) of Phase~A path and sink node (blue) of Phase~B path.
This indicated that the malicious content can be first written to the agent memory in a chat with the attacker; then be retrieved back in another chat with the user.
Therefore, the CEP is identified as the function call to the user-chat interface. The full incident and its backtracking process are detailed in Appendix~\ref{subsec:case-memory}.
For simpler single-phase incidents such as the motivating GPT-Researcher MC-Poison~(KB) case, the analyst can directly identify the CEP by examining the function-level paths returned by a variable-level query
(detailed in Appendix~\ref{subsec:case-kb}).

\begin{figure*}[htbp]
    \centering
    \vspace{-1mm}
    \begin{subfigure}{\textwidth}
        \centering
        \includegraphics[width=\linewidth]{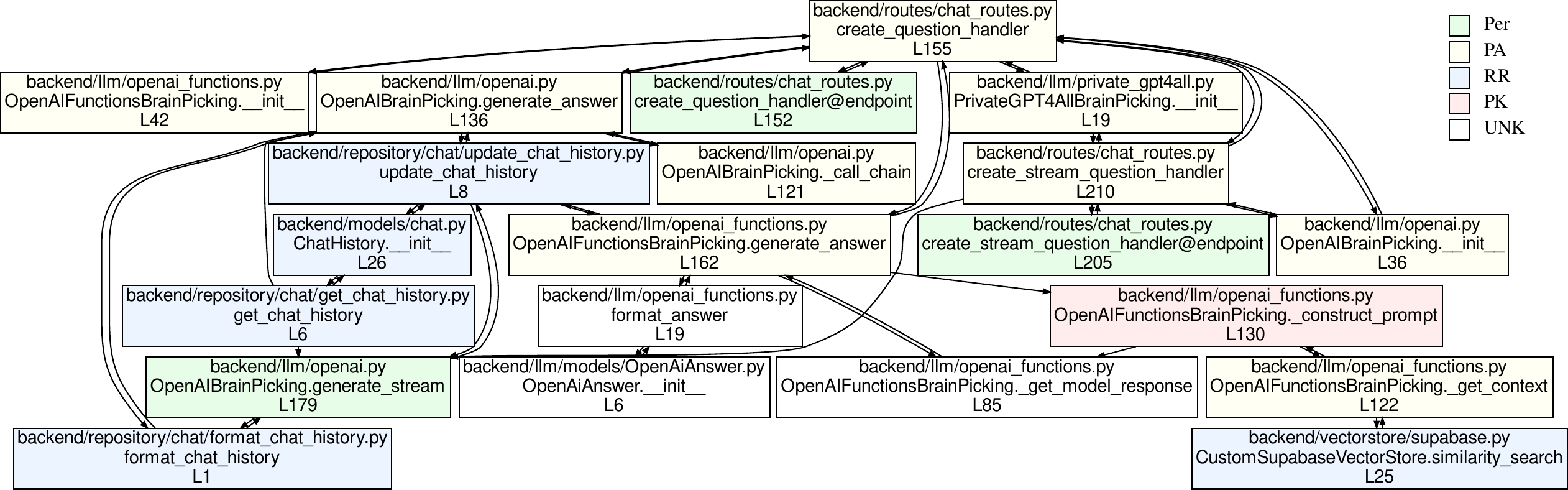}
        \vspace{-5mm}
        \caption{Region-level query result of Quivr MC-Poison (M) incident (phase A). Note that the node IDs are eliminated for simplicity.}
        \label{fig:case-study-region}
    \end{subfigure}

    \vspace{1mm}

    \begin{subfigure}{\textwidth}
        \centering
        \includegraphics[width=\linewidth]{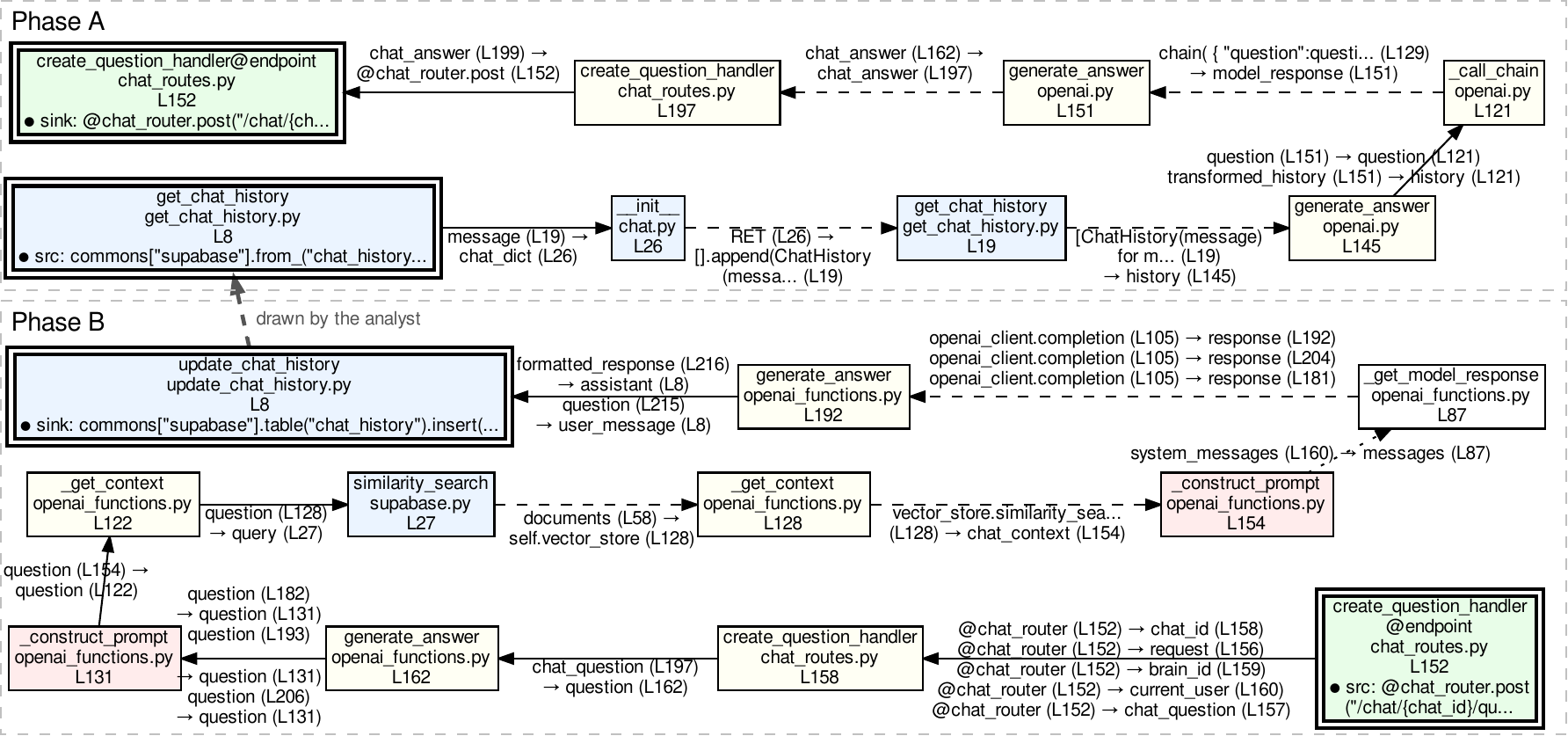}
        \vspace{-4.5mm}
        \caption{Function-level paths returned by two variable-level queries.
Each node shows the function name, the file name of the enclosing Python file, and the line number of the first taint-path variable within that function.
A solid edge denotes a call from the enclosing function to the callee; a dashed edge denotes a return from the callee back to the caller.
Each edge label shows the mapping from a variable in the source function to a variable in the target function, with the line number indicating the location of that variable in the corresponding Python file.}
        \label{fig:case-study-variable}
    \end{subfigure}

    \vspace{-3mm}
    \caption{Backtracking the CEP for the Quivr MC-Poison (M) incident.}
    \label{fig:case-study}
    \vspace{-4mm}
\end{figure*}

\subsection{Common Practices for Graph Inspection}\label{subsec:graph-inspect}
We report the common practices in step~(iii) graph inspection, which replaces the time-consuming source code inspection:

\textit{(a)~Not all nodes in the resulting graph of a region-level query require inspection.}
The graph contains nodes from the regions specified in the query as well as nodes from other regions that appear because they lie on the taint paths between the queried regions.
Each node is colored by its region, so the analyst usually focuses only on nodes whose colors correspond to the queried regions and skips the rest.
For example, in the Quivr MC-Poison~(M) incident (phase~A), the region-level query returns a graph of 21 function nodes, but only the 3 Per nodes (green) and 5 RR nodes (blue) match the queried regions and need examination (Figure~\ref{fig:case-study}(a)).

\textit{(b)~Among the relevant nodes, it requires examining only a small number (usually less than 10) of function names and an even smaller number (usually 2 or 3) of variable-plane taint subgraphs
corresponding to the functions.}
As in Figure~\ref{fig:case-study}(a), to identify the taint sink, the analyst examined the names of 3 green nodes of {\tt Per} region, found \texttt{create\_question\_handler} relevant to the incident, and clicked it to reveal a variable-plane taint subgraph
of a single node (Figure~\ref{fig:cs1-click} in Appendix) that is identified as the sink candidate.
The analyst then takes a similar process to identify the taint source candidate from the 5 blue nodes of {\tt RR} region.
\textbf{The entire process requires only examining 8 function names and 2 taint subgraphs, which is a huge manual effort reduction compared to 2,615 LoC of Quivr's source code inspection (Table~\ref{tab:subjects}).}
\tool also provides features that help the analyst handle larger graphs more efficiently, which are detailed in Appendix~\ref{subsec:case-kb} (Figure~\ref{fig:cs2-a}) and Appendix~\ref{subsec:case-compound} (Figure~\ref{fig:cs4-region-query}).

\vspace{-1mm}
\subsection{Notable Findings on Individual Applications}
\label{subsec:case-study-unique}
We report other notable observations from the case studies.

\textit{(a) An agentic loop within an execution does not prevent \tool from identifying the CEP, as long as both the CEP and the function call corresponding to the attack effect lie outside the loop.}
In the SWE-agent IPI incident, for example, each iteration of the agent loop in the attack execution queries the LLM for the next tool to invoke, executes that tool, and feeds the result back.
However, since the loop appears in the middle of the returned taint path as pass-through functions and static taint analysis tracks data flow through variables in the source code, the taint propagation from source to sink remains intact, and \tool correctly identifies the CEP.

\textit{(b) When the source and sink functions are generic utilities shared by two or more agent functionalities in the application, specifying an intermediate variable as a waypoint is helpful in constraining the returned paths to the incident-relevant subset.}
In both aider and MetaGPT, a single pair of file read/write functions (\texttt{read\_text}/\texttt{write\_text} in aider; \texttt{aread}/\texttt{awrite} in MetaGPT) serves as the major file I/O interface across the entire codebase.
A standard source-to-sink query between these functions returns a large set of taint paths between file-read and file-write in the application, the majority of which are unrelated to the incident.
We observe a good practice for getting a more incident-related result, i.e., identifying an intermediate function from the region-level query result and specifying a variable within it as a waypoint.
These functions are identifiable because trajectory logs record semantically meaningful steps (e.g., named commands and task-level actions) that map directly to function-plane nodes in the \cldg.

\textit{(c) \tool can handle an attack incident rooted in inter-agent communication manipulation.}
In MetaGPT, the identified CEP (\texttt{aread}) reads a specification file that the upstream architect agent planned and wrote to disk (the file is compromised by an attacker); the downstream engineer agent reads malicious data and generates code according to a manipulated task.
The coordination mechanism between agents (i.e., shared files on disk) functions exactly as designed.
However, due to a lack of verification for the artifacts produced by a previous execution, the IPI happened across agent boundaries through the application's normal coordination mechanism.

\section{Evaluation}
\label{sec:evaluation}

\subsection{Research Questions (RQs)}
\label{subsec:rqs}

We evaluate \tool on the same set of security incidents as in Section~\ref{sec:case-studies} to answer the following research questions:
(RQ1) Can analysts use \tool to correctly identify the CEPs and the corresponding paths of the eight security incidents?
(RQ2) How accurately does the region inference algorithm assign functions to regions?
(RQ3) To what extent does \tool reduce the portion of the \cldg that analysts must inspect compared to using static taint analysis alone?
(RQ4) What are the execution time and memory costs of \tool?
(RQ5) How much developer knowledge is needed for region inference?
(RQ6) How useful is the LLM-generated report?

\vspace{-1mm}
\subsection{Experimental Setup}
\label{subsec:setup}

All experiments are conducted on a machine with an Intel Core Ultra 9 285 CPU, 128\,GB RAM, running Ubuntu 24.04.
We use the human-validated
\kb (Section~\ref{subsec:dev-kb}) for all RQ evaluations.

\vspace{-1mm}
\subsection{CEP + Path Identification Correctness (RQ1)}
\label{subsec:rq1}

{\bf Success Criteria.}
A diagnosis is considered successful \emph{iff} the ground truth CEP and corresponding taint propagation path (the one matching the actual attack execution) defined in Section~\ref{subsec:cep-def} are identified.

\noindent{\bf Results.}
As shown in Table~\ref{tab:case-study}, an analyst using \tool correctly identified the CEPs of all eight incidents, each having a single CEP candidate.
The analyst can identify the CEPs by submitting two queries in five incidents and four queries in the other three incidents.

\noindent{\bf Comparison with Existing Tools.}
We also evaluate existing static taint analyzers, i.e., Joern~\cite{JoernBugHunters2025}, CodeQL~\cite{CodeQL2025} and Pysa~\cite{PysaStaticAnalysis}, on the same incidents.
Since they lack the intermediate abstraction compared to \tool, we assume the tools are provided with the ground-truth source (i.e., the CEP) and sink as input and evaluate
if
analysts can use the tools to verify the correctness of the CEP.
Specifically, for Joern and CodeQL: \emph{Success} means the correct taint propagation path is identified and \emph{Fail} means the opposite;
For Pysa: \emph{Success} means at least one reported function call lies on
the correct taint propagation path; \emph{Fail} means the expected function on the correct path is not reported.

\begin{table}[htbp]
\footnotesize
\centering
\setlength{\tabcolsep}{2pt}
\renewcommand{\arraystretch}{0.9}
\vspace{-2mm}
\caption{Diagnostic outcomes of Joern, CodeQL, and Pysa}
\vspace{-3mm}
\label{tab:baseline}
\begin{tabular}{
  l
  @{\hspace{10pt}}
  p{0.36\columnwidth}
  @{\hspace{-2pt}}
  ccc
}
\hline
\textbf{Application} &
\textbf{Attack} &
\textbf{Joern} &
\textbf{CodeQL} &
\textbf{Pysa} \\
\hline
GPT-Researcher
  & IPI                    & Fail & Fail & Fail \\
  & MC-Poison (KB)         & Fail & Fail & Fail \\
Quivr
  & MC-Poison (M)          & Fail & Fail & Fail \\
  & MC-Poison (KB)         & Fail & Fail & Fail \\
  & IPI + MC-Poison (KB)   & Fail & Fail & Fail \\
aider
  & IPI                    & Fail & Fail & Fail \\
MetaGPT
  & IPI                    & Fail & Fail & Success \\
SWE-agent
  & IPI                    & Fail & Fail & Fail \\
\hline
\end{tabular}
\vspace{-2mm}
\end{table}

\textit{As shown in Table~\ref{tab:baseline}, the tools fail on most incidents, primarily due to unresolved indirect call targets, missing data dependencies through language constructs, API calls, and first-writes-later-reads of object attributes, files, or databases (Appendix~\ref{apdx:tool-failures}).}
For example, in SWE-agent IPI, Joern fails to propagate taint through nested dictionary accesses.
In Quivr MC-Poison~(M), CodeQL fails to propagate taint from an LLM API call argument to its returned value because it does not model the corresponding library.
In GPT-Researcher MC-Poison~(KB), Pysa fails to resolve the target of an application function call, interrupting taint propagation from a variable used within the callee to a returned variable in the caller.

\begin{table*}[htbp]\footnotesize
\centering
\setlength{\tabcolsep}{7.7pt}
\renewcommand{\arraystretch}{0.9}
\vspace{-1mm}
\caption{Inspection-scope reduction via region-level queries.}
\label{tab:rq4}
\vspace{-3mm}
\begin{tabular}{llcccc}
\hline
\textbf{Application} &
\textbf{Attack Class [(Subclass)]} &
\textbf{Region-level Query} &
\textbf{$|V_q|$} &
\textbf{$|V_{\mathrm{var}}|$} &
\textbf{$|V_q|/|V_{\mathrm{var}}|$} \\
\hline
GPT-Researcher
& IPI
& \texttt{CHECK (PK DEPENDS\_ON PA) AND (PA DEPENDS\_ON PK)}
& 2,605 & 7,284 & 35.76\% \\
& MC-Poison (KB)
& \texttt{CHECK (PK DEPENDS\_ON RR) AND (PA DEPENDS\_ON PK)}
& 2,213 & 7,284 & 30.38\% \\
\hline
\multirow{6}{*}{Quivr}
& \multirow{2}{*}{MC-Poison (M)}
& \texttt{CHECK (PK DEPENDS\_ON Per) AND (RR DEPENDS\_ON PK)}
& 157 & 2,018 & 7.78\% \\
&
& \texttt{CHECK (PK DEPENDS\_ON RR) AND (Per DEPENDS\_ON PK)}
& 131 & 2,018 & 6.49\% \\
\cline{2-6}
& \multirow{2}{*}{MC-Poison (KB)}
& \texttt{CHECK RR DEPENDS\_ON RR}
& 321 & 2,018 & 15.91\% \\
&
& \texttt{CHECK (PK DEPENDS\_ON RR) AND (PA DEPENDS\_ON PK)}
& 224 & 2,018 & 11.10\% \\
\cline{2-6}
& \multirow{2}{*}{IPI + MC-Poison (KB)}
& \texttt{CHECK RR DEPENDS\_ON PA}
& 378 & 2,018 & 18.73\% \\
&
& \texttt{CHECK (PK DEPENDS\_ON RR) AND (Per DEPENDS\_ON PK)}
& 131 & 2,018 & 6.49\% \\
\hline
aider
& IPI
& \texttt{CHECK (PK DEPENDS\_ON PA) AND (PA DEPENDS\_ON PK)}
& 4,649 & 13,820 & 33.64\% \\
\hline
MetaGPT
& IPI
& \texttt{CHECK (PK DEPENDS\_ON PA) AND (PA DEPENDS\_ON PK)}
& 3,749 & 19,135 & 19.59\% \\
\hline
SWE-agent
& IPI
& \texttt{CHECK (PK DEPENDS\_ON PA) AND (PA DEPENDS\_ON PK)}
& 775 & 8,105 & 9.56\% \\
\hline
\end{tabular}
\vspace{-2mm}
\end{table*}

\textit{The limited data-flow information in Pysa's output is insufficient for the analyst to identify the CEPs.}
For example, in MetaGPT IPI incident, {\tt Pysa} reports four function calls, of which only one (\texttt{WriteCode.run} at \texttt{write\_code.py:142}) lies on the correct path.
When comparing against the path of 14 functions
returned by \tool, \texttt{WriteCode.run} is the 4th function on the path: 3 functions from the CEP \texttt{aread} and 10 functions from the attack effect \texttt{awrite}.
Therefore, the analyst would need extra analysis to identify the CEP.

\subsection{Region Inference Accuracy (RQ2)}
\label{subsec:rq2}

Table~\ref{tab:rq2-main} reports region inference accuracy per application. For each function in $V_{\mathrm{fn}}$, the algorithm classifies it into one of the four regions (excluding UNK).
The description of the region set and the ground truth region distribution across the applications can be found in Appendix~\ref{apdx:region-set}.

\begin{table}[htbp]\footnotesize
\centering
\setlength{\tabcolsep}{8pt}
\renewcommand{\arraystretch}{0.85}
\vspace{-1mm}
\caption{Region inference accuracy per application.}
\label{tab:rq2-main}
\vspace{-3mm}
\begin{tabular}{lrrrrr}
\hline
\textbf{Application} &
\textbf{TP} &
\textbf{FP} &
\textbf{FN} &
\textbf{Precision} &
\textbf{Recall} \\
\hline
GPT-Researcher  & 196 & 48 & 42 & 0.8033 & 0.8235 \\
Quivr     & 50 & 12 & 9 & 0.8065 & 0.8475 \\
aider     & 229 & 100 & 95 & 0.6960 & 0.7068 \\
MetaGPT   & 571 & 101 & 78 & 0.8497 & 0.8798 \\
SWE-agent & 140 & 20 & 20 & 0.8750 & 0.8750 \\
\hline
\end{tabular}
\vspace{-3mm}
\end{table}

Across all five applications, precision and recall range from 0.70 to 0.88.
{\tt aider} exhibits the lowest accuracy because the coding copilot provides a set of commands to use interactively (e.g., adding files, prompting
the LLM, committing edits) rather than the application
chaining them automatically.
\tool produces short, disjoint taint paths for each {\tt aider} command, and labels propagated from an API call along one taint path (detailed in Section~\ref{subsec:region-inference}) cannot reach functions that belong to a different command's taint path.

\subsection{Inspection Scope Reduced by Region-level Queries (RQ3)}
\label{subsec:rq3}

Region-level queries can significantly reduce the graph inspection scope for the analyst.
We measure this by comparing $|V_q|$, the number of variable-plane nodes in the region-level query result, against $|V_{\mathrm{var}}|$, the full variable plane of the \cldg.
Lower $|V_q|/|V_{\mathrm{var}}|$ means the analyst encounters a smaller fraction of the taint graph.
As shown in Table~\ref{tab:rq4}, across all incidents, region-level queries reduce the variable plane to between 6.49\% and 35.76\% of its original size. We note that:

\textit{(a)~PK is a key factor in this reduction and appears in 9 of the 11 region-level queries.}
This is because PK is where data from other regions converges to be assembled into the LLM prompt, and it typically contains very few functions (e.g., 1 in Quivr, 2 in aider, 3 in SWE-agent; more details in Appendix~\ref{apdx:region-set}).
When a query involves PK, the returned graph is limited to function nodes that are transitively connected to this small set of PK functions, which substantially reduces the number of variable-plane nodes the analyst must inspect.

\textit{(b)~The reduction ratio also depends on the size of the regions involved in the query, which varies across applications.}
The IPI incidents across the four applications use the same query \texttt{CHECK (PK DEPENDS\_ON PA) AND (PA DEPENDS\_ON PK)}, yet the ratios range from 9.56\% (SWE-agent) to 35.76\% (GPT-Researcher).
Regarding the ratio difference, we observe that a larger PA in an application often gives a larger result graph: PA contains 133 of 244 functions in GPT-Researcher but only 64 of 160 in SWE-agent (more details in Appendix~\ref{apdx:region-set}).

\subsection{Performance of \tool (RQ4)}
\label{subsec:rq4}

Table~\ref{tab:performance} reports the execution time and peak memory usage (Mem.) for each incident.
Three stages of \tool execution time are reported separately: \cldg construction (CLDG), region inference (Infer.), and the queries evaluation time of each application averaged on corresponding incidents ($\bar{Q}$).

\begin{table}[htbp]\footnotesize
\centering
\setlength{\tabcolsep}{6.3pt}
\renewcommand{\arraystretch}{0.9}
\vspace{-2mm}
\caption{Performance of \tool.}
\label{tab:performance}
\vspace{-3mm}
\begin{tabular}{lcccc}
\hline
\textbf{Application} &
\textbf{CLDG (s)} &
\textbf{Infer.\ (s)} &
\textbf{$\bar{Q}$ (ms)} &
\textbf{Mem.\ (GB)} \\
\hline
GPT-Researcher & 19.20 & 0.29 & 36.00 & 2.51 \\
Quivr          & 6.38 & 0.03 & 5.63 & 2.56 \\
aider          & 109.09 & 1.97 & 89.38 & 5.57 \\
MetaGPT        & 72.15 & 0.77 & 67.50 & 7.00 \\
SWE-agent      & 18.74 & 0.14 & 9.38 & 4.27 \\
\hline
\end{tabular}
\vspace{-3mm}
\end{table}

\cldg construction dominates execution time, accounting for over 95\% of the total in every application.
\cldg construction and region inference are one-time offline costs.
The set of query evaluations is the only per-incident cost, and they were completed in under 90\,ms across all five applications on average.
Peak memory remains under 7\,GB.

\subsection{Required Developer Knowledge Amount (RQ5)}
\label{subsec:rq5}

Table~\ref{tab:developer-overhead} reports the number of application functions and the numbers of \texttt{EXT}/\texttt{INT} labels, with the total number of labels also expressed as a percentage of application functions.
Across the five applications, only 2.4\%--8.9\% of functions receive labels, indicating that \tool requires only a small amount of application-specific knowledge from developers.

\begin{table}[htbp]\footnotesize
\centering
\setlength{\tabcolsep}{4pt}
\renewcommand{\arraystretch}{0.9}
\vspace{-2mm}
\caption{Application functions and labels received.}
\label{tab:developer-overhead}
\vspace{-3mm}
\begin{tabular}{lccccc}
\hline
\textbf{Application} &
\textbf{App. funcs} &
\textbf{EXT} &
\textbf{INT} &
\textbf{Total} &
\textbf{\%} \\
\hline
GPT-Researcher & 607   & 35 & 6  & 41  & 6.8 \\
Quivr          & 246   & 7  & 15 & 22  & 8.9 \\
aider          & 837   & 51 & 3  & 54  & 6.5 \\
MetaGPT        & 2,872 & 84 & 35 & 119 & 4.1 \\
SWE-agent      & 713   & 17 & 0  & 17  & 2.4 \\
\hline
\end{tabular}
\vspace{-5mm}
\end{table}

\subsection{Utility of LLM-Generated Report (RQ6)}
\label{subsec:rq6}

\noindent{\bf Human Study Setup.}
We recruit four participants with different levels of agent-application knowledge and Python familiarity to examine whether the EXT/INT report benefits users with different backgrounds.
Each participant completes two 10-minute tests: manual labeling (M) and LLM report's label validation (L).
In the M test, participants follow the developer workflow described in Section~\ref{subsec:dev-kb}.
In the L test, participants are provided with the LLM-generated report.
We prepare two non-overlapping sets, A and B, each containing 200 functions from the five applications, so that participants do not encounter the same function twice.
We set up a one-to-one mapping between the four participants and the four test sequences: $M_A\!\to\!L_B$, $L_A\!\to\!M_B$, $M_B\!\to\!L_A$, and $L_B\!\to\!M_A$, where subscripts identify the function sets.
This counterbalancing
reduces the influence of the test order and differences between the function sets.

Before the tests, we provide a labeling guide and 15 functions for practice, followed by 15 functions to check their understanding.
We also provide an example showing participants how to use the report’s explanations and code references to validate its proposed labels.
We measure speed as the number of functions labeled during the given 10 minutes and accuracy in terms of the reference answers, including decisions that neither label is needed.

\noindent{\bf Results.}
Table~\ref{tab:rq6} reports speed changes relative to M ($\Delta$Speed), accuracy in L ($\mathrm{Acc.}_L$), and accuracy differences between L and M in percentage points ($\Delta$Acc.).
All four participants achieve higher labeling speed and accuracy in the L test.
Speed increases by 15.4--75.0\%, while accuracy improves by 8.3--50.4 percentage points, reaching 88.9--100\%.

\begin{table}[htbp]\footnotesize
\centering
\setlength{\tabcolsep}{2pt}
\renewcommand{\arraystretch}{0.9}
\vspace{-3mm}
\caption{Effect of LLM reports on function labeling.}
\label{tab:rq6}
\vspace{-3mm}
\begin{tabular*}{\columnwidth}{@{\extracolsep{\fill}}l|ccc|rrr@{}}
\hline
\textbf{ID} &
\textbf{First} &
\textbf{Agent} &
\textbf{Python} &
\textbf{$\Delta$Speed} &
\textbf{Acc.$_L$} &
\textbf{$\Delta$Acc.(pp)} \\
\hline
P1 & $M_A$ & Low  & Med. & $+15.4\%$ & $88.9\%$  & $+50.4$ \\
P2 & $L_A$ & Med. & Low  & $+55.6\%$ & $92.9\%$  & $+32.9$ \\
P3 & $M_B$ & High & High & $+26.7\%$ & $92.1\%$  & $+22.1$ \\
P4 & $L_B$ & Low  & High & $+75.0\%$ & $100.0\%$ & $+8.3$  \\
\hline
\end{tabular*}
\vspace{-5.5mm}
\end{table}
\section{Discussion}
\label{sec:discussion}

\noindent{\bf Framework-Induced Indirect Call Resolution.}
Our static analysis cannot resolve framework-induced indirect call targets (e.g., LangChain~\cite{LangChain2025} tool dispatch).
This causes the corresponding taint propagation path to be missing from the \cldg.
Addressing this limitation requires either framework-specific modeling or dynamic analysis, which is future work.

\noindent\textbf{Agents with An LLM-driven Agentic Loop.}
Autonomous agents such as OpenClaw~\cite{OpenClaw2026} and Hermes~\cite{Hermes2026} are driven by an agentic loop in which, for each iteration, the LLM decides at runtime which agent functionality to invoke next.
For such agents, static analysis alone cannot determine the CEP function call (e.g., which tool introduced the malicious data), except for the case discussed in Section~\ref{subsec:case-study-unique}, Finding~(b).
This remains an area for future research.

\section{Related Work}
\label{sec:related}
\FloatBarrier{}

{\bf Root Cause Analysis for Software Systems.}
Diagnosing the root cause of exploits and failures in software systems is the first step to repairing programs.
Most existing works are based on backtracking~\cite{xuPostmortemProgramAnalysis2017,cuiRETracerTriagingCrashes2016,xuCREDALLocatingMemory2016,changFirmrcaPostFuzzingAnalysis2025}, reconstruction~\cite{cuiREPTReverseDebugging2018,geReverseDebuggingKernel2020,yagemannARCUSSymbolicRoot2021,heFreeWillAutomaticallyDiagnosing2022}, or statistical analysis~\cite{blazytkoAURORAStatisticalCrash2020,xuRacingNegativeForce2024,parkBenzenePracticalRoot2024,kasikciFailureSketchingTechnique2015,kasikciLazyDiagnosisInProduction2017}.
These approaches focus on diagnosing the buggy code location(s) triggering memory safety violations~\cite{xuPostmortemProgramAnalysis2017,cuiRETracerTriagingCrashes2016,xuCREDALLocatingMemory2016,changFirmrcaPostFuzzingAnalysis2025,cuiREPTReverseDebugging2018,yagemannARCUSSymbolicRoot2021,heFreeWillAutomaticallyDiagnosing2022,blazytkoAURORAStatisticalCrash2020,xuRacingNegativeForce2024,parkBenzenePracticalRoot2024} and data races~\cite{kasikciFailureSketchingTechnique2015,kasikciLazyDiagnosisInProduction2017}.
However, in agent application security incidents, the application code itself is not buggy in terms of condition expressions and statement orders. The problem is that the application lacks validation or sanitization at the boundary between untrusted external information sources and the application. Consequently, the attacker-controlled data is entered through a legitimate API call and propagated through the pipeline unimpeded.

\noindent{\bf Causality Analysis for APT Attacks.}
A large body of work investigates APT attacks by constructing attack provenance graphs from OS-level audit logs~\cite{kingBacktrackingIntrusions2003,milajerdiHOLMESRealTimeAPT2019,gaoAiqlEnablingEfficient,goyalRCAIDEmbeddingRoot2024}. Subsequent work refines the granularity of logs by partitioning long-running processes into finer execution units~\cite{maProTracerPracticalProvenance2016,maMPIMultiplePerspective2017,jiRAINRefinableAttack2017} or by fusing application-layer logs with system audit data to recover application semantics~\cite{hassanOmegaLogHighfidelityAttack2020}.
The logs and audit data used in these systems are sufficient for analyzing traditional APT attacks. In agent applications, however, analysts rely on trajectory logs and static analysis on program code. Trajectory logs record agent behaviors at the agent functionality abstraction layer but miss the app-specific
function call logging needed for pinpointing the CEP function calls.
Static analysis does not provide
the agent-functionality semantics that analysts need for reasoning.
The bridging between the agent functionalities and the program code is difficult due to a semantic gap.
Our work addresses this by implementing an intermediate abstraction that connects them.

\noindent{\bf Agent Application Attacks and Defenses.}
Agent attacks have targeted a wide range of commercial and open-source agents \cite{rehbergerTrustNoAI2024}, most notably IPI attacks \cite{greshakeNotWhatYouve2023,zhanInjecAgentBenchmarkingIndirect2024,debenedettiAgentDojoDynamicEnvironment2024,liuFormalizingBenchmarkingPrompt2024} and MC-Poison attacks \cite{chenAgentPoisonRedteamingLLM2024,wangBadAgentInsertingActivating2024}.
In real-world settings, adversaries often employ compound attacks that combine both methods across different stages of agent execution \cite{zhangAgentSecurityBench2025}.

Defenses have been proposed at multiple layers: data-instruction separation-based LLM hardening~\cite{chenStruQDefendingPrompt2025,chenSecAlignDefendingPrompt2025,wallaceInstructionHierarchyTraining2024} and prompt hardening~\cite{hinesDefendingIndirectPrompt2024} against IPI attacks, memory writes/retrieval-stage filtering against MC-Poison attacks \cite{xiangCertifiablyRobustRAG2026,weiAMemGuardProactiveDefense2025,sunilMemoryPoisoningAttack2026}, and application-level taint tracking and policy enforcement at runtime \cite{debenedettiDefeatingPromptInjections2025,wuIsolateGPTExecutionIsolation2025,costaSecuringAIAgents2025,shiProgentSecuringAIAgents2025,jiTamingVariousPrivilege2026}.
All these defenses, however, are preventive in nature and can still be bypassed. When an attack does succeed, we still need to backtrack the initial compromise of the incident.
In addition, detection of IPI attacks and MC-Poison attacks has been investigated in \cite{kimPromptFlowIntegrity2025,zhangAgentSentryMitigatingIndirect2026} and \cite{tanRevPRAGRevealingPoisoning2025,weiAMemGuardProactiveDefense2025}, respectively. However, detection and diagnosis (through backtracking) are two different problems.
\vspace{-5mm}
\section{Conclusion}
\label{sec:conclusion}
\FloatBarrier{}

We present \tool, a backtracking tool for security incidents in agent applications.
\tool implements a new intermediate abstraction as \cldg, which connects region-inferred functions with variable-level taint subgraphs, and provides the associated query language that
allows analysts to identify the CEPs with a small number of queries.
\tool constructs the region-to-variable mapping in the \cldg by rule-guided region inference, which only takes a small set of developer-provided labels.
Case studies on five applications and eight diverse attack incidents show that an analyst using \tool can correctly identify the CEPs of all incidents.

\bibliographystyle{ACM-Reference-Format}
\bibliography{paper}


\begin{thebibliography}{82}

%%% the Web address.

\ifx \showCODEN    \undefined \def \showCODEN     #1{\unskip}     \fi
\ifx \showISBNx    \undefined \def \showISBNx     #1{\unskip}     \fi
\ifx \showISBNxiii \undefined \def \showISBNxiii  #1{\unskip}     \fi
\ifx \showISSN     \undefined \def \showISSN      #1{\unskip}     \fi
\ifx \showLCCN     \undefined \def \showLCCN      #1{\unskip}     \fi
\ifx \shownote     \undefined \def \shownote      #1{#1}          \fi
\ifx \showarticletitle \undefined \def \showarticletitle #1{#1}   \fi
\ifx \showURL      \undefined \def \showURL       {\relax}        \fi
% invisible to TeX
\providecommand\bibfield[2]{#2}
\providecommand\bibinfo[2]{#2}
\providecommand\natexlab[1]{#1}
\providecommand\showeprint[2][]{arXiv:#2}

\bibitem[5ir(2025)]%
        {5ireCve202558357}
 \bibinfo{year}{2025}\natexlab{}.
\newblock \bibinfo{title}{5ire - {{CVE}}-2025-58357}.
\newblock
  \bibinfo{howpublished}{\url{https://nvd.nist.gov/vuln/detail/cve-2025-58357}}.
\newblock
\newblock
\shownote{(accessed 2026-02-01)}.


\bibitem[Aut(2025)]%
        {AutoGPT2025}
 \bibinfo{year}{2025}\natexlab{}.
\newblock \bibinfo{title}{{{AutoGPT}}}.
\newblock \bibinfo{howpublished}{\url{https://agpt.co/}}.
\newblock
\newblock
\shownote{(accessed 2025-09-01)}.


\bibitem[Cod(2025a)]%
        {CodePropertyGraph}
 \bibinfo{year}{2025}\natexlab{a}.
\newblock \bibinfo{title}{Code Property Graph Specification}.
\newblock \bibinfo{howpublished}{\url{https://cpg.joern.io/}}.
\newblock
\newblock
\shownote{(accessed 2025-02-07)}.


\bibitem[Cod(2025b)]%
        {CodeQL2025}
 \bibinfo{year}{2025}\natexlab{b}.
\newblock \bibinfo{title}{{{CodeQL}}}.
\newblock \bibinfo{howpublished}{\url{https://codeql.github.com/}}.
\newblock
\newblock
\shownote{(accessed 2025-09-01)}.


\bibitem[Cur(2025)]%
        {CursorCve202554130}
 \bibinfo{year}{2025}\natexlab{}.
\newblock \bibinfo{title}{Cursor - {{CVE}}-2025-54130}.
\newblock
  \bibinfo{howpublished}{\url{https://nvd.nist.gov/vuln/detail/CVE-2025-54130}}.
\newblock
\newblock
\shownote{(accessed 2026-02-01)}.


\bibitem[GPT(2025)]%
        {GPTResearcher2024}
 \bibinfo{year}{2025}\natexlab{}.
\newblock \bibinfo{title}{{{GPT}} Researcher}.
\newblock \bibinfo{howpublished}{\url{https://gptr.dev}}.
\newblock
\newblock
\shownote{(accessed 2026-09-01)}.


\bibitem[Joe(2025)]%
        {JoernBugHunters2025}
 \bibinfo{year}{2025}\natexlab{}.
\newblock \bibinfo{title}{Joern}.
\newblock \bibinfo{howpublished}{\url{https://joern.io/}}.
\newblock
\newblock
\shownote{(accessed 2025-09-01)}.


\bibitem[Lan(2025)]%
        {LangChain2025}
 \bibinfo{year}{2025}\natexlab{}.
\newblock \bibinfo{title}{{{LangChain}}}.
\newblock \bibinfo{howpublished}{\url{https://www.langchain.com}}.
\newblock
\newblock
\shownote{(accessed 2025-09-01)}.


\bibitem[Neu(2025)]%
        {NeuronCve202567510}
 \bibinfo{year}{2025}\natexlab{}.
\newblock \bibinfo{title}{Neuron - {{CVE}}-2025-67510}.
\newblock
  \bibinfo{howpublished}{\url{https://nvd.nist.gov/vuln/detail/CVE-2025-67510}}.
\newblock
\newblock
\shownote{(accessed 2026-02-01)}.


\bibitem[Qui(2025)]%
        {Quivr2024}
 \bibinfo{year}{2025}\natexlab{}.
\newblock \bibinfo{title}{Quivr}.
\newblock \bibinfo{howpublished}{\url{https://core.quivr.com/en/latest/}}.
\newblock
\newblock
\shownote{(accessed 2025-09-01)}.


\bibitem[Smo(2025)]%
        {Smolagents}
 \bibinfo{year}{2025}\natexlab{}.
\newblock \bibinfo{title}{{{SmolAgents}}}.
\newblock
  \bibinfo{howpublished}{\url{https://huggingface.co/docs/smolagents/en/index}}.
\newblock
\newblock
\shownote{(accessed 2025-09-01)}.


\bibitem[Aid(2026)]%
        {Aider2026}
 \bibinfo{year}{2026}\natexlab{}.
\newblock \bibinfo{title}{Aider}.
\newblock \bibinfo{howpublished}{\url{https://github.com/Aider-AI/aider}}.
\newblock
\newblock
\shownote{(accessed 2026-04-13)}.


\bibitem[Her(2026)]%
        {Hermes2026}
 \bibinfo{year}{2026}\natexlab{}.
\newblock \bibinfo{title}{{{Hermes Agent}}}.
\newblock \bibinfo{howpublished}{\url{https://hermes-agent.nousresearch.com/}}.
\newblock
\newblock
\shownote{(accessed 2026-02-01)}.


\bibitem[Met(2026)]%
        {MetaGPT2026}
 \bibinfo{year}{2026}\natexlab{}.
\newblock \bibinfo{title}{{{MetaGPT}}}.
\newblock
  \bibinfo{howpublished}{\url{https://github.com/FoundationAgents/MetaGPT}}.
\newblock
\newblock
\shownote{(accessed 2026-04-13)}.


\bibitem[Ope(2026)]%
        {OpenClaw2026}
 \bibinfo{year}{2026}\natexlab{}.
\newblock \bibinfo{title}{{{OpenClaw}}}.
\newblock \bibinfo{howpublished}{\url{https://openclaw.ai/}}.
\newblock
\newblock
\shownote{(accessed 2026-02-01)}.


\bibitem[Fac(2026)]%
        {FacebookPyrefly2026}
 \bibinfo{year}{2026}\natexlab{}.
\newblock \bibinfo{title}{Pyrefly}.
\newblock \bibinfo{howpublished}{\url{https://github.com/facebook/pyrefly}}.
\newblock
\newblock
\shownote{(accessed 2026-02-18)}.


\bibitem[Pys(2026)]%
        {PysaStaticAnalysis}
 \bibinfo{year}{2026}\natexlab{}.
\newblock \bibinfo{title}{Pysa (Static Analysis for Security)}.
\newblock
  \bibinfo{howpublished}{\url{https://pyre-check.org/docs/pysa-basics/}}.
\newblock
\newblock
\shownote{(accessed 2026-03-31)}.


\bibitem[SWE(2026)]%
        {SWEagent2026}
 \bibinfo{year}{2026}\natexlab{}.
\newblock \bibinfo{title}{{{SWE-agent}}}.
\newblock \bibinfo{howpublished}{\url{https://github.com/SWE-agent/SWE-agent}}.
\newblock
\newblock
\shownote{(accessed 2026-04-13)}.


\bibitem[Blazytko et~al\mbox{.}(2020)]%
        {blazytkoAURORAStatisticalCrash2020}
\bibfield{author}{\bibinfo{person}{Tim Blazytko}, \bibinfo{person}{Moritz
  Schl{\"o}gel}, \bibinfo{person}{Cornelius Aschermann}, \bibinfo{person}{Ali
  Abbasi}, \bibinfo{person}{Joel Frank}, \bibinfo{person}{Simon W{\"o}rner},
  {and} \bibinfo{person}{Thorsten Holz}.} \bibinfo{year}{2020}\natexlab{}.
\newblock \showarticletitle{{{AURORA}}: Statistical Crash Analysis for
  Automated Root Cause Explanation}. In \bibinfo{booktitle}{\emph{Proceedings
  of the 29th {{USENIX Security Symposium}}}}. \bibinfo{address}{Virtual
  Event}, \bibinfo{pages}{235--252}.
\newblock


\bibitem[Chang et~al\mbox{.}(2025)]%
        {changFirmrcaPostFuzzingAnalysis2025}
\bibfield{author}{\bibinfo{person}{Boyu Chang}, \bibinfo{person}{Binbin Zhao},
  \bibinfo{person}{Qiao Zhang}, \bibinfo{person}{Peiyu Liu},
  \bibinfo{person}{Yuan Tian}, \bibinfo{person}{Raheem Beyah}, {and}
  \bibinfo{person}{Shouling Ji}.} \bibinfo{year}{2025}\natexlab{}.
\newblock \showarticletitle{{{FirmRCA}}: {{Towards Post-Fuzzing Analysis}} on
  {{ARM Embedded Firmware}} with {{Efficient Event-Based Fault Localization}}}.
  In \bibinfo{booktitle}{\emph{Proceedings of the 46th {{IEEE Symposium}} on
  {{Security}} and {{Privacy}}}}. \bibinfo{address}{San Francisco, CA, USA},
  \bibinfo{pages}{3783--3800}.
\newblock


\bibitem[Chen et~al\mbox{.}(2025a)]%
        {chenStruQDefendingPrompt2025}
\bibfield{author}{\bibinfo{person}{Sizhe Chen}, \bibinfo{person}{Julien Piet},
  \bibinfo{person}{Chawin Sitawarin}, {and} \bibinfo{person}{David Wagner}.}
  \bibinfo{year}{2025}\natexlab{a}.
\newblock \showarticletitle{{{StruQ}}: Defending Against Prompt Injection with
  Structured Queries}. In \bibinfo{booktitle}{\emph{Proceedings of the 34th
  {{USENIX Security Symposium}}}}. \bibinfo{address}{Seattle, WA, USA},
  \bibinfo{pages}{2383--2400}.
\newblock


\bibitem[Chen et~al\mbox{.}(2025b)]%
        {chenSecAlignDefendingPrompt2025}
\bibfield{author}{\bibinfo{person}{Sizhe Chen}, \bibinfo{person}{Arman
  Zharmagambetov}, \bibinfo{person}{Saeed Mahloujifar},
  \bibinfo{person}{Kamalika Chaudhuri}, \bibinfo{person}{David Wagner}, {and}
  \bibinfo{person}{Chuan Guo}.} \bibinfo{year}{2025}\natexlab{b}.
\newblock \showarticletitle{{{SecAlign}}: {{Defending Against Prompt
  Injection}} with {{Preference Optimization}}}. In
  \bibinfo{booktitle}{\emph{Proceedings of the 2025 {{ACM SIGSAC Conference}}
  on {{Computer}} and {{Communications Security}}}}. \bibinfo{address}{Taipei,
  Taiwan}, \bibinfo{pages}{2833--2847}.
\newblock


\bibitem[Chen et~al\mbox{.}(2024)]%
        {chenAgentPoisonRedteamingLLM2024}
\bibfield{author}{\bibinfo{person}{Zhaorun Chen}, \bibinfo{person}{Zhen Xiang},
  \bibinfo{person}{Chaowei Xiao}, \bibinfo{person}{Dawn Song}, {and}
  \bibinfo{person}{Bo Li}.} \bibinfo{year}{2024}\natexlab{}.
\newblock \showarticletitle{{{AgentPoison}}: Red-teaming {{LLM}} Agents via
  Poisoning Memory or Knowledge Bases}. In
  \bibinfo{booktitle}{\emph{Proceedings of the 38th {{Annual Conference}} on
  {{Neural Information Processing Systems}}}}. \bibinfo{address}{Vancouver, BC,
  Canada}, \bibinfo{pages}{130185--130213}.
\newblock


\bibitem[Chocron et~al\mbox{.}(2026)]%
        {chocronWhoOwnsThis2026}
\bibfield{author}{\bibinfo{person}{Ruben Chocron}, \bibinfo{person}{Doron
  Jonathan~Ben Chayim}, \bibinfo{person}{Eyal Lenga}, \bibinfo{person}{Gilad
  Gressel}, \bibinfo{person}{Alina Oprea}, {and} \bibinfo{person}{Yisroel
  Mirsky}.} \bibinfo{year}{2026}\natexlab{}.
\newblock \bibinfo{title}{Who Owns This Agent? {{Tracing AI}} Agents Back to
  Their Owners}.
\newblock \bibinfo{howpublished}{\url{http://arxiv.org/abs/2605.16035}}.
\newblock


\bibitem[Costa et~al\mbox{.}(2025)]%
        {costaSecuringAIAgents2025}
\bibfield{author}{\bibinfo{person}{Manuel Costa}, \bibinfo{person}{Boris
  K{\"o}pf}, \bibinfo{person}{Aashish Kolluri}, \bibinfo{person}{Andrew
  Paverd}, \bibinfo{person}{Mark Russinovich}, \bibinfo{person}{Ahmed Salem},
  \bibinfo{person}{Shruti Tople}, \bibinfo{person}{Lukas Wutschitz}, {and}
  \bibinfo{person}{Santiago {Zanella-B{\'e}guelin}}.}
  \bibinfo{year}{2025}\natexlab{}.
\newblock \bibinfo{title}{Securing {{AI Agents}} with {{Information-Flow
  Control}}}.
\newblock \bibinfo{howpublished}{\url{https://arxiv.org/abs/2505.23643}}.
\newblock


\bibitem[Cui et~al\mbox{.}(2018)]%
        {cuiREPTReverseDebugging2018}
\bibfield{author}{\bibinfo{person}{Weidong Cui}, \bibinfo{person}{Xinyang Ge},
  \bibinfo{person}{Baris Kasikci}, \bibinfo{person}{Ben Niu},
  \bibinfo{person}{Upamanyu Sharma}, \bibinfo{person}{Ruoyu Wang}, {and}
  \bibinfo{person}{Insu Yun}.} \bibinfo{year}{2018}\natexlab{}.
\newblock \showarticletitle{{{REPT}}: Reverse Debugging of Failures in Deployed
  Software}. In \bibinfo{booktitle}{\emph{Proceedings of the 13th {{USENIX
  Symposium}} on {{Operating Systems Design}} and {{Implementation}}}}.
  \bibinfo{address}{Carlsbad, CA, USA}, \bibinfo{pages}{17--32}.
\newblock


\bibitem[Cui et~al\mbox{.}(2016)]%
        {cuiRETracerTriagingCrashes2016}
\bibfield{author}{\bibinfo{person}{Weidong Cui}, \bibinfo{person}{Marcus
  Peinado}, \bibinfo{person}{Sang~Kil Cha}, \bibinfo{person}{Yanick
  Fratantonio}, {and} \bibinfo{person}{Vasileios~P. Kemerlis}.}
  \bibinfo{year}{2016}\natexlab{}.
\newblock \showarticletitle{{{RETracer}}: Triaging Crashes by Reverse Execution
  from Partial Memory Dumps}. In \bibinfo{booktitle}{\emph{Proceedings of the
  38th {{International Conference}} on {{Software Engineering}}}}.
  \bibinfo{address}{Austin, TX, USA}, \bibinfo{pages}{820--831}.
\newblock


\bibitem[Debenedetti et~al\mbox{.}(2025)]%
        {debenedettiDefeatingPromptInjections2025}
\bibfield{author}{\bibinfo{person}{Edoardo Debenedetti}, \bibinfo{person}{Ilia
  Shumailov}, \bibinfo{person}{Tianqi Fan}, \bibinfo{person}{Jamie Hayes},
  \bibinfo{person}{Nicholas Carlini}, \bibinfo{person}{Daniel Fabian},
  \bibinfo{person}{Christoph Kern}, \bibinfo{person}{Chongyang Shi},
  \bibinfo{person}{Andreas Terzis}, {and} \bibinfo{person}{Florian
  Tram{\`e}r}.} \bibinfo{year}{2025}\natexlab{}.
\newblock \bibinfo{title}{Defeating Prompt Injections by Design}.
\newblock \bibinfo{howpublished}{\url{https://arxiv.org/abs/2503.18813}}.
\newblock


\bibitem[Debenedetti et~al\mbox{.}(2024)]%
        {debenedettiAgentDojoDynamicEnvironment2024}
\bibfield{author}{\bibinfo{person}{Edoardo Debenedetti}, \bibinfo{person}{Jie
  Zhang}, \bibinfo{person}{Mislav Balunovic}, \bibinfo{person}{Luca
  {Beurer-Kellner}}, \bibinfo{person}{Marc Fischer}, {and}
  \bibinfo{person}{Florian Tram{\`e}r}.} \bibinfo{year}{2024}\natexlab{}.
\newblock \showarticletitle{{{AgentDojo}}: A Dynamic Environment to Evaluate
  Prompt Injection Attacks and Defenses for {{LLM}} Agents}. In
  \bibinfo{booktitle}{\emph{Proceedings of the 38th {{Conference}} on {{Neural
  Information Processing Systems Datasets}} and {{Benchmarks Track}}}}.
  \bibinfo{address}{Vancouver, BC, Canada}, \bibinfo{pages}{82895--82920}.
\newblock


\bibitem[Dong et~al\mbox{.}(2025)]%
        {dongPracticalMemoryInjection2025}
\bibfield{author}{\bibinfo{person}{Shen Dong}, \bibinfo{person}{Shaochen Xu},
  \bibinfo{person}{Pengfei He}, \bibinfo{person}{Yige Li},
  \bibinfo{person}{Jiliang Tang}, \bibinfo{person}{Tianming Liu},
  \bibinfo{person}{Hui Liu}, {and} \bibinfo{person}{Zhen Xiang}.}
  \bibinfo{year}{2025}\natexlab{}.
\newblock \bibinfo{title}{Memory Injection Attacks on {{LLM}} Agents via
  {Query-Only} Interaction}.
\newblock \bibinfo{howpublished}{\url{https://arxiv.org/abs/2503.03704}}.
\newblock


\bibitem[Gao et~al\mbox{.}(2018)]%
        {gaoAiqlEnablingEfficient}
\bibfield{author}{\bibinfo{person}{Peng Gao}, \bibinfo{person}{Xusheng Xiao},
  \bibinfo{person}{Zhichun Li}, \bibinfo{person}{Kangkook Jee},
  \bibinfo{person}{Fengyuan Xu}, \bibinfo{person}{Sanjeev~R. Kulkarni}, {and}
  \bibinfo{person}{Prateek Mittal}.} \bibinfo{year}{2018}\natexlab{}.
\newblock \showarticletitle{{{AIQL}}: Enabling Efficient Attack Investigation
  from System Monitoring Data}. In \bibinfo{booktitle}{\emph{Proceedings of the
  2018 {{USENIX Annual Technical Conference}}}}. \bibinfo{address}{Boston, MA,
  USA}, \bibinfo{pages}{113--126}.
\newblock


\bibitem[Gartner(2025)]%
        {GartnerPredicts402025}
\bibfield{author}{\bibinfo{person}{Gartner}.} \bibinfo{year}{2025}\natexlab{}.
\newblock \bibinfo{title}{Gartner Predicts 40\% of Enterprise Apps Will Feature
  Task-Specific {{AI}} Agents by 2026, up from Less than 5\% in 2025}.
\newblock
  \bibinfo{howpublished}{\url{https://www.gartner.com/en/newsroom/press-releases/2025-08-26-gartner-predicts-40-percent-of-enterprise-apps-will-feature-task-specific-ai-agents-by-2026-up-from-less-than-5-percent-in-2025}}.
\newblock
\newblock
\shownote{(accessed 2026-02-01)}.


\bibitem[Ge et~al\mbox{.}(2020)]%
        {geReverseDebuggingKernel2020}
\bibfield{author}{\bibinfo{person}{Xinyang Ge}, \bibinfo{person}{Ben Niu},
  {and} \bibinfo{person}{Weidong Cui}.} \bibinfo{year}{2020}\natexlab{}.
\newblock \showarticletitle{Reverse Debugging of Kernel Failures in Deployed
  Systems}. In \bibinfo{booktitle}{\emph{Proceedings of the 2020 {{USENIX
  Annual Technical Conference}}}}. \bibinfo{address}{Virtual Event},
  \bibinfo{pages}{281--292}.
\newblock


\bibitem[Goodin(2025)]%
        {GoodinGeminiMemory2025}
\bibfield{author}{\bibinfo{person}{Dan Goodin}.}
  \bibinfo{year}{2025}\natexlab{}.
\newblock \bibinfo{title}{New hack uses prompt injection to corrupt {Gemini}'s
  long-term memory}.
\newblock
  \bibinfo{howpublished}{\url{https://arstechnica.com/security/2025/02/new-hack-uses-prompt-injection-to-corrupt-geminis-long-term-memory/}}.
\newblock
\newblock
\shownote{(accessed 2026-06-04)}.


\bibitem[Goyal et~al\mbox{.}(2024)]%
        {goyalRCAIDEmbeddingRoot2024}
\bibfield{author}{\bibinfo{person}{Akul Goyal}, \bibinfo{person}{Gang Wang},
  {and} \bibinfo{person}{Adam Bates}.} \bibinfo{year}{2024}\natexlab{}.
\newblock \showarticletitle{R-{{CAID}}: Embedding Root Cause Analysis within
  Provenance-Based Intrusion Detection}. In
  \bibinfo{booktitle}{\emph{Proceedings of the 45th {{IEEE Symposium}} on
  {{Security}} and {{Privacy}}}}. \bibinfo{address}{San Francisco, CA, USA},
  \bibinfo{pages}{3515--3532}.
\newblock


\bibitem[Greshake et~al\mbox{.}(2023)]%
        {greshakeNotWhatYouve2023}
\bibfield{author}{\bibinfo{person}{Kai Greshake}, \bibinfo{person}{Sahar
  Abdelnabi}, \bibinfo{person}{Shailesh Mishra}, \bibinfo{person}{Christoph
  Endres}, \bibinfo{person}{Thorsten Holz}, {and} \bibinfo{person}{Mario
  Fritz}.} \bibinfo{year}{2023}\natexlab{}.
\newblock \showarticletitle{Not What You've Signed Up For: Compromising
  Real-World {{LLM-Integrated}} Applications with Indirect Prompt Injection}.
  In \bibinfo{booktitle}{\emph{Proceedings of the 16th {{ACM Workshop}} on
  {{Artificial Intelligence}} and {{Security}}}}. \bibinfo{address}{Copenhagen,
  Denmark}, \bibinfo{pages}{79--90}.
\newblock


\bibitem[Hassan et~al\mbox{.}(2020)]%
        {hassanOmegaLogHighfidelityAttack2020}
\bibfield{author}{\bibinfo{person}{Wajih~Ul Hassan},
  \bibinfo{person}{Mohammad~A. Noureddine}, \bibinfo{person}{Pubali Datta},
  {and} \bibinfo{person}{Adam Bates}.} \bibinfo{year}{2020}\natexlab{}.
\newblock \showarticletitle{{{OmegaLog}}: High-Fidelity Attack Investigation
  via Transparent Multi-Layer Log Analysis}. In
  \bibinfo{booktitle}{\emph{Proceedings of the 2020 {{Network}} and
  {{Distributed System Security Symposium}}}}. \bibinfo{address}{San Diego, CA,
  USA}.
\newblock


\bibitem[He et~al\mbox{.}(2022)]%
        {heFreeWillAutomaticallyDiagnosing2022}
\bibfield{author}{\bibinfo{person}{Liang He}, \bibinfo{person}{Hong Hu},
  \bibinfo{person}{Purui Su}, \bibinfo{person}{Yan Cai}, {and}
  \bibinfo{person}{Zhenkai Liang}.} \bibinfo{year}{2022}\natexlab{}.
\newblock \showarticletitle{{{FreeWill}}: Automatically Diagnosing
  Use-after-free Bugs via Reference Miscounting Detection on Binaries}. In
  \bibinfo{booktitle}{\emph{Proceedings of the 31st {{USENIX Security
  Symposium}}}}. \bibinfo{address}{Boston, MA, USA},
  \bibinfo{pages}{2497--2512}.
\newblock


\bibitem[Hines et~al\mbox{.}(2024)]%
        {hinesDefendingIndirectPrompt2024}
\bibfield{author}{\bibinfo{person}{Keegan Hines}, \bibinfo{person}{Gary Lopez},
  \bibinfo{person}{Matthew Hall}, \bibinfo{person}{Federico Zarfati},
  \bibinfo{person}{Yonatan Zunger}, {and} \bibinfo{person}{Emre Kıcıman}.}
  \bibinfo{year}{2024}\natexlab{}.
\newblock \showarticletitle{Defending against Indirect Prompt Injection Attacks
  with Spotlighting}. In \bibinfo{booktitle}{\emph{Proceedings of the 7th
  Conference on Applied Machine Learning for Information Security}}.
  \bibinfo{address}{Arlington, VA, USA}, \bibinfo{pages}{48--62}.
\newblock


\bibitem[Ji et~al\mbox{.}(2017)]%
        {jiRAINRefinableAttack2017}
\bibfield{author}{\bibinfo{person}{Yang Ji}, \bibinfo{person}{Sangho Lee},
  \bibinfo{person}{Evan Downing}, \bibinfo{person}{Weiren Wang},
  \bibinfo{person}{Mattia Fazzini}, \bibinfo{person}{Taesoo Kim},
  \bibinfo{person}{Alessandro Orso}, {and} \bibinfo{person}{Wenke Lee}.}
  \bibinfo{year}{2017}\natexlab{}.
\newblock \showarticletitle{{{RAIN}}: {{Refinable Attack Investigation}} with
  {{On-demand Inter-Process Information Flow Tracking}}}. In
  \bibinfo{booktitle}{\emph{Proceedings of the 2017 {{ACM SIGSAC Conference}}
  on {{Computer}} and {{Communications Security}}}}. \bibinfo{address}{Dallas,
  TX, USA}, \bibinfo{pages}{377--390}.
\newblock


\bibitem[Ji et~al\mbox{.}(2026)]%
        {jiTamingVariousPrivilege2026}
\bibfield{author}{\bibinfo{person}{Zimo Ji}, \bibinfo{person}{Daoyuan Wu},
  \bibinfo{person}{Wenyuan Jiang}, \bibinfo{person}{Pingchuan Ma},
  \bibinfo{person}{Zongjie Li}, \bibinfo{person}{Yudong Gao},
  \bibinfo{person}{Shuai Wang}, {and} \bibinfo{person}{Yingjiu Li}.}
  \bibinfo{year}{2026}\natexlab{}.
\newblock \bibinfo{title}{Taming Various Privilege Escalation in {{LLM}}-Based
  Agent Systems: A Mandatory Access Control Framework}.
\newblock \bibinfo{howpublished}{\url{http://arxiv.org/abs/2601.11893}}.
\newblock


\bibitem[Kasikci et~al\mbox{.}(2017)]%
        {kasikciLazyDiagnosisInProduction2017}
\bibfield{author}{\bibinfo{person}{Baris Kasikci}, \bibinfo{person}{Weidong
  Cui}, \bibinfo{person}{Xinyang Ge}, {and} \bibinfo{person}{Ben Niu}.}
  \bibinfo{year}{2017}\natexlab{}.
\newblock \showarticletitle{Lazy {{Diagnosis}} of {{In-Production Concurrency
  Bugs}}}. In \bibinfo{booktitle}{\emph{Proceedings of the 26th {{Symposium}}
  on {{Operating Systems Principles}}}}. \bibinfo{address}{Shanghai, China},
  \bibinfo{pages}{582--598}.
\newblock


\bibitem[Kasikci et~al\mbox{.}(2015)]%
        {kasikciFailureSketchingTechnique2015}
\bibfield{author}{\bibinfo{person}{Baris Kasikci}, \bibinfo{person}{Benjamin
  Schubert}, \bibinfo{person}{Cristiano Pereira}, \bibinfo{person}{Gilles
  Pokam}, {and} \bibinfo{person}{George Candea}.}
  \bibinfo{year}{2015}\natexlab{}.
\newblock \showarticletitle{Failure Sketching: A Technique for Automated Root
  Cause Diagnosis of in-Production Failures}. In
  \bibinfo{booktitle}{\emph{Proceedings of the 25th {{Symposium}} on
  {{Operating Systems Principles}}}}. \bibinfo{address}{Monterey, CA, USA},
  \bibinfo{pages}{344--360}.
\newblock


\bibitem[Kim et~al\mbox{.}(2025)]%
        {kimPromptFlowIntegrity2025}
\bibfield{author}{\bibinfo{person}{Juhee Kim}, \bibinfo{person}{Woohyuk Choi},
  {and} \bibinfo{person}{Byoungyoung Lee}.} \bibinfo{year}{2025}\natexlab{}.
\newblock \bibinfo{title}{Prompt Flow Integrity to Prevent Privilege Escalation
  in {{LLM}} Agents}.
\newblock \bibinfo{howpublished}{\url{https://arxiv.org/abs/2503.15547}}.
\newblock


\bibitem[King and Chen(2003)]%
        {kingBacktrackingIntrusions2003}
\bibfield{author}{\bibinfo{person}{Samuel~T. King} {and}
  \bibinfo{person}{Peter~M. Chen}.} \bibinfo{year}{2003}\natexlab{}.
\newblock \showarticletitle{Backtracking Intrusions}. In
  \bibinfo{booktitle}{\emph{Proceedings of the 19th {{ACM Symposium}} on
  {{Operating Systems Principles}}}}. \bibinfo{address}{Bolton Landing, NY,
  USA}, \bibinfo{pages}{223--236}.
\newblock


\bibitem[Liu et~al\mbox{.}(2025)]%
        {liuAdvancesChallengesFoundation2025}
\bibfield{author}{\bibinfo{person}{Bang Liu}, \bibinfo{person}{Xinfeng Li},
  \bibinfo{person}{Jiayi Zhang}, \bibinfo{person}{Jinlin Wang},
  \bibinfo{person}{Tanjin He}, \bibinfo{person}{Sirui Hong},
  \bibinfo{person}{Hongzhang Liu}, \bibinfo{person}{Shaokun Zhang},
  \bibinfo{person}{Kaitao Song}, \bibinfo{person}{Kunlun Zhu},
  \bibinfo{person}{Yuheng Cheng}, \bibinfo{person}{Suyuchen Wang},
  \bibinfo{person}{Xiaoqiang Wang}, \bibinfo{person}{Yuyu Luo},
  \bibinfo{person}{Haibo Jin}, \bibinfo{person}{Peiyan Zhang},
  \bibinfo{person}{Ollie Liu}, \bibinfo{person}{Jiaqi Chen},
  \bibinfo{person}{Huan Zhang}, \bibinfo{person}{Zhaoyang Yu},
  \bibinfo{person}{Haochen Shi}, \bibinfo{person}{Boyan Li},
  \bibinfo{person}{Dekun Wu}, \bibinfo{person}{Fengwei Teng},
  \bibinfo{person}{Xiaojun Jia}, \bibinfo{person}{Jiawei Xu},
  \bibinfo{person}{Jinyu Xiang}, \bibinfo{person}{Yizhang Lin},
  \bibinfo{person}{Tianming Liu}, \bibinfo{person}{Tongliang Liu},
  \bibinfo{person}{Yu Su}, \bibinfo{person}{Huan Sun}, \bibinfo{person}{Glen
  Berseth}, \bibinfo{person}{Jianyun Nie}, \bibinfo{person}{Ian Foster},
  \bibinfo{person}{Logan Ward}, \bibinfo{person}{Qingyun Wu},
  \bibinfo{person}{Yu Gu}, \bibinfo{person}{Mingchen Zhuge},
  \bibinfo{person}{Xinbing Liang}, \bibinfo{person}{Xiangru Tang},
  \bibinfo{person}{Haohan Wang}, \bibinfo{person}{Jiaxuan You},
  \bibinfo{person}{Chi Wang}, \bibinfo{person}{Jian Pei},
  \bibinfo{person}{Qiang Yang}, \bibinfo{person}{Xiaoliang Qi}, {and}
  \bibinfo{person}{Chenglin Wu}.} \bibinfo{year}{2025}\natexlab{}.
\newblock \bibinfo{title}{Advances and Challenges in Foundation Agents: From
  Brain-Inspired Intelligence to Evolutionary, Collaborative, and Safe
  Systems}.
\newblock \bibinfo{howpublished}{\url{http://arxiv.org/abs/2504.01990}}.
\newblock


\bibitem[Liu et~al\mbox{.}(2024a)]%
        {liuImprovingIndirectcallAnalysis2024}
\bibfield{author}{\bibinfo{person}{Dinghao Liu}, \bibinfo{person}{Shouling Ji},
  \bibinfo{person}{Kangjie Lu}, {and} \bibinfo{person}{Qinming He}.}
  \bibinfo{year}{2024}\natexlab{a}.
\newblock \showarticletitle{Improving Indirect-Call Analysis in {{LLVM}} with
  Type and Data-Flow {{Co-Analysis}}}. In \bibinfo{booktitle}{\emph{Proceedings
  of the 33rd {{USENIX Security Symposium}}}}. \bibinfo{address}{Philadelphia,
  PA, USA}, \bibinfo{pages}{5895--5912}.
\newblock


\bibitem[Liu et~al\mbox{.}(2024b)]%
        {liuFormalizingBenchmarkingPrompt2024}
\bibfield{author}{\bibinfo{person}{Yupei Liu}, \bibinfo{person}{Yuqi Jia},
  \bibinfo{person}{Runpeng Geng}, \bibinfo{person}{Jinyuan Jia}, {and}
  \bibinfo{person}{Neil~Zhenqiang Gong}.} \bibinfo{year}{2024}\natexlab{b}.
\newblock \showarticletitle{Formalizing and Benchmarking Prompt Injection
  Attacks and Defenses}. In \bibinfo{booktitle}{\emph{Proceedings of the 33rd
  {{USENIX Security Symposium}}}}. \bibinfo{address}{Philadelphia, PA, USA},
  \bibinfo{pages}{1831--1847}.
\newblock


\bibitem[Lyons(2025)]%
        {LyonsAtlasVuln2025}
\bibfield{author}{\bibinfo{person}{Jessica Lyons}.}
  \bibinfo{year}{2025}\natexlab{}.
\newblock \bibinfo{title}{{Atlas} vuln lets crims inject malicious prompts
  {ChatGPT} won't forget between sessions}.
\newblock
  \bibinfo{howpublished}{\url{https://www.theregister.com/special-features/2025/10/27/atlas-vuln-allows-malicious-memory-injection-into-chatgpt/741516}}.
\newblock
\newblock
\shownote{(accessed 2026-06-04)}.


\bibitem[Ma et~al\mbox{.}(2017)]%
        {maMPIMultiplePerspective2017}
\bibfield{author}{\bibinfo{person}{Shiqing Ma}, \bibinfo{person}{Juan Zhai},
  \bibinfo{person}{Fei Wang}, \bibinfo{person}{Kyu~Hyung Lee},
  \bibinfo{person}{Xiangyu Zhang}, {and} \bibinfo{person}{Dongyan Xu}.}
  \bibinfo{year}{2017}\natexlab{}.
\newblock \showarticletitle{{{MPI}}: Multiple Perspective Attack Investigation
  with Semantic Aware Execution Partitioning}. In
  \bibinfo{booktitle}{\emph{Proceedings of the 26th {{USENIX Security
  Symposium}}}}. \bibinfo{address}{Vancouver, BC, Canada},
  \bibinfo{pages}{1111--1128}.
\newblock


\bibitem[Ma et~al\mbox{.}(2016)]%
        {maProTracerPracticalProvenance2016}
\bibfield{author}{\bibinfo{person}{Shiqing Ma}, \bibinfo{person}{Xiangyu
  Zhang}, {and} \bibinfo{person}{Dongyan Xu}.} \bibinfo{year}{2016}\natexlab{}.
\newblock \showarticletitle{{{ProTracer}}: Towards Practical Provenance Tracing
  by Alternating between Logging and Tainting}. In
  \bibinfo{booktitle}{\emph{Proceedings of the 2016 {{Network}} and
  {{Distributed System Security Symposium}}}}. \bibinfo{address}{San Diego, CA,
  USA}.
\newblock


\bibitem[Ma et~al\mbox{.}(2025)]%
        {maDiagnosingFailureRoot2025}
\bibfield{author}{\bibinfo{person}{Xuyan Ma}, \bibinfo{person}{Xiaofei Xie},
  \bibinfo{person}{Yawen Wang}, \bibinfo{person}{Junjie Wang},
  \bibinfo{person}{Boyu Wu}, \bibinfo{person}{Mingyang Li}, {and}
  \bibinfo{person}{Qing Wang}.} \bibinfo{year}{2025}\natexlab{}.
\newblock \bibinfo{title}{Demystifying the Lifecycle of Failures in
  Platform-Orchestrated Agentic Workflows}.
\newblock \bibinfo{howpublished}{\url{https://arxiv.org/abs/2509.23735}}.
\newblock


\bibitem[Milajerdi et~al\mbox{.}(2019)]%
        {milajerdiHOLMESRealTimeAPT2019}
\bibfield{author}{\bibinfo{person}{Sadegh~M. Milajerdi}, \bibinfo{person}{Rigel
  Gjomemo}, \bibinfo{person}{Birhanu Eshete}, \bibinfo{person}{R. Sekar}, {and}
  \bibinfo{person}{V.N. Venkatakrishnan}.} \bibinfo{year}{2019}\natexlab{}.
\newblock \showarticletitle{{{HOLMES}}: {{Real-Time APT Detection}} through
  {{Correlation}} of {{Suspicious Information Flows}}}. In
  \bibinfo{booktitle}{\emph{Proceedings of the 40th {{IEEE Symposium}} on
  {{Security}} and {{Privacy}}}}. \bibinfo{address}{San Francisco, CA, USA},
  \bibinfo{pages}{1137--1152}.
\newblock


\bibitem[{OWASP GenAI Security Project}(2025)]%
        {editorOWASPTop10}
\bibfield{author}{\bibinfo{person}{{OWASP GenAI Security Project}}.}
  \bibinfo{year}{2025}\natexlab{}.
\newblock \bibinfo{title}{{{OWASP}} Top 10 for Agentic Applications for 2026}.
\newblock
  \bibinfo{howpublished}{\url{https://genai.owasp.org/resource/owasp-top-10-for-agentic-applications-for-2026/}}.
\newblock
\newblock
\shownote{(accessed 2026-02-04)}.


\bibitem[Park et~al\mbox{.}(2024)]%
        {parkBenzenePracticalRoot2024}
\bibfield{author}{\bibinfo{person}{Younggi Park}, \bibinfo{person}{Hwiwon Lee},
  \bibinfo{person}{Jinho Jung}, \bibinfo{person}{Hyungjoon Koo}, {and}
  \bibinfo{person}{Huy~Kang Kim}.} \bibinfo{year}{2024}\natexlab{}.
\newblock \showarticletitle{Benzene: A Practical Root Cause Analysis System
  with an Under-Constrained State Mutation}. In
  \bibinfo{booktitle}{\emph{Proceedings of the 45th {{IEEE Symposium}} on
  {{Security}} and {{Privacy}}}}. \bibinfo{address}{San Francisco, CA, USA},
  \bibinfo{pages}{1865--1883}.
\newblock


\bibitem[Rehberger(2024)]%
        {rehbergerTrustNoAI2024}
\bibfield{author}{\bibinfo{person}{Johann Rehberger}.}
  \bibinfo{year}{2024}\natexlab{}.
\newblock \showarticletitle{Trust No {{AI}}: Prompt Injection Along The {{CIA}}
  Security Triad}. In \bibinfo{booktitle}{\emph{{{BlackHat Europe}} 2024}}.
  \bibinfo{address}{London, United Kingdom}.
\newblock


\bibitem[Rehberger~(wunderwuzzi)(2024)]%
        {ChatGPTMemory}
\bibfield{author}{\bibinfo{person}{Johann Rehberger~(wunderwuzzi)}.}
  \bibinfo{year}{2024}\natexlab{}.
\newblock \bibinfo{title}{Spyware Injection Into Your ChatGPT’s Long-Term
  Memory (SpAIware)}.
\newblock
  \bibinfo{howpublished}{\url{https://embracethered.com/blog/posts/2024/chatgpt-macos-app-persistent-data-exfiltration/}}.
\newblock
\newblock
\shownote{(accessed 2026-02-01)}.


\bibitem[Rehberger~(wunderwuzzi)(2025)]%
        {ClaudeCodeIPI}
\bibfield{author}{\bibinfo{person}{Johann Rehberger~(wunderwuzzi)}.}
  \bibinfo{year}{2025}\natexlab{}.
\newblock \bibinfo{title}{Claude Code: Data Exfiltration with DNS
  (CVE-2025-55284)}.
\newblock
  \bibinfo{howpublished}{\url{https://embracethered.com/blog/posts/2025/claude-code-exfiltration-via-dns-requests/}}.
\newblock
\newblock
\shownote{(accessed 2026-02-01)}.


\bibitem[Shi et~al\mbox{.}(2025)]%
        {shiProgentSecuringAIAgents2025}
\bibfield{author}{\bibinfo{person}{Tianneng Shi}, \bibinfo{person}{Jingxuan
  He}, \bibinfo{person}{Zhun Wang}, \bibinfo{person}{Hongwei Li},
  \bibinfo{person}{Linyu Wu}, \bibinfo{person}{Wenbo Guo}, {and}
  \bibinfo{person}{Dawn Song}.} \bibinfo{year}{2025}\natexlab{}.
\newblock \bibinfo{title}{Progent: Securing {{AI}} Agents with Privilege
  Control}.
\newblock \bibinfo{howpublished}{\url{http://arxiv.org/abs/2504.11703}}.
\newblock


\bibitem[Sunil et~al\mbox{.}(2026)]%
        {sunilMemoryPoisoningAttack2026}
\bibfield{author}{\bibinfo{person}{Balachandra~Devarangadi Sunil},
  \bibinfo{person}{Isheeta Sinha}, \bibinfo{person}{Piyush Maheshwari},
  \bibinfo{person}{Shantanu Todmal}, \bibinfo{person}{Shreyan Mallik}, {and}
  \bibinfo{person}{Shuchi Mishra}.} \bibinfo{year}{2026}\natexlab{}.
\newblock \bibinfo{title}{Memory Poisoning Attack and Defense on Memory Based
  {{LLM}}-Agents}.
\newblock \bibinfo{howpublished}{\url{http://arxiv.org/abs/2601.05504}}.
\newblock


\bibitem[Tan et~al\mbox{.}(2025)]%
        {tanRevPRAGRevealingPoisoning2025}
\bibfield{author}{\bibinfo{person}{Xue Tan}, \bibinfo{person}{Hao Luan},
  \bibinfo{person}{Mingyu Luo}, \bibinfo{person}{Xiaoyan Sun},
  \bibinfo{person}{Ping Chen}, {and} \bibinfo{person}{Jun Dai}.}
  \bibinfo{year}{2025}\natexlab{}.
\newblock \showarticletitle{{{RevPRAG}}: Revealing Poisoning Attacks in
  Retrieval-Augmented Generation through {{LLM}} Activation Analysis}. In
  \bibinfo{booktitle}{\emph{Proceedings of the {{Findings}} of the
  {{Association}} for {{Computational Linguistics}}: {{EMNLP}} 2025}}.
  \bibinfo{address}{Suzhou, China}, \bibinfo{pages}{12999--13011}.
\newblock


\bibitem[Wallace et~al\mbox{.}(2024)]%
        {wallaceInstructionHierarchyTraining2024}
\bibfield{author}{\bibinfo{person}{Eric Wallace}, \bibinfo{person}{Kai Xiao},
  \bibinfo{person}{Reimar Leike}, \bibinfo{person}{Lilian Weng},
  \bibinfo{person}{Johannes Heidecke}, {and} \bibinfo{person}{Alex Beutel}.}
  \bibinfo{year}{2024}\natexlab{}.
\newblock \bibinfo{title}{The Instruction Hierarchy: Training {{LLMs}} to
  Prioritize Privileged Instructions}.
\newblock \bibinfo{howpublished}{\url{https://arxiv.org/abs/2404.13208}}.
\newblock


\bibitem[Wang et~al\mbox{.}(2024)]%
        {wangBadAgentInsertingActivating2024}
\bibfield{author}{\bibinfo{person}{Yifei Wang}, \bibinfo{person}{Dizhan Xue},
  \bibinfo{person}{Shengjie Zhang}, {and} \bibinfo{person}{Shengsheng Qian}.}
  \bibinfo{year}{2024}\natexlab{}.
\newblock \showarticletitle{{{BadAgent}}: Inserting and Activating Backdoor
  Attacks in {{LLM}} Agents}. In \bibinfo{booktitle}{\emph{Proceedings of the
  62nd {{Annual Meeting}} of the {{Association}} for {{Computational
  Linguistics}}}}. \bibinfo{address}{Bangkok, Thailand},
  \bibinfo{pages}{9811--9827}.
\newblock


\bibitem[Wei et~al\mbox{.}(2025)]%
        {weiAMemGuardProactiveDefense2025}
\bibfield{author}{\bibinfo{person}{Qianshan Wei}, \bibinfo{person}{Tengchao
  Yang}, \bibinfo{person}{Yaochen Wang}, \bibinfo{person}{Xinfeng Li},
  \bibinfo{person}{Lijun Li}, \bibinfo{person}{Zhenfei Yin},
  \bibinfo{person}{Yi Zhan}, \bibinfo{person}{Thorsten Holz},
  \bibinfo{person}{Zhiqiang Lin}, {and} \bibinfo{person}{XiaoFeng Wang}.}
  \bibinfo{year}{2025}\natexlab{}.
\newblock \bibinfo{title}{A-{{MemGuard}}: A Proactive Defense Framework for
  {{LLM-Based}} Agent Memory}.
\newblock \bibinfo{howpublished}{\url{http://arxiv.org/abs/2510.02373}}.
\newblock


\bibitem[Wu et~al\mbox{.}(2024)]%
        {wuAutoGenEnablingNextgen2024}
\bibfield{author}{\bibinfo{person}{Qingyun Wu}, \bibinfo{person}{Gagan Bansal},
  \bibinfo{person}{Jieyu Zhang}, \bibinfo{person}{Yiran Wu},
  \bibinfo{person}{Beibin Li}, \bibinfo{person}{Erkang Zhu},
  \bibinfo{person}{Li Jiang}, \bibinfo{person}{Xiaoyun Zhang},
  \bibinfo{person}{Shaokun Zhang}, \bibinfo{person}{Jiale Liu},
  \bibinfo{person}{Ahmed~Hassan Awadallah}, \bibinfo{person}{Ryen~W. White},
  \bibinfo{person}{Doug Burger}, {and} \bibinfo{person}{Chi Wang}.}
  \bibinfo{year}{2024}\natexlab{}.
\newblock \showarticletitle{{{AutoGen}}: Enabling Next-Gen {{LLM}} Applications
  via Multi-Agent Conversations}. In \bibinfo{booktitle}{\emph{Proceedings of
  the 1st {{Conference}} on {{Language Modeling}}}}.
  \bibinfo{address}{Philadelphia, PA, USA}.
\newblock


\bibitem[Wu et~al\mbox{.}(2025)]%
        {wuIsolateGPTExecutionIsolation2025}
\bibfield{author}{\bibinfo{person}{Yuhao Wu}, \bibinfo{person}{Franziska
  Roesner}, \bibinfo{person}{Tadayoshi Kohno}, \bibinfo{person}{Ning Zhang},
  {and} \bibinfo{person}{Umar Iqbal}.} \bibinfo{year}{2025}\natexlab{}.
\newblock \showarticletitle{{{IsolateGPT}}: An Execution Isolation Architecture
  for {{LLM-Based}} Agentic Systems}. In \bibinfo{booktitle}{\emph{Proceedings
  of the 32nd {{Annual Network}} and {{Distributed System Security
  Symposium}}}}. \bibinfo{address}{San Diego, CA, USA}.
\newblock


\bibitem[Xia et~al\mbox{.}(2024)]%
        {xiaDEEPTYPERefiningIndirect2024}
\bibfield{author}{\bibinfo{person}{Tianrou Xia}, \bibinfo{person}{Hong Hu},
  {and} \bibinfo{person}{Dinghao Wu}.} \bibinfo{year}{2024}\natexlab{}.
\newblock \showarticletitle{{{DEEPTYPE}}: Refining Indirect Call Targets with
  Strong Multi-layer Type Analysis}. In \bibinfo{booktitle}{\emph{Proceedings
  of the 33rd {{USENIX Security Symposium}}}}. \bibinfo{address}{Philadelphia,
  PA, USA}, \bibinfo{pages}{5877--5894}.
\newblock


\bibitem[Xiang et~al\mbox{.}(2024)]%
        {xiangCertifiablyRobustRAG2026}
\bibfield{author}{\bibinfo{person}{Chong Xiang}, \bibinfo{person}{Tong Wu},
  \bibinfo{person}{Zexuan Zhong}, \bibinfo{person}{David Wagner},
  \bibinfo{person}{Danqi Chen}, {and} \bibinfo{person}{Prateek Mittal}.}
  \bibinfo{year}{2024}\natexlab{}.
\newblock \bibinfo{title}{Certifiably {{Robust RAG}} against {{Retrieval
  Corruption}}}.
\newblock \bibinfo{howpublished}{\url{https://arxiv.org/abs/2405.15556}}.
\newblock


\bibitem[Xu et~al\mbox{.}(2024)]%
        {xuRacingNegativeForce2024}
\bibfield{author}{\bibinfo{person}{Dandan Xu}, \bibinfo{person}{Di Tang},
  \bibinfo{person}{Yi Chen}, \bibinfo{person}{XiaoFeng Wang},
  \bibinfo{person}{Kai Chen}, \bibinfo{person}{Haixu Tang}, {and}
  \bibinfo{person}{Longxing Li}.} \bibinfo{year}{2024}\natexlab{}.
\newblock \showarticletitle{Racing on the Negative Force: Efficient
  Vulnerability Root-Cause Analysis through Reinforcement Learning on
  Counterexamples}. In \bibinfo{booktitle}{\emph{Proceedings of the 33rd
  {{USENIX Security Symposium}}}}. \bibinfo{address}{Philadelphia, PA, USA},
  \bibinfo{pages}{4229--4246}.
\newblock


\bibitem[Xu et~al\mbox{.}(2016)]%
        {xuCREDALLocatingMemory2016}
\bibfield{author}{\bibinfo{person}{Jun Xu}, \bibinfo{person}{Dongliang Mu},
  \bibinfo{person}{Ping Chen}, \bibinfo{person}{Xinyu Xing},
  \bibinfo{person}{Pei Wang}, {and} \bibinfo{person}{Peng Liu}.}
  \bibinfo{year}{2016}\natexlab{}.
\newblock \showarticletitle{{{CREDAL}}: {{Towards Locating}} a {{Memory
  Corruption Vulnerability}} with {{Your Core Dump}}}. In
  \bibinfo{booktitle}{\emph{Proceedings of the 2016 {{ACM SIGSAC Conference}}
  on {{Computer}} and {{Communications Security}}}}. \bibinfo{address}{Vienna,
  Austria}, \bibinfo{pages}{529--540}.
\newblock


\bibitem[Xu et~al\mbox{.}(2017)]%
        {xuPostmortemProgramAnalysis2017}
\bibfield{author}{\bibinfo{person}{Jun Xu}, \bibinfo{person}{Dongliang Mu},
  \bibinfo{person}{Xinyu Xing}, \bibinfo{person}{Peng Liu},
  \bibinfo{person}{Ping Chen}, {and} \bibinfo{person}{Bing Mao}.}
  \bibinfo{year}{2017}\natexlab{}.
\newblock \showarticletitle{{{POMP}}: {{Postmortem}} Program Analysis with
  Hardware-Enhanced Post-Crash Artifacts}. In
  \bibinfo{booktitle}{\emph{Proceedings of the 26th {{USENIX Security
  Symposium}}}}. \bibinfo{address}{Vancouver, BC, Canada},
  \bibinfo{pages}{17--32}.
\newblock


\bibitem[Yagemann et~al\mbox{.}(2021)]%
        {yagemannARCUSSymbolicRoot2021}
\bibfield{author}{\bibinfo{person}{Carter Yagemann}, \bibinfo{person}{Matthew
  Pruett}, \bibinfo{person}{Simon~P. Chung}, \bibinfo{person}{Kennon Bittick},
  \bibinfo{person}{Brendan Saltaformaggio}, {and} \bibinfo{person}{Wenke Lee}.}
  \bibinfo{year}{2021}\natexlab{}.
\newblock \showarticletitle{{{ARCUS}}: Symbolic Root Cause Analysis of Exploits
  in Production Systems}. In \bibinfo{booktitle}{\emph{Proceedings of the 30th
  {{USENIX Security Symposium}}}}. \bibinfo{address}{Virtual Event},
  \bibinfo{pages}{1989--2006}.
\newblock


\bibitem[Yin et~al\mbox{.}(2007)]%
        {yinPanoramaCapturingSystemwide2007}
\bibfield{author}{\bibinfo{person}{Heng Yin}, \bibinfo{person}{Dawn Song},
  \bibinfo{person}{Manuel Egele}, \bibinfo{person}{Christopher Kruegel}, {and}
  \bibinfo{person}{Engin Kirda}.} \bibinfo{year}{2007}\natexlab{}.
\newblock \showarticletitle{Panorama: Capturing System-Wide Information Flow
  for Malware Detection and Analysis}. In \bibinfo{booktitle}{\emph{Proceedings
  of the 14th {{ACM Conference}} on {{Computer}} and {{Communications
  Security}}}}. \bibinfo{address}{Alexandria, VA, USA},
  \bibinfo{pages}{116--127}.
\newblock


\bibitem[Yuan et~al\mbox{.}(2012a)]%
        {yuanBeConservativeEnhancing2012}
\bibfield{author}{\bibinfo{person}{Ding Yuan}, \bibinfo{person}{Soyeon Park},
  \bibinfo{person}{Peng Huang}, \bibinfo{person}{Yang Liu},
  \bibinfo{person}{Michael~M. Lee}, \bibinfo{person}{Xiaoming Tang},
  \bibinfo{person}{Yuanyuan Zhou}, {and} \bibinfo{person}{Stefan Savage}.}
  \bibinfo{year}{2012}\natexlab{a}.
\newblock \showarticletitle{Be Conservative: Enhancing Failure Diagnosis with
  Proactive Logging}. In \bibinfo{booktitle}{\emph{Proceedings of the 10th
  {{USENIX Symposium}} on {{Operating Systems Design}} and
  {{Implementation}}}}. \bibinfo{address}{Hollywood, CA},
  \bibinfo{pages}{293--306}.
\newblock


\bibitem[Yuan et~al\mbox{.}(2012b)]%
        {yuanImprovingSoftwareDiagnosability2012}
\bibfield{author}{\bibinfo{person}{Ding Yuan}, \bibinfo{person}{Jing Zheng},
  \bibinfo{person}{Soyeon Park}, \bibinfo{person}{Yuanyuan Zhou}, {and}
  \bibinfo{person}{Stefan Savage}.} \bibinfo{year}{2012}\natexlab{b}.
\newblock \showarticletitle{Improving {{Software Diagnosability}} via {{Log
  Enhancement}}}.
\newblock \bibinfo{journal}{\emph{ACM Transactions on Computer Systems}}
  \bibinfo{volume}{30}, \bibinfo{number}{1} (\bibinfo{year}{2012}),
  \bibinfo{pages}{4:1--4:28}.
\newblock


\bibitem[Zhan et~al\mbox{.}(2024)]%
        {zhanInjecAgentBenchmarkingIndirect2024}
\bibfield{author}{\bibinfo{person}{Qiusi Zhan}, \bibinfo{person}{Zhixiang
  Liang}, \bibinfo{person}{Zifan Ying}, {and} \bibinfo{person}{Daniel Kang}.}
  \bibinfo{year}{2024}\natexlab{}.
\newblock \showarticletitle{{{InjecAgent}}: {{Benchmarking Indirect Prompt
  Injections}} in {{Tool-Integrated Large Language Model Agents}}}. In
  \bibinfo{booktitle}{\emph{Proceedings of the {{Findings}} of the
  {{Association}} for {{Computational Linguistics}}: {{ACL}} 2024}}.
  \bibinfo{address}{Bangkok, Thailand}, \bibinfo{pages}{10471--10506}.
\newblock


\bibitem[Zhang et~al\mbox{.}(2025a)]%
        {zhangAgentSecurityBench2025}
\bibfield{author}{\bibinfo{person}{Hanrong Zhang}, \bibinfo{person}{Jingyuan
  Huang}, \bibinfo{person}{Kai Mei}, \bibinfo{person}{Yifei Yao},
  \bibinfo{person}{Zhenting Wang}, \bibinfo{person}{Chenlu Zhan},
  \bibinfo{person}{Hongwei Wang}, {and} \bibinfo{person}{Yongfeng Zhang}.}
  \bibinfo{year}{2025}\natexlab{a}.
\newblock \showarticletitle{Agent Security Bench ({{ASB}}): Formalizing and
  Benchmarking Attacks and Defenses in {{LLM-based}} Agents}. In
  \bibinfo{booktitle}{\emph{Proceedings of the 13th {{International
  Conference}} on {{Learning Representations}}}}. \bibinfo{address}{Singapore}.
\newblock


\bibitem[Zhang et~al\mbox{.}(2025b)]%
        {zhangWhichAgentCauses2025}
\bibfield{author}{\bibinfo{person}{Shaokun Zhang}, \bibinfo{person}{Ming Yin},
  \bibinfo{person}{Jieyu Zhang}, \bibinfo{person}{Jiale Liu},
  \bibinfo{person}{Zhiguang Han}, \bibinfo{person}{Jingyang Zhang},
  \bibinfo{person}{Beibin Li}, \bibinfo{person}{Chi Wang},
  \bibinfo{person}{Huazheng Wang}, \bibinfo{person}{Yiran Chen}, {and}
  \bibinfo{person}{Qingyun Wu}.} \bibinfo{year}{2025}\natexlab{b}.
\newblock \showarticletitle{Which Agent Causes Task Failures and When? {On}
  Automated Failure Attribution of {{LLM}} Multi-Agent Systems}. In
  \bibinfo{booktitle}{\emph{Proceedings of the 42nd {{International
  Conference}} on {{Machine Learning}}}}. \bibinfo{address}{Vancouver, Canada},
  \bibinfo{pages}{76583--76599}.
\newblock


\bibitem[Zhang et~al\mbox{.}(2026)]%
        {zhangAgentSentryMitigatingIndirect2026}
\bibfield{author}{\bibinfo{person}{Tian Zhang}, \bibinfo{person}{Yiwei Xu},
  \bibinfo{person}{Juan Wang}, \bibinfo{person}{Keyan Guo},
  \bibinfo{person}{Xiaoyang Xu}, \bibinfo{person}{Bowen Xiao},
  \bibinfo{person}{Quanlong Guan}, \bibinfo{person}{Jinlin Fan},
  \bibinfo{person}{Jiawei Liu}, \bibinfo{person}{Zhiquan Liu}, {and}
  \bibinfo{person}{Hongxin Hu}.} \bibinfo{year}{2026}\natexlab{}.
\newblock \bibinfo{title}{{{AgentSentry}}: Mitigating Indirect Prompt Injection
  in {{LLM}} Agents via Temporal Causal Diagnostics and Context Purification}.
\newblock \bibinfo{howpublished}{\url{https://arxiv.org/abs/2602.22724}}.
\newblock


\bibitem[Zhao et~al\mbox{.}(2017)]%
        {zhaoLog20FullyAutomated2017}
\bibfield{author}{\bibinfo{person}{Xu Zhao}, \bibinfo{person}{Kirk Rodrigues},
  \bibinfo{person}{Yu Luo}, \bibinfo{person}{Michael Stumm},
  \bibinfo{person}{Ding Yuan}, {and} \bibinfo{person}{Yuanyuan Zhou}.}
  \bibinfo{year}{2017}\natexlab{}.
\newblock \showarticletitle{Log20: {{Fully Automated Optimal Placement}} of
  {{Log Printing Statements}} under {{Specified Overhead Threshold}}}. In
  \bibinfo{booktitle}{\emph{Proceedings of the 26th {{Symposium}} on
  {{Operating Systems Principles}}}}. \bibinfo{address}{Shanghai, China},
  \bibinfo{pages}{565--581}.
\newblock


\bibitem[Zhu et~al\mbox{.}(2025)]%
        {zhuWhereLLMAgents2025}
\bibfield{author}{\bibinfo{person}{Kunlun Zhu}, \bibinfo{person}{Zijia Liu},
  \bibinfo{person}{Bingxuan Li}, \bibinfo{person}{Muxin Tian},
  \bibinfo{person}{Yingxuan Yang}, \bibinfo{person}{Jiaxun Zhang},
  \bibinfo{person}{Pengrui Han}, \bibinfo{person}{Qipeng Xie},
  \bibinfo{person}{Fuyang Cui}, \bibinfo{person}{Weijia Zhang},
  \bibinfo{person}{Xiaoteng Ma}, \bibinfo{person}{Xiaodong Yu},
  \bibinfo{person}{Gowtham Ramesh}, \bibinfo{person}{Jialian Wu},
  \bibinfo{person}{Zicheng Liu}, \bibinfo{person}{Pan Lu},
  \bibinfo{person}{James Zou}, {and} \bibinfo{person}{Jiaxuan You}.}
  \bibinfo{year}{2025}\natexlab{}.
\newblock \bibinfo{title}{Where {{LLM}} Agents Fail and How They can Learn From
  Failures}.
\newblock \bibinfo{howpublished}{\url{http://arxiv.org/abs/2509.25370}}.
\newblock


\bibitem[Zou et~al\mbox{.}(2025)]%
        {zouPoisonedRAGKnowledgeCorruption2025}
\bibfield{author}{\bibinfo{person}{Wei Zou}, \bibinfo{person}{Runpeng Geng},
  \bibinfo{person}{Binghui Wang}, {and} \bibinfo{person}{Jinyuan Jia}.}
  \bibinfo{year}{2025}\natexlab{}.
\newblock \showarticletitle{{{PoisonedRAG}}: Knowledge Corruption Attacks to
  Retrieval-Augmented Generation of Large Language Models}. In
  \bibinfo{booktitle}{\emph{Proceedings of the 34th {{USENIX Security
  Symposium}}}}. \bibinfo{address}{Seattle, WA, USA},
  \bibinfo{pages}{3827--3844}.
\newblock


\end{thebibliography}

\appendix

\clearpage
\section{Configured Region Set}
\label{apdx:region-set}
{\bf Region set.}
Our observations, consistent with prior work~\cite{zhuWhereLLMAgents2025}, show that program variables serving the same agent functionality cluster naturally into five regions as follows:
\begin{itemize}
    \item \textbf{Perceiving (Per):}
    Variables that receive input from external sources where the application does not initiate the interaction (e.g., user queries, incoming agent messages, sensor readings).
    \item \textbf{Planning \& Acting (PA):}
    Variables that store LLM-generated responses and variables involved in application-initiated interactions with external sources (e.g., tool invocations, outgoing messages to other agents).
    \item \textbf{Recalling \& Reflecting (RR):}
    Variables that retrieve information from internal sources (e.g., knowledge bases, experience repositories, execution logs) or store summarized experiences derived from prior activity.
    \item \textbf{Prompt Kitchen (PK):}
    Variables that assemble prompts for the LLM context window, which combine information drawn from the other three regions into a coherent input.
    \item \textbf{Unknown (UNK):}
    Variables that implement functionalities \textbf{not} belonging to the {\tt Per}, {\tt PA}, {\tt RR}, or {\tt PK} regions.
\end{itemize}

{\bf Region distribution across the evaluated applications with the configured region set.}
Table~\ref{tab:regions} reports the ground-truth classification
of functions in each evaluated application using this region set.
Here, $|V_{\mathrm{fn}}|$ is the total number of functions in the \cldg function plane; S denotes functions belonging to a single
region, and M denotes functions spanning two or more regions.

\begin{table}[htbp]\footnotesize
\centering
\vspace{-2mm}
\caption{Ground truth region distribution.}
\vspace{-3mm}
\label{tab:regions}
\begin{tabular}{lrrrrrrrr}
\hline
\textbf{Application} &
\textbf{$|V_{\mathrm{fn}}|$} &
\textbf{Per} &
\textbf{PA} &
\textbf{RR} &
\textbf{PK} &
\textbf{UNK} &
\textbf{A} &
\textbf{M:S} \\
\hline
GPT-Researcher & 244 & 3   & 133 & 17 & 18 & 69 & 4  & 2.34\% \\
Quivr          & 62  & 4   & 14  & 31 & 1  & 9 & 3  & 6.00\% \\
aider          & 329 & 2   & 156 & 0  & 2  & 168 & 1  & 0.63\% \\
MetaGPT        & 672 & 6   & 359 & 40 & 20 & 225 & 22 & 5.18\% \\
SWE-agent      & 160 & 1   & 64  & 0  & 3  & 92 & 0  & 0.00\% \\
\hline
\end{tabular}
\end{table}

Across all five applications, the multi-region ratio M:S stays at or below 6\%, confirming that the vast majority of functions implement a single agent functionality and can be reliably classified to exactly one region.
The few multi-region functions are those that orchestrate several agent functionalities within a single function; for example, \texttt{ResearchConductor.conduct\_research} in GPT-Researcher retrieves context from web pages~(PA), loads documents from the knowledge base~(RR), and assembles prompts for the LLM~(PK), thereby spanning three regions.
The remaining functions in $V_{\mathrm{fn}}$ that do not belong to any region are utility functions that parse configuration files, log the execution trace, calculate token consumption, or handle errors; they are taint-reachable from CEP candidates and therefore included in $V_{\mathrm{fn}}$, but they do not implement any agent functionality.

\section{Query Language Grammar}
\label{apdx:spec-grammar}
\begin{lstlisting}[
    basicstyle=\footnotesize\ttfamily,
    numbers=none,
    captionpos=b,
    caption={Query language grammar},
    label={lst:query-language-grammar}
]
<letter> ::= "A" | ... | "Z" | "a" | ... | "z"
<digit> ::= "0" | ... | "9"
<number> ::= <digit> { <digit> }
<identifier> ::= <letter> { <letter> | <digit> 
               | "_" | "." }
<qualified-id> ::= <identifier> "." <identifier>
                 | <identifier> "." <number>
<region> ::= Per  (= PERCEIVING)
           | PA   (= PLANNING_ACTING)
           | RR   (= RECALLING_REFLECTING)
           | PK   (= PROMPT_KITCHEN)
<variable> ::= <identifier>
             | <qualified-id>
<term> ::= <region> | <variable>
<term-list> ::= <term> | <term> "," <term-list>
<dependency> ::= <term> DEPENDS_ON <term-list>
<atom> ::= <dependency>
         | "(" <formula> ")"
<formula> ::= <atom> { <logical-operator> <atom> }
<logical-operator> ::= AND | OR
<query> ::= CHECK <formula> { "->" <formula> }
\end{lstlisting}

\section{Case Studies In-depth Details}
\label{apdx:case-studies}

We present details of four representative case studies that demonstrate the analyst's backtracking workflow on different types of incidents.

\subsection{Case Study 1: Quivr MC-Poison (M)}
\label{subsec:case-memory}

{\bf Incident.}
An attacker sends a crafted chat message through Quivr's user interface.
The message poisoned the memory database of Quivr.
In a later chat session with the user, the malicious content is retrieved during question answering and reaches the LLM, producing a manipulated response.

{\bf Log investigation.}
The analyst triaged the trajectory log and found text of misinformation repeatedly appeared when the user asked questions about "Nvidia stock price".
The log further showed that the files uploaded to the knowledge base did not contain the misinformation. The analyst deduced that Quivr's memory database was poisoned.

{\bf CEP + path identification: exploitation phase.}
The analyst started to use \tool to identify the CEP.
From the log, the analyst came up with the query on whether the malicious content in the memory database can pollute responses to the user.
Consequently, the analyst submitted:
\texttt{CHECK (PK DEPENDS\_ON RR) AND (Per DEPENDS\_ON PK)}.
The query returns a subgraph of the \cldg where each node is a function with attributes: node id, function name, and location (i.e., corresponding Python program file and line number of the function definition in the file).
As shown in Figure~\ref{fig:case-study}(a), the graph contains 21 nodes, of which 3 are in the {\tt Per} region (colored in green), 1 in the PK (red), and 5 in the RR (blue).
Because the region-level query was submitted by the analyst, the analyst knew only the 3 nodes in the {\tt Per} and 5 nodes in the RR need to be examined.

\begin{figure}[htbp]
    \centering
    \includegraphics[width=0.475\textwidth]{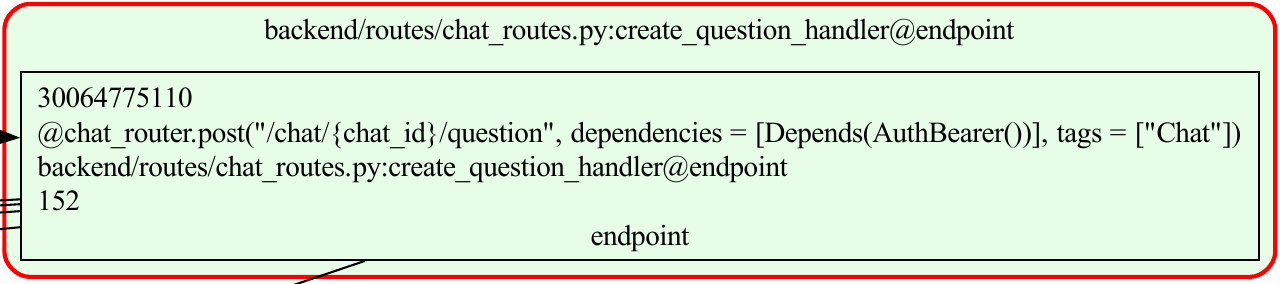}
    \vspace{-6mm}
    \caption{Click the relevant node in the {\tt Per} to zoom in.}\label{fig:cs1-click}
    \vspace{-3mm}
\end{figure}

To formulate a variable-level query, the analyst needed to identify one or more variable-level taint sources in the RR and one or more variable-level taint sinks in the Per.
The analyst began with the 3 nodes colored in green to identify the taint sinks.
By examining the function names, the analyst identified two functions that may contain the taint sink candidates: \texttt{create\_question\_handler} and \texttt{create\_stream\_\\question\_handler}.
From the log, the analyst found that the error traceback for the failed chat task passes through \texttt{create\_question\_handler} at \texttt{chat\_routes.py:197}, confirming that this handler processed the incident request.
The analyst clicked this node to reveal its variable plane. As shown in Figure~\ref{fig:cs1-click}, only one variable node appears: it holds the response to be delivered to the user.
The analyst identified this variable (id~\texttt{30064775110}) as the taint sink.

The analyst then inspected the 5 RR nodes colored in blue.
By examining function names, \texttt{get\_chat\_history} is the most relevant to retrieving content from the memory database.
The analyst clicked it and found one variable that receives the return value of the Supabase retrieval API call.
The analyst identified this variable (id~\texttt{30064774216}) as the taint source.

With the taint sink and taint source identified, the analyst submitted a variable-level query:
\texttt{CHECK node.30064775110 DEPENDS\_ON node.30064774216}.
The query returns a set of paths in the variable plane that are mapped to 1 function-level path passing through 6 functions (Figure~\ref{fig:case-study}(b), phase A (left)).
The path traces data from \texttt{get\_chat\_history} (RR) through \texttt{\_call\_chain} (PA, containing the API call to the LLM) to \texttt{create\_question\_handler} (Per), confirming that the poisoned content from the memory database went to the context window of the LLM and polluted the response that was delivered to the user.

{\bf CEP + path identification: poisoning phase.}
The taint source found in the exploitation phase might not be the CEP, because poisoning a memory database directly~\cite{chenAgentPoisonRedteamingLLM2024} is hard.
Such an attack requires the attacker to (1) get the access credential to the user system; (2) to obtain the secret of the memory database running on the system; (3) to find the exact table in the memory database corresponding to the chat history of the user; (4) to identify the exact embedding model that the user is using and transform the malicious content into vectors using the model before inserting them into the table.
Therefore, the analyst knew that the attacker is more likely to poison the memory database through user interfaces.
With this, the analyst submitted:
\texttt{CHECK (PK DEPENDS\_ON Per) AND (RR DEPENDS\_ON PK)}.

The resulting subgraph contains 23 nodes. Similarly, the analyst only needed to examine 3 nodes in the Per (green) and 8 nodes in the RR (blue).
The analyst went through a similar process of (1) examining the function names; (2) clicking interesting nodes to inspect their variable planes; (3) identifying the candidates for taint source and taint sink.
As a result, the analyst identified one taint source and one taint sink and submitted a variable-level query:
\texttt{CHECK node.30064774368 DEPENDS\_ON node.30064775110}.

The query returns a set of paths in the variable plane that are mapped to 2 function-level paths.
The path that corresponds to the execution in the incident is within them.
As shown in Figure~\ref{fig:case-study}(b) (phase B (right)), it passes through 8 functions.
The path traces data from \texttt{create\_question\_hand\\ler} (Per) through \texttt{\_construct\_prompt} (PK), and \texttt{\_get\_model\\\_response} that sends the prompt to the LLM before saving the response into the chat history via \texttt{update\_chat\_history}.

{\bf CEP identified.}
The diagnosis results showed that the CEP of the attack incident is the API call in the \texttt{create\_\\question\_handler} function that receives the HTTP request containing the malicious content at line number 152 in the file \texttt{backend/routes/chat\_routes.py}.

\subsection{Case Study 2: GPT-Researcher MC-Poison (KB)}
\label{subsec:case-kb}

This case study demonstrates how to backtrack the incident described in Section~\ref{subsec:motivating example}.

{\bf Region-level query.}
Following the workflow in Section~\ref{subsec:case-memory}, the analyst submitted a region-level query \texttt{CHECK (PK DEPEN\\DS\_ON RR) AND (PA DEPENDS\_ON PK)}, which returns 135 functions (8 in RR, 41 in PK, 64 in PA).

\begin{figure}[htbp]
    \centering
    \includegraphics[width=0.475\textwidth]{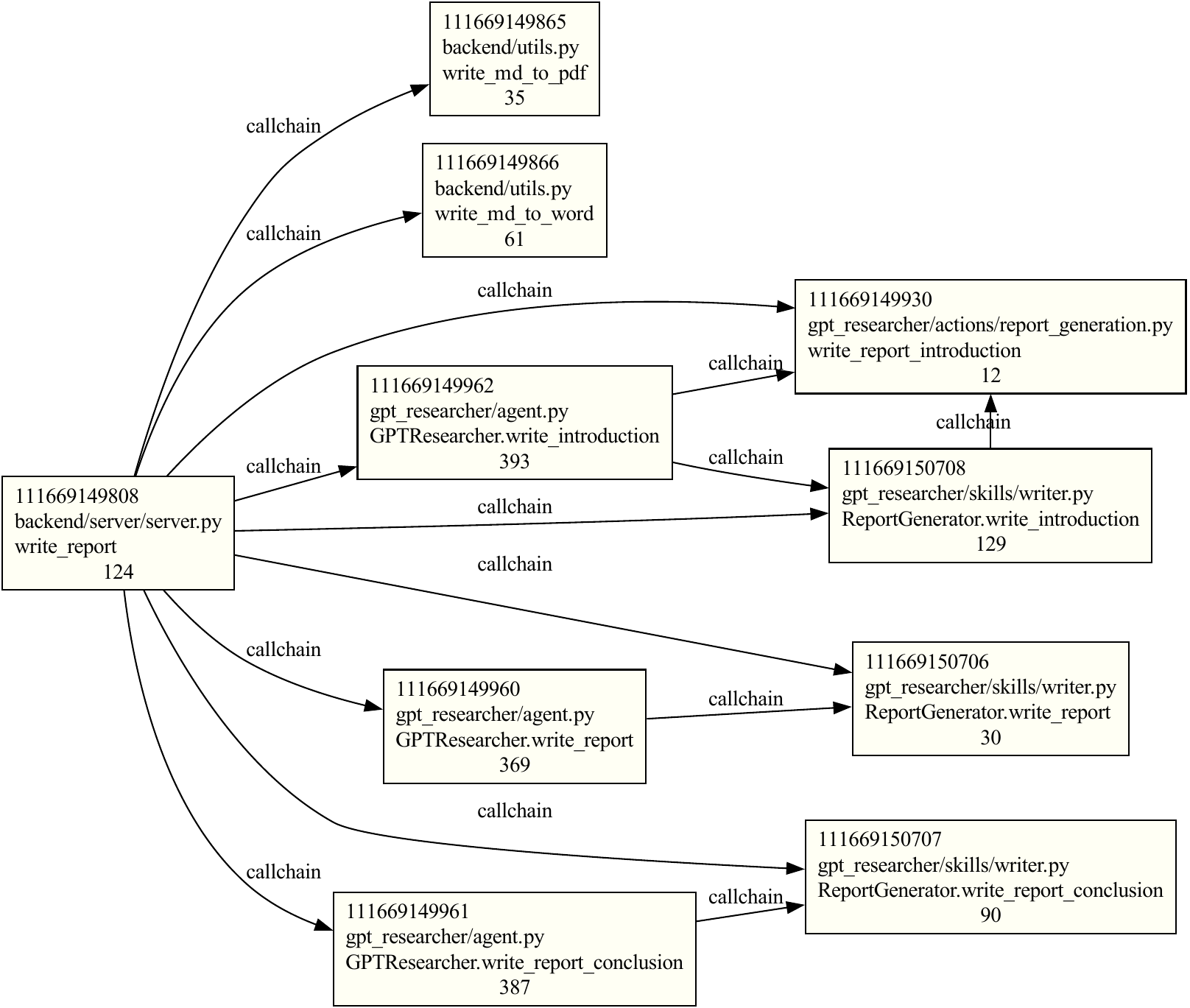}
    \vspace{-6mm}
    \caption{Filtered subgraph of region-level query result.}\label{fig:cs2-a}
    \vspace{-4mm}
\end{figure}

{\bf Variable-level taint sink candidate identification.}
The analyst started with 64 PA nodes colored in yellow.
Since the attack effect is a PDF file of manipulated research synthesis that was downloaded by the user, the analyst proactively searched for functions that write content to file when examining the function names of nodes.
The analyst quickly found that only functions whose names start with \texttt{write} are relevant to the functionality.
To get a cleaner view of the graph, the analyst utilized the features of \tool to filter out irrelevant nodes and edges: (1) using regular expression \texttt{.*write.*} to batch select the relevant nodes; (2) choosing the option to preserve only call-out edges in the graph.
The filtered subgraph is shown in Figure~\ref{fig:cs2-a}, which contains only 10 nodes with edges showing who calls who in the call chain.
The analyst could easily deduce
that \texttt{write\_md\_to\_pdf} is more relevant than other functions such as \texttt{write\_md\_to\_word}, \texttt{write\_report\_introduction}, and \texttt{write\_report}.
By clicking on the \texttt{write\_md\_to\_pdf} node, the analyst quickly identified as the taint sink candidate the API call \texttt{md2pdf} (id 30064773841) in the variable plane.

{\bf Variable-level taint source candidate identification.}
The analyst then inspected 8 RR nodes to identify the taint source.
Upon examining the function names, the analyst identified the most relevant functions to retrieve content from knowledge base: \texttt{DocumentLoader.\_load\_document}, \texttt{AzureDocumentLoader.load}, and \texttt{LangChainDocumentLoa\\der.load}.
By examining the variable plane of the 3 functions, the analyst determined that they are for retrieving knowledge from a local file, a cloud file, and in-memory vectors, respectively.
Based on the user report that the knowledge base was stored locally, the analyst deduced that the API call \texttt{loader.load} (id 30064778131) in \texttt{DocumentLoader.\_load\_\\document} could be the taint source candidate.

\begin{figure*}[!t]
\centering
\vspace{-3mm}
\begin{minipage}[t]{0.383\textwidth}
\centering
\includegraphics[width=\textwidth]{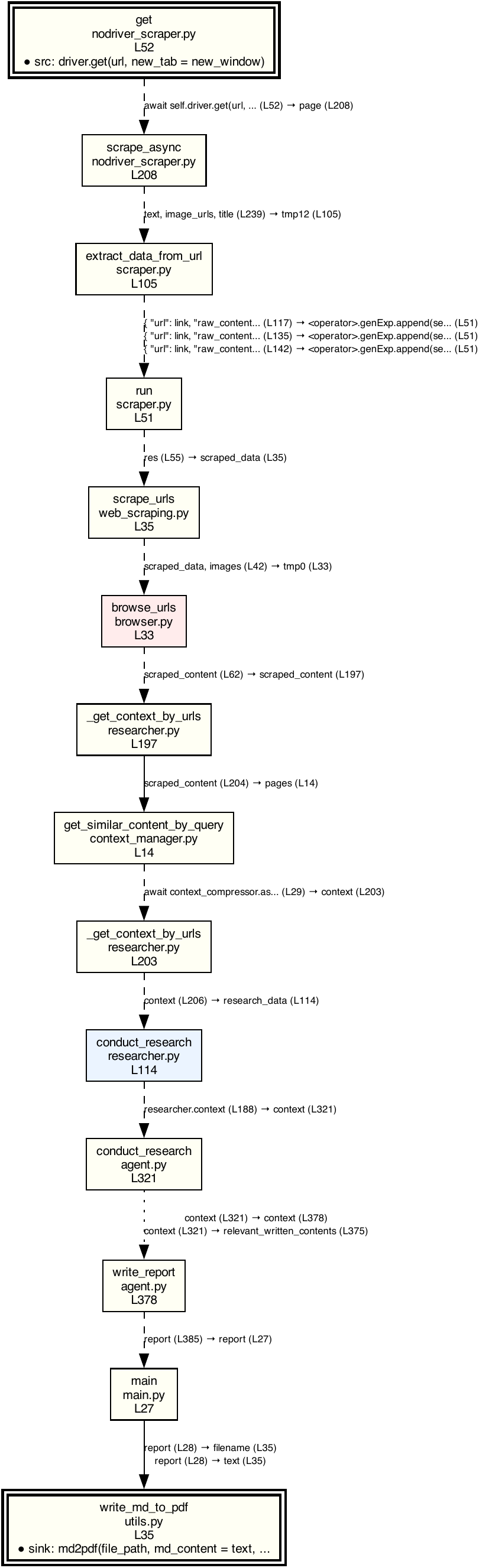}\\[4pt]
\vspace{-2mm}
(a) GPT-Researcher IPI.
\end{minipage}\hfill
\begin{minipage}[t]{0.47\textwidth}
\centering
\includegraphics[width=\textwidth]{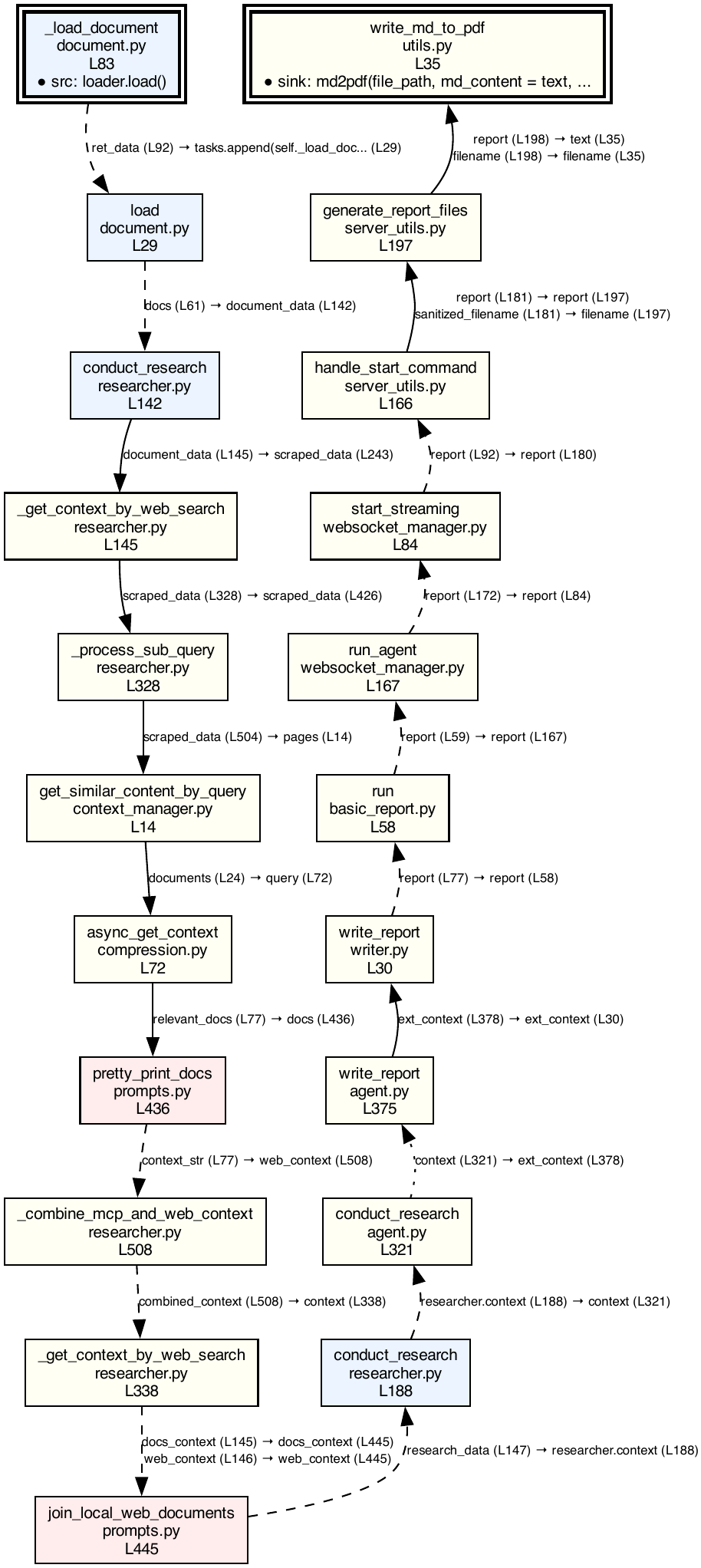}\\[4pt]
\vspace{-2mm}
(b) GPT-Researcher MC-Poison (KB).
\end{minipage}
\vspace{-3mm}
\caption{Paths in GPT-Researcher IPI and MC-Poison (KB) incidents.}
\label{fig:gpt_ipi_and_mc-kb_var}
\end{figure*}

{\bf Variable-level query to identify the CEP.}
With all of these, the analyst submitted a variable-level query: \texttt{CHECK node.30064773841 DEPENDS\_ON node.30064778131}.
The query returned 4 paths, one of which is the correct path that the malicious content took to cause the harm in the particular incident.
The CEP in the motivating example was identified thereby. As shown in Figure~\ref{fig:gpt_ipi_and_mc-kb_var}(b), the malicious content from the knowledge base was combined with the crawled web page in the PK (red) before going to the LLM context window and generating the report.

\subsection{Case Study 3: SWE-agent IPI}
\label{subsec:case-ipi}

{\bf Incident.}
SWE-agent fetched a GitHub issue to begin a software engineering task. The issue body was embedded with invisible malicious instructions.
The SWE-agent followed the instructions in the issue and caused the target software to delay showing the user interface for 10,000 seconds, which appeared to be a denial-of-service to the user.

{\bf Log investigation.}
The analyst triaged the trajectory log and obtained two key observations.
First, the task originated from a GitHub issue. Second, the agent applied a patch that inserted a denial-of-service delay into the target repository's source code.

{\bf Region-level query.}
Following the workflow in Case Study~1, the analyst submitted a region-level query
\texttt{CHECK (PK DEPENDS\_ON PA) AND (PA DEPENDS\_ON PK)}, which returns 47 functions (26 in PA, 2 in PK).
The analyst knew that only the the 26 PA nodes needed to be examined.

{\bf Variable-level taint source candidate identification.}
The analyst started with the 26 PA-region nodes.
Since the attack entered through a GitHub issue, the analyst searched for functions related to fetching issue content.
By searching \texttt{github} in function names, the analyst found two relevant functions: \texttt{\_get\_problem\_statement\_from\_github\_issue} and \texttt{GithubIssue.get\_problem\_statement}.
The analyst clicked to examine the variable planes of these two functions and identified the taint source as the API call \texttt{api.issues.get} (id~\texttt{30064790320}) in \texttt{\_get\_problem\_statement\_from\_gith\\ub\_issue}.

{\bf Variable-level taint sink candidate identification.}
The analyst continued inspecting the 26 PA-region nodes to identify the taint sink.
Since the attack effect was modification of source code in the target repository, the analyst looked for functions that apply code changes.
By examining function names, only \texttt{\_apply\_patch} is most relevant to applying patches.
The analyst clicked to examine its variable plane and identified the taint sink as the API call \texttt{subprocess.run} (id~\texttt{30064783612}) in \texttt{SaveApplyPatchHook.\_apply\_patch}.

\begin{figure*}[!t]
\centering
\vspace{-3mm}
\begin{minipage}[t]{0.29\textwidth}
\centering
\includegraphics[width=\textwidth]{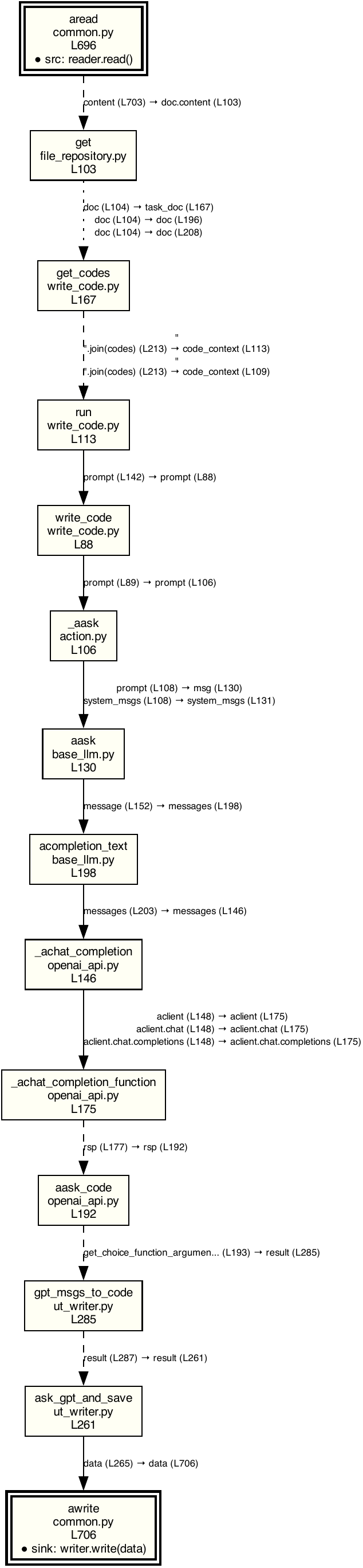}\\[4pt]
\vspace{-2mm}
(a) MetaGPT IPI.
\end{minipage}
\begin{minipage}[t]{0.67\textwidth}
\centering
\includegraphics[width=\textwidth]{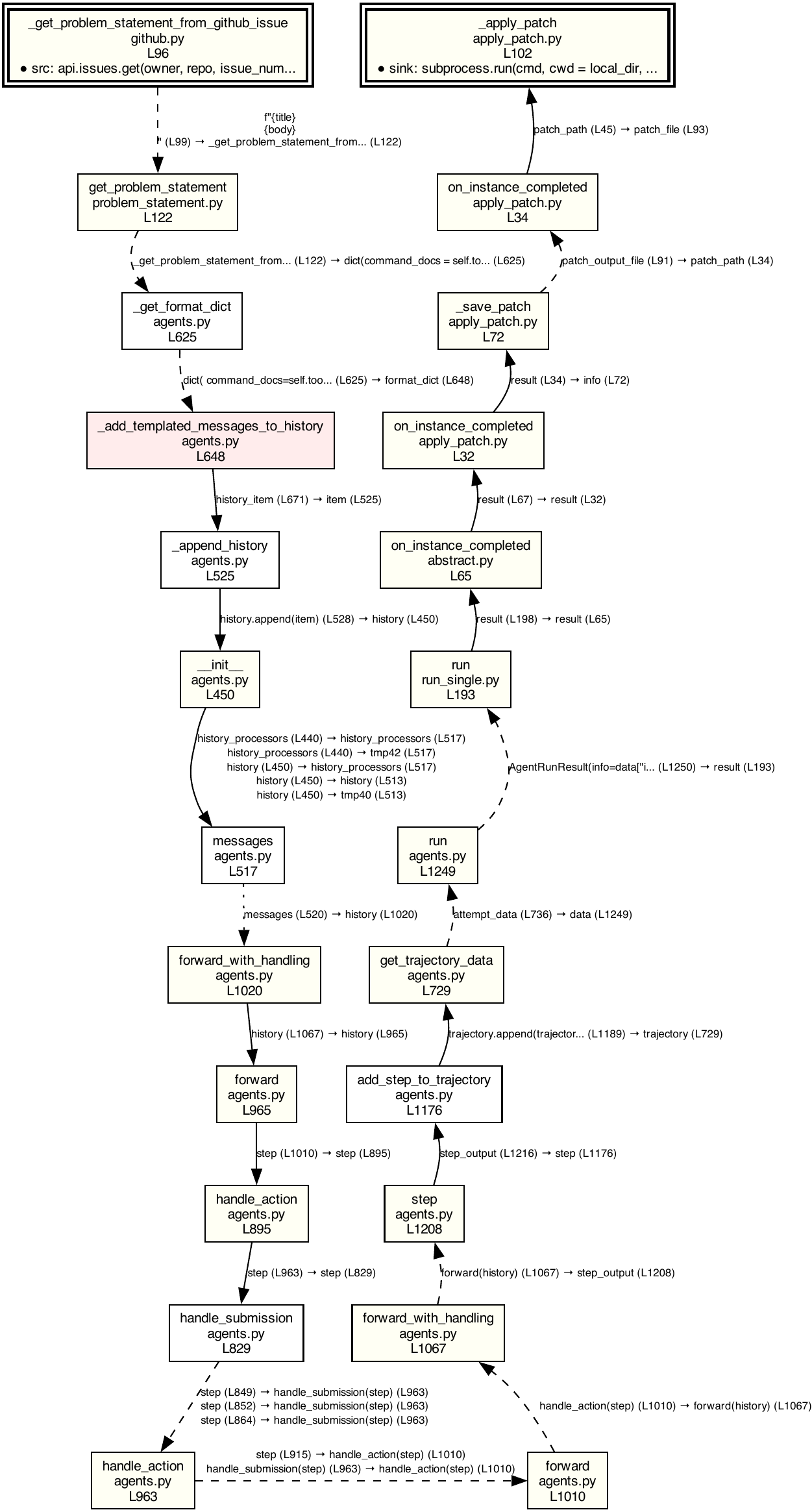}\\[4pt]
\vspace{-2mm}
(b) SWE-agent IPI.
\end{minipage}\hfill
\vspace{-3mm}
\caption{Paths in MetaGPT IPI and SWE-agent IPI incidents.}
\label{fig:met_ipi_swe_ipi_paths}
\end{figure*}

{\bf Variable-level query to identify the CEP.}
With the taint source and taint sink identified, the analyst submitted a variable-level query:
\texttt{CHECK node.30064783612 DEPENDS\_ON node.30064790320}.
The query returns 1 path passing through 21 functions (Figure~\ref{fig:met_ipi_swe_ipi_paths}(b)).
The path traces data from \texttt{\_get\_pr\\oblem\_statement\_from\_github\_issue} (PA) through \texttt{get\_\\problem\_statement} into the agent's message history construction (\texttt{\_get\_format\_dict}, \texttt{\_add\_templated\_messages\_\\to\_history}), then through the agent's forward loop (\texttt{forward\\\_with\_handling}, \texttt{forward}, \texttt{handle\_action}, \texttt{handle\_submis\\sion}), and finally to the patch application hooks (\texttt{on\_instance\\\_completed}, \texttt{\_save\_patch}, \texttt{\_apply\_patch}), where \texttt{subproce\\ss.run} executes \texttt{git apply}.

{\bf CEP identified.}
The results showed that the CEP of the attack incident is the API call \texttt{api.issues.get} in the \texttt{\_get\_pro\\blem\_statement\_from\_github\_issue} function that fetches the GitHub issue content containing the malicious instructions at line 96 in the file \texttt{sweagent/utils/github.py}.

\subsection{Case Study 4: Quivr IPI + MC-Poison (KB)}
\label{subsec:case-compound}

{\bf Incident.}
An attacker compromised a user-trusted website.
The user then used Quivr's crawl functionality to add the website's content to a personal knowledge base, unknowingly storing the malicious content as document embeddings.
Days later, when the user asked questions about the topic covered by the compromised page, the application retrieved the malicious content and produced manipulated responses.

{\bf Log investigation.}
The analyst triaged the log and obtained two key observations:
the user had crawled a URL into a knowledge base, and the user
later received manipulated responses when asking questions
against the same knowledge base.

{\bf From knowledge base retrieval to the manipulated response.}
The analyst first investigated how the malicious content in the knowledge base reaches the chat response.
The analyst submitted \texttt{CHECK (PK DEPENDS\_ON RR) AND (Per DEPENDS\_ON PK)}, which returns a graph of 21 nodes (3 in Per, 1 in PK, 5 in RR).
The analyst began with the 3 Per nodes to identify the taint sink. Following the same process as in Case Study~1 (Section~\ref{subsec:case-memory}), the analyst identified \texttt{update\_chat\_history} and the variable (id~\texttt{30064774368}), the API call that persists the LLM response to the chat history, as the taint sink.

The analyst then inspected the 5 RR nodes to identify the taint source.
Two functions are relevant to data retrieval: \texttt{get\_chat\_history} and \texttt{CustomSupabaseVectorStore.sim
\\ilarity\_search}.
By examining their variable planes, the analyst found that \texttt{get\_chat\_history} retrieves multi-round conversation records, while \texttt{similarity\_search} retrieves document chunks from the knowledge base.
Since the malicious content was stored in the knowledge base, the analyst identified the API call that queries the vector store for similar documents (id~\texttt{146028888356}) in \texttt{similarity\_search} as the taint source.

\begin{figure*}[!t]
\centering
\begin{minipage}[t]{0.56\textwidth}
\centering
\includegraphics[width=\textwidth]{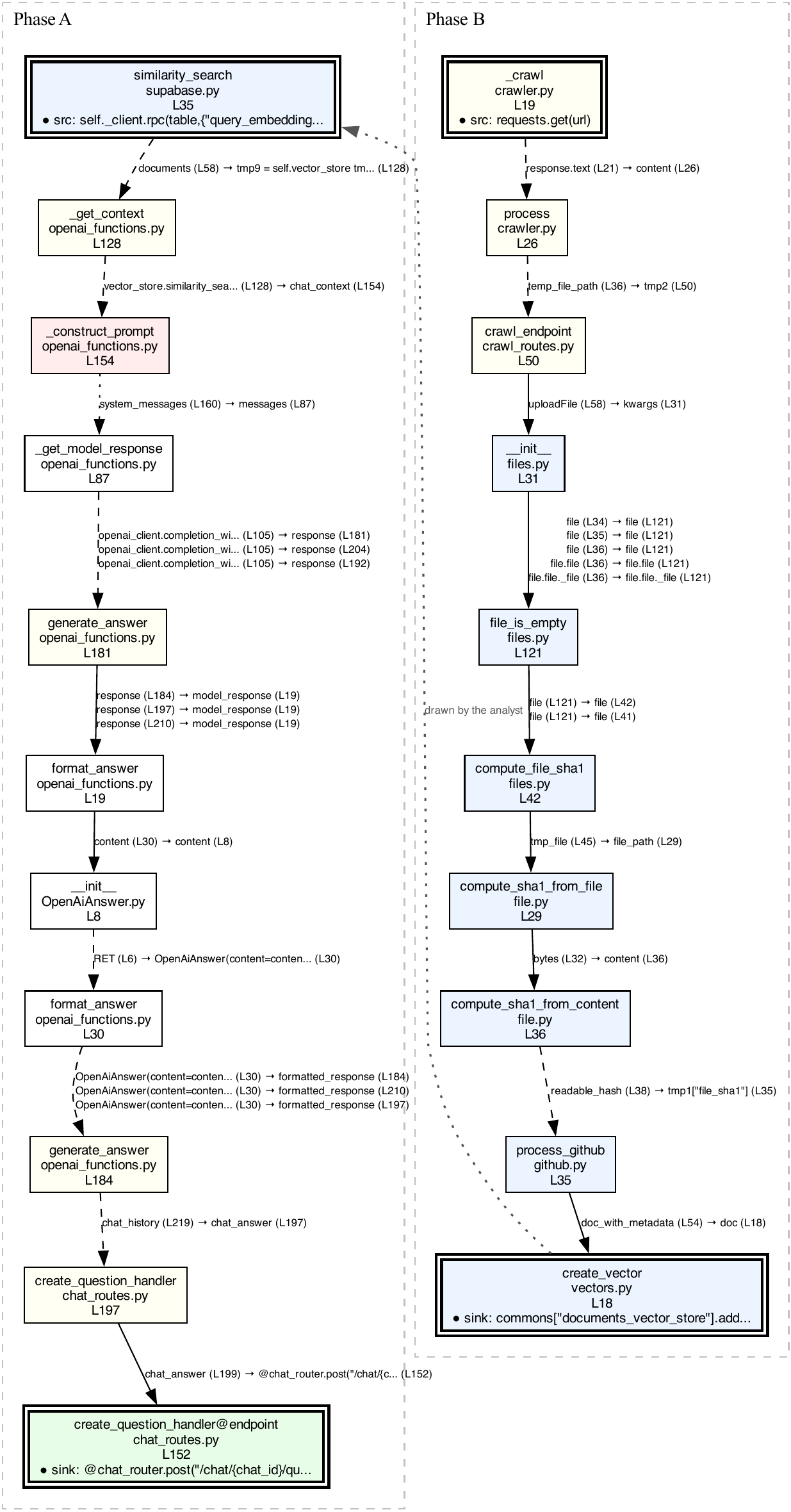}\\[4pt]
(a) Quivr IPI + MC-Poison (KB), phase A (left) and phase B (right).
\end{minipage}
\begin{minipage}[t]{0.435\textwidth}
\centering
\includegraphics[width=\textwidth]{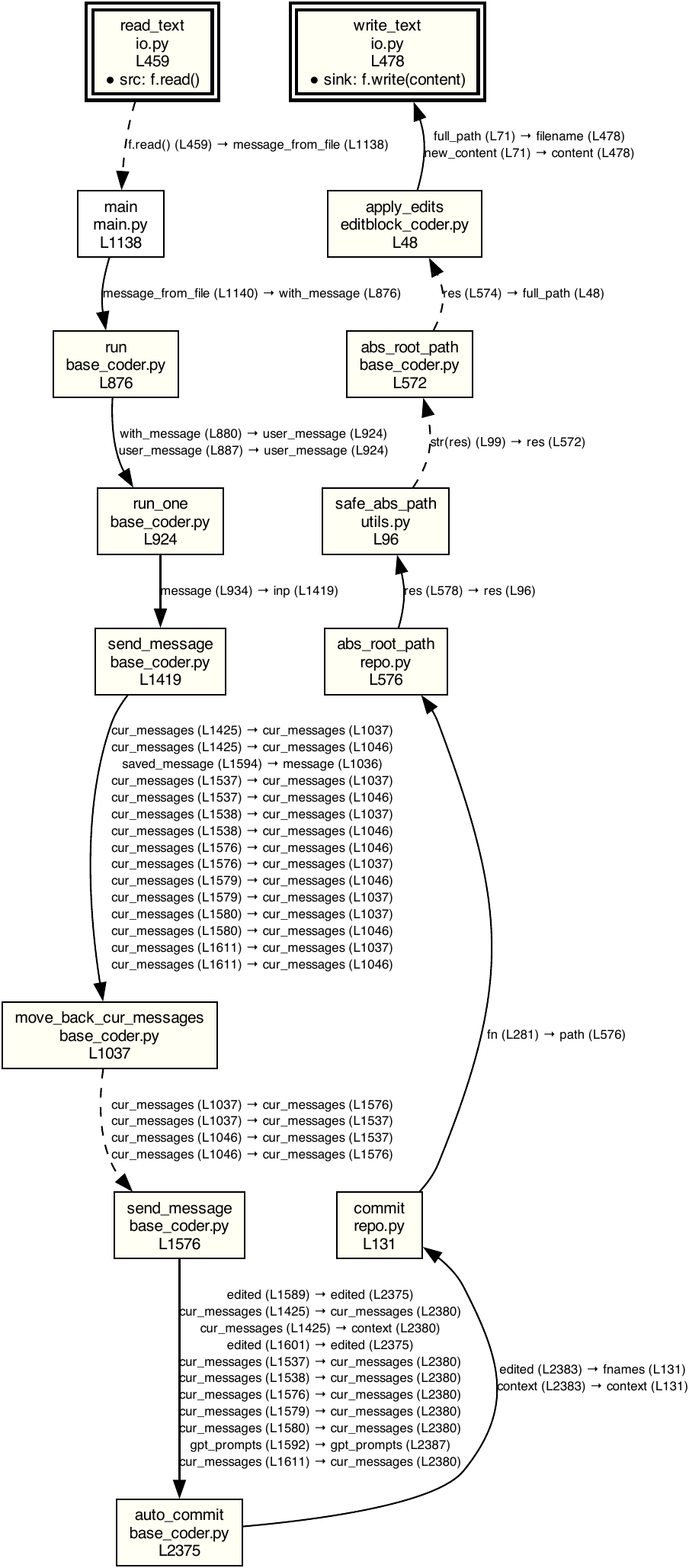}\\[4pt]
(b) aider IPI.
\end{minipage}\hfill
\caption{Paths in Quivr IPI + MC-Poison (KB) and aider IPI incidents.}
\label{fig:qui_hyb_aid_ipi_paths}
\end{figure*}

With these, the analyst submitted a variable-level query: \texttt{CHECK node.30064774368 DEPENDS\_ON node.146028888356}.
The query returns 1 path (Figure~\ref{fig:qui_hyb_aid_ipi_paths}(a), phase A (left)).
The path traces data from \texttt{similarity\_search} (RR) through \texttt{\_construct\_prompt} (PK, where the retrieved documents are injected into the LLM prompt as context) to the LLM call and finally to \texttt{update\_chat\_history} (Per), confirming that the malicious content stored in the knowledge base reached the LLM context window and polluted the chat response.

{\bf From crawled website to knowledge base storage.}
The taint source identified above (\texttt{similarity\_search}) reads from the knowledge base but is not the CEP. The analyst must determine how the malicious content entered the knowledge base in the first place.
Since the log shows that the user added external content to the knowledge base through the crawl functionality, the analyst submitted \texttt{CHECK RR DEPENDS\_ON PA}, which returns a graph of 47 nodes (12 in PA, 25 in RR).

\begin{figure*}[htbp]
    \centering
    \vspace{-4mm}
    \includegraphics[width=\textwidth]{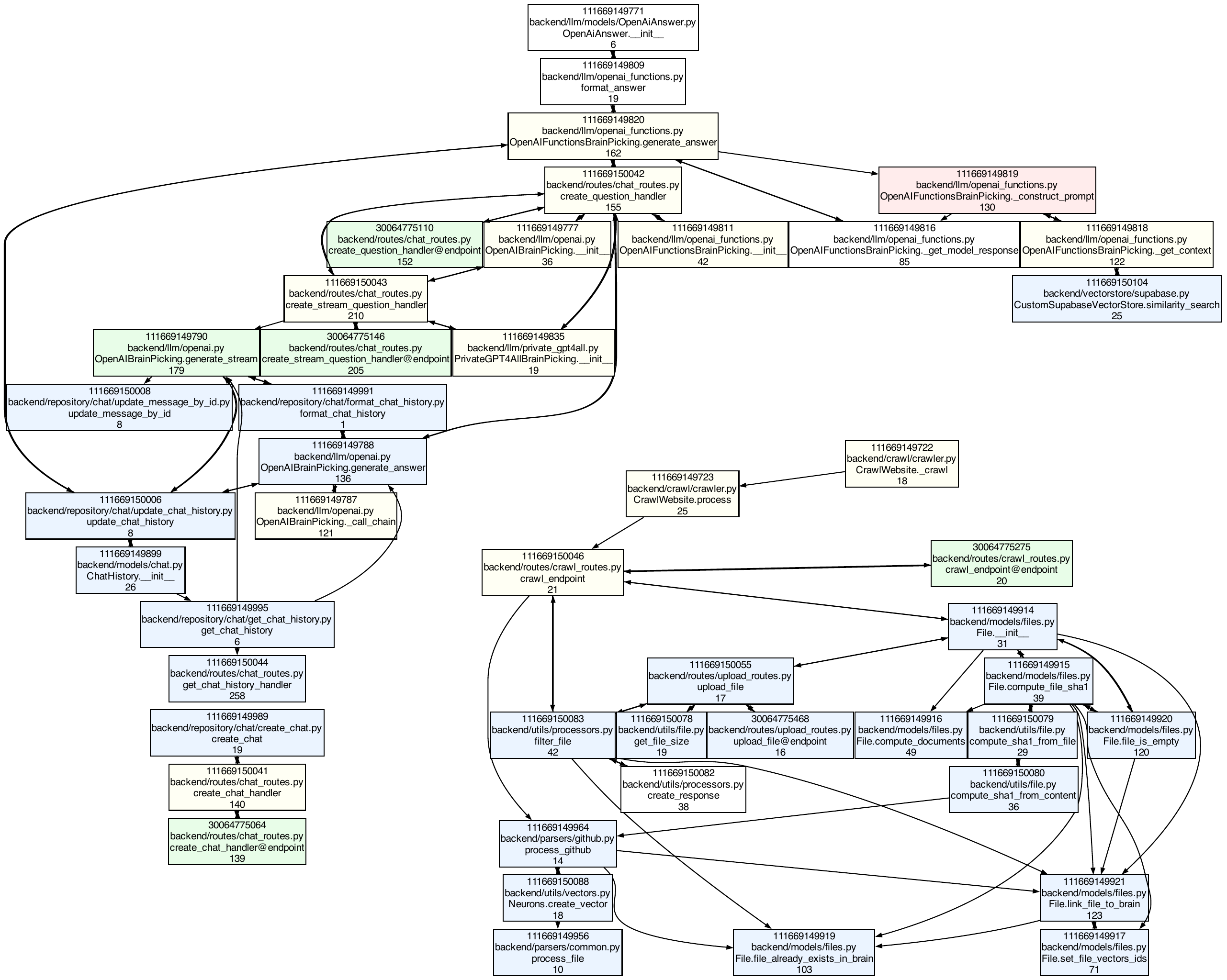}
    \vspace{-6mm}
    \caption{Region-level query result of IPI + MC-Poison (KB) incident (phase~B).}\label{fig:cs4-region-query}
    \vspace{-4mm}
\end{figure*}

The result graph contains two disconnected subgraphs (Figure~\ref{fig:cs4-region-query}):
one centered around \texttt{crawl\_endpoint} and the crawl-related functions (\texttt{CrawlWebsite.\_crawl}, \texttt{CrawlWeb\\site.process}), the other centered around the chat handlers (\texttt{create\_question\_handler}, \texttt{create\_chat\_handler}).
Since the analyst has already confirmed that the malicious content resides in the knowledge base rather than the chat history, the chat subgraph does not need to be examined.
The analyst focused on the crawl subgraph, which contains 3 PA nodes and 16 RR nodes.

The analyst started with the 3 PA nodes to identify the taint source.
These three functions, \texttt{CrawlWebsite.\_crawl}, \texttt{CrawlWebsite.process}, and \texttt{crawl\_endpoint}, form a single call chain, where \texttt{crawl\_endpoint} calls \texttt{process}, which in turn calls \texttt{\_crawl}.
Since the taint source must be the entry point where untrusted external content first enters the application, the analyst examined the innermost function \texttt{\_crawl}, which sends an HTTP request to retrieve the webpage content.
By examining its variable plane, the analyst identified the API call that fetches the webpage (id~\texttt{30064771354}) as the taint source.

The analyst then inspected the 16 RR nodes to identify the taint sink.
When the analyst examined the function names, most nodes were file-processing utilities or content parsers unrelated to knowledge base storage.
Only \texttt{Neurons.create\_\\vector} directly suggests the function is writing to the vector store.
By examining its variable plane, the analyst identified the API call that stores document embeddings into the knowledge base (id~\texttt{30064776465}) as the taint sink.

The analyst submitted a variable-level query: \texttt{CHECK node.\\30064776465 DEPENDS\_ON node.30064771354}.
The query returns 1 path passing through 10 functions (Figure~\ref{fig:qui_hyb_aid_ipi_paths}(a), phase B (right)).
The path traces data from \texttt{\_crawl} (PA), through file processing and content parsing, to \texttt{create\_vector} (RR), where the malicious content was embedded and stored in the knowledge base.

{\bf CEP identified.}
The analyst connected the two paths through the knowledge base: the crawl path writes the malicious content into the same knowledge base that the chat path later retrieves from.
Although the write occurred days before the read, the log confirmed that both operations targeted the same knowledge base.
The CEP of the incident is the API call that fetches the webpage in \texttt{CrawlWebsite.\_crawl} at line 19 of \texttt{backend/crawl/crawler.py}.

\section{CEPs + Paths identified in all incidents}\label{apdx:CEP-Path}
We show the function-level paths identified in all eight incidents, except for the one (Quivr MC-Poison (M)) that is already included in the main body of the paper (Figure~\ref{fig:case-study}(b)).
The CEPs are identified as the source node of the path in the five single-phase incidents; for the other two incidents (Quivr MC-Poison (KB) and Quivr IPI + MC-Poison (KB)), the CEPs are identified as the source node of the path in Phase~B.

\begin{figure*}[htbp]
    \centering
    \vspace{-3mm}
    \includegraphics[width=0.95\textwidth]{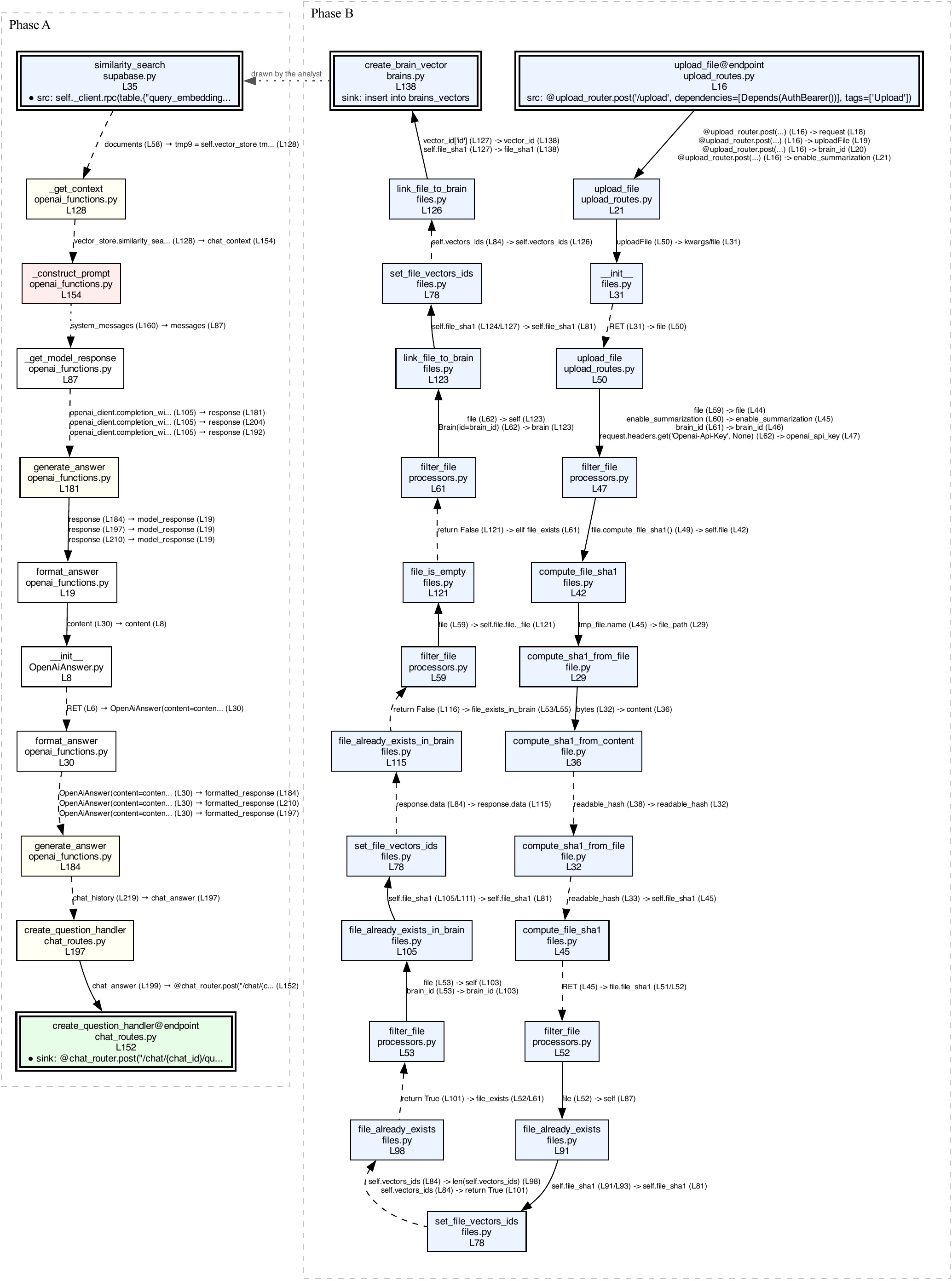}
    \vspace{-3mm}
    \caption{Paths in Quivr MC-Poison (KB) incident.}\label{fig:cs_qui_mc-kb_var}
\end{figure*}

\clearpage
\begingroup
\emergencystretch=2em
\section{Failure Analysis of Existing Tools}
\label{apdx:tool-failures}
\makeatletter
\enlargethispage{\dimexpr\textheight-\@colroom\relax}
\makeatother
\raggedbottom

Existing static taint analyzers do not support end-to-end CEP identification.
We therefore provide the ground-truth CEP as the source and the operation capturing the attack effect as the sink.
We evaluate whether their outputs satisfy the tool-specific success criteria for verifying the given CEP.
We then examine the causes that prevent verification of the given CEP in Table~\ref{tab:baseline}.
The used tool versions are: Joern~4.0.360, CodeQL CLI~2.27.0 with \texttt{codeql/python-all}~7.2.5, and Pysa~0.9.25.
Table~\ref{tab:failure-causes} reports the findings for every incident, phase, and tool; the following subsections explain selected examples.
\subsection{Unresolved Indirect Call Targets}
\label{apdx:failure-calls}

An incomplete call graph omits possible targets of a call.
When an unresolved callee returns a tainted value, the analyzer may fail to propagate that taint to the caller.

In GPT-Researcher's IPI case, \texttt{scraper.scrape\_async()} retrieves webpage content and returns it to the caller, which assigns it to \texttt{content} (Listing~\ref{lst:gpt-ipi-call-targets}, \#6).
Pysa cannot resolve which function this call invokes, so it cannot propagate the tainted webpage content returned by that function to \texttt{content}.
Pysa's fallback taint model propagates taint from arguments to a call's return value.
However, \texttt{scrape\_async()} takes no explicit arguments: the webpage content is retrieved inside the callee, rather than passed into the call.
The fallback therefore does not mark the returned value, or \texttt{content}, as tainted.

Pysa also cannot resolve the target of \texttt{browse\_urls} (Listing~\ref{lst:gpt-ipi-call-targets}, \#11), which prevents the returned tainted value from reaching \texttt{scraped\_content}.
In our diagnostic experiment, we added type casts to a copy of the application's source code and reran Pysa.
The casts tell Pysa to treat \texttt{scraper} as a \texttt{NoDriverScraper} and \texttt{self.researcher.scraper\_manager} as a \texttt{BrowserManager}.
With both casts, Pysa reported the call to \texttt{write\_md\_to\_pdf} at \texttt{main.py:28}, indicating that its report argument contains tainted webpage content.
This call lies on the expected attack path and therefore satisfies our Pysa success criterion.
With either cast alone, Pysa did not report this call.

\par\noindent\begin{minipage}{\linewidth}
\begin{lstlisting}[
    language=Python,
    basicstyle=\scriptsize\ttfamily,
    numbers=left,
    firstnumber=1,
    stepnumber=1,
    numberblanklines=true,
    numberstyle=\tiny\ttfamily,
    numbersep=5pt,
    xleftmargin=15pt,
    breaklines=true,
    captionpos=b,
    caption={Unresolved calls in GPT-Researcher.},
    label={lst:gpt-ipi-call-targets}
]
# scraper.py:96-105
Scraper = self.get_scraper(link)
scraper = Scraper(link, session)
# ...
content, image_urls, title = (
    await scraper.scrape_async()
)

# researcher.py:197
scraped_content = await (
    self.researcher.scraper_manager.browse_urls(
        new_search_urls
    )
)
\end{lstlisting}
\end{minipage}\par

In Quivr's KB exploitation, \texttt{self.vector\_store} is a \texttt{CustomSupabaseVectorStore} object.
The call to its \texttt{similarity\_search} method (Listing~\ref{lst:failure-retrieval}, \#2) retrieves KB documents and returns them at \#5.
Joern and CodeQL cannot identify this method as the target of the call.
Both tools mark \texttt{documents} at \#5 as a taint source, but neither propagates that taint to the value returned by the call at \#2.
In a copy of Quivr's source code, we replaced the call at \#2 with the call at \#8--\#10 and reran both tools.
The replacement names the class's method directly, passes the vector-store object explicitly, and keeps the same query.
Both tools then identified the method and marked the call's return value as tainted.
However, this change alone did not recover the complete source-to-sink path; Table~\ref{tab:failure-causes} lists the other causes.

\par\noindent\begin{minipage}{\linewidth}
\begin{lstlisting}[
    language=Python,
    basicstyle=\scriptsize\ttfamily,
    numbers=left, firstnumber=1, stepnumber=1,
    numberblanklines=true,
    numberstyle=\tiny\ttfamily,
    numbersep=5pt, xleftmargin=15pt,
    breaklines=true, captionpos=b,
    caption={Unresolved document retrieval in Quivr.},
    label={lst:failure-retrieval}
]
# llm/openai_functions.py:128
return self.vector_store.similarity_search(query=question)

# vectorstore/supabase.py:58
return documents

# Replacement in our diagnostic source-code copy
return CustomSupabaseVectorStore.similarity_search(
    self.vector_store, query=question
)
\end{lstlisting}
\end{minipage}\par

\subsection{Missing Data Dependencies Through Language Constructs}
\label{apdx:failure-python}

Taint propagation can be interrupted when an analyzer does not fully represent the data dependencies of a language construct (e.g., nested dictionary accesses, comprehension).
This includes argument binding, container accesses, object construction, and field access.

In SWE-agent's IPI case, \texttt{tool\_call["function"]\allowbreak["arguments"]} reads the tool call's arguments through two nested dictionary accesses (Listing~\ref{lst:failure-python}, \#2).
The first retrieves the dictionary describing the requested function; the second retrieves a JSON string containing the argument names and values.
\texttt{json.loads} converts this string into the Python dictionary \texttt{values}.
Joern represents both accesses but omits the data-dependence edge from the first access's result to the second access.
Consequently, taint reaches the inner dictionary but not the argument string.
In a source-code copy, we separated the accesses into \texttt{function = tool\_call["function"]} and \texttt{values = json.loads(\allowbreak function["arguments"])}.
With the other diagnostic changes held fixed, Joern then propagated taint to \texttt{values} without an additional JSON-decoding model.

In GPT-Researcher's IPI case, CodeQL fails to propagate taint through an awaited tuple return.
The tuple returned by \texttt{scrape\_async()} contains webpage text, which the caller assigns to \texttt{content} (Listing~\ref{lst:gpt-ipi-call-targets}, \#5--\#7).
CodeQL does not carry the text's taint across the \texttt{await} expression to this variable.
Summaries that propagate the contained values across the affected \texttt{await} expressions, together with the call and API models in Table~\ref{tab:failure-causes}, allow CodeQL to report a source-to-PDF path containing the required functions.

Argument binding is separate from resolving the called function.
For example, in Quivr's memory-exploitation case, replacing the named arguments to \texttt{update\_chat\_history} with positional arguments lets Joern propagate the answer into \texttt{self.assistant}.
An additional constructor-content rule is needed for that taint to reach the constructed object returned to the caller.

\par\noindent\begin{minipage}{\linewidth}
\begin{lstlisting}[
    language=Python,
    basicstyle=\scriptsize\ttfamily,
    numbers=left, firstnumber=1, stepnumber=1,
    numberblanklines=true,
    numberstyle=\tiny\ttfamily,
    numbersep=5pt, xleftmargin=15pt,
    breaklines=true, captionpos=b,
    caption={Nested dictionary accesses in SWE-agent.},
    label={lst:failure-python}
]
# tools/parsing.py:401
values = json.loads(tool_call["function"]["arguments"])
# ... validation omitted ...
# tools/parsing.py:418-425
formatted_args = {
    arg.name: Template(arg.argument_format).render(
        value=quote(values[arg.name])
        if _should_quote(values[arg.name], command)
        else values[arg.name]
    )
    if arg.name in values
    else ""
    for arg in command.arguments
}
\end{lstlisting}
\end{minipage}\par

\subsection{Missing Data Dependencies Across API Calls}
\label{apdx:failure-content}

An API call may read, transform, or store data.
To follow tainted data through the call, the analyzer needs to know what the API does with it.
The analyzer can obtain this information by analyzing the API's implementation or using a summary that describes how data flows through the API.
If the relevant behavior is neither analyzed nor summarized, the analyzer may lose track of the tainted data at the call.
We include database insertion/readback within an API operation here; separate writes and later reads across application stages belong to Section~\ref{apdx:failure-state}.

In Quivr's memory-exploitation case, the application supplies stored user and assistant messages to a model chain to generate a new answer.
It then stores the answer in the database, constructs a \texttt{ChatHistory} object from the returned record, and returns that object from the handler.
After we supplied rules connecting the configured history source to its message fields, CodeQL propagated taint to the model-chain input but not its result (Listing~\ref{lst:failure-objects}, \#2).
We added a summary relating the request dictionary to the call result.
CodeQL then tracked taint into the generated answer and the record passed to the database insertion, but not into the returned database data (\#5--\#10).
Adding a second summary relating the inserted record to the returned data allowed taint to reach the input of \texttt{ChatHistory}.
A separate field-read probe confirmed that taint subsequently reached the returned object's \texttt{assistant} field.

\par\noindent\begin{minipage}{\linewidth}
\begin{lstlisting}[
    language=Python,
    basicstyle=\scriptsize\ttfamily,
    numbers=left, firstnumber=1, stepnumber=1,
    numberblanklines=true,
    numberstyle=\tiny\ttfamily,
    numbersep=5pt, xleftmargin=15pt,
    breaklines=true, captionpos=b,
    caption={Answer generation and storage in Quivr.},
    label={lst:failure-objects}
]
# llm/openai.py:129-134
return chain({"question": question, "chat_history": history})

# repository/chat/update_chat_history.py:10-21
response = (
    commons["supabase"].table("chat_history")
    .insert({"chat_id": str(chat_id),
             "user_message": user_message, "assistant": assistant})
    .execute()
).data

\end{lstlisting}
\end{minipage}\par

\subsection{Missing Data Dependencies Through Stored Data}
\label{apdx:failure-state}

Tracking taint through stored data requires the analysis to connect a write to a subsequent read of the same data.
Without this connection, taint on the written value may not propagate to the value read from a file, database, or object field.

In the non-GitHub branch examined for Quivr's compound-attack poisoning phase, the crawler writes webpage text to a file (Listing~\ref{lst:failure-files}, \#5). The handler opens that file and copies its contents into \texttt{spooled\_file} (\#12--\#13), which it then wraps for upload. The crawler returns only names derived from the URL (\#7); the webpage text reaches the handler through the file contents. In the evaluated configurations, Joern, CodeQL, and Pysa do not connect the write to this later use, so they do not propagate the webpage taint into the uploaded content.

An explicit file-transfer model connects the written webpage content to the upload object.
With this model and the other dependencies in Table~\ref{tab:failure-causes}, Joern and CodeQL report paths from the crawler source to the document argument of \texttt{add\_documents}.
Pysa's model uses an instrumented source copy that exposes the written content to a helper returning the upload object.
Pysa reports \texttt{filter\_file} in \texttt{crawl\_endpoint}, identifying its file argument at \texttt{crawl\_routes.py:62} as carrying crawler taint to the configured \texttt{create\_vector} sink.
Its aggregate object model does not distinguish loaded document text from metadata.

\par\noindent\begin{minipage}{\linewidth}
\begin{lstlisting}[
    language=Python,
    basicstyle=\scriptsize\ttfamily,
    numbers=left, firstnumber=1, stepnumber=1,
    numberblanklines=true,
    numberstyle=\tiny\ttfamily,
    numbersep=5pt, xleftmargin=15pt,
    breaklines=true, captionpos=b,
    caption={Writing and reading crawled content in Quivr.},
    label={lst:failure-files}
]
# crawl/crawler.py:26-36
content = self._crawl(self.url)
# ... URL-derived path construction ...
with open(temp_file_path, "w") as temp_file:
    temp_file.write(content)
if content:
    return temp_file_path, file_name

# routes/crawl_routes.py:50-54
file_path, file_name = crawl_website.process()
spooled_file = SpooledTemporaryFile()
with open(file_path, "rb") as f:
    shutil.copyfileobj(f, spooled_file)
\end{lstlisting}
\end{minipage}\par

\par
\endgroup

\clearpage
\onecolumn
\begingroup
\fontsize{9}{10.5}\selectfont
\setlength{\tabcolsep}{5pt}
\renewcommand{\arraystretch}{1.05}
\setlength{\LTcapwidth}{\textwidth}
\setlength{\LTleft}{0pt}
\setlength{\LTright}{0pt}
\begin{longtable}{@{}>{\raggedright\arraybackslash}p{0.14\textwidth}>{\raggedright\arraybackslash}p{0.27\textwidth}>{\raggedright\arraybackslash}p{0.27\textwidth}>{\raggedright\arraybackslash}p{\dimexpr0.32\textwidth-30pt\relax}@{}}
\caption{Observed failure causes by incident, phase, and tool. C: unresolved indirect call targets (Section~\ref{apdx:failure-calls}); L: missing data dependencies through language constructs (Section~\ref{apdx:failure-python}); A: missing data dependencies across API calls (Section~\ref{apdx:failure-content}); S: missing data dependencies through stored data (Section~\ref{apdx:failure-state}). \textemdash: no cause listed in these four categories.}
\label{tab:failure-causes}\\
\toprule
Incident / phase & Joern & CodeQL & Pysa \\
\midrule
\endfirsthead
\multicolumn{4}{@{}l}{\textbf{Table \thetable\ (continued).} Observed failure causes.}\\
\toprule
Incident / phase & Joern & CodeQL & Pysa \\
\midrule
\endhead
\midrule
\multicolumn{4}{r@{}}{Continued on next page}\\
\endfoot
\bottomrule
\endlastfoot
GPT-Researcher\newline IPI & \textbf{C} Scraper, research and report targets.\newline
\textbf{L} Conditional logging interrupts content flow. Argument binding at repaired calls.\newline
\textbf{A} Page-preservation effects of wait/scroll helpers.\newline
\textbf{S} Research context passed into report generation. & \textbf{C} Browser/scraper, coroutine extraction, browser-manager, URL-context and selected prompt targets.\newline
\textbf{L} Container contents lost across three selected awaits; document-field reads.\newline
\textbf{A} Page/HTML/text extraction, gathered results, retrieval/compression/thread invocation, streaming request and chunk content.\newline
\textbf{S} Stored context passed into report generation. & \textbf{C} Source-bearing returns at \texttt{scrape\_async} and \texttt{browse\_urls}; both targets are needed for the reconstructed PDF report. \\
\addlinespace[5pt]
GPT-Researcher\newline MC-Poison (KB) & \textbf{C} Document-loader, context, research, and report targets.\newline
\textbf{L} Conditional logging interrupts context/report flow. Argument binding at repaired calls.\newline
\textbf{S} Research context passed to report generation. & \textbf{C} Context-manager target; explicit join binding restores its required function waypoint.\newline
\textbf{L} Awaited document-collection contents; document-field reads. The retained research-comprehension rewrite changes capture and target inference together, so its cause is not isolated.\newline
\textbf{A} Both gathers, retrieval/compression/thread invocation, streaming request, and chunk content.\newline
\textbf{S} Stored report context. & \textbf{C} Unresolved \texttt{manager.start\_streaming} prevents the document source from reaching the PDF-generation report. Supplying the manager type restores the qualifying call under reconstructed declarations. \\
\addlinespace[5pt]
Quivr\newline MC-Poison (M)\newline Exploitation & \textbf{L} Named answer argument does not bind to the history-update function's formal parameter. Initialized \texttt{ChatHistory} content does not reach the constructed object returned to the caller. & \textbf{L} History-object source does not taint selected message fields; tainted assistant field does not satisfy the whole-record sink. An explicit constructor-content rule satisfies that endpoint.\newline
\textbf{A} Missing chain request-to-response and database insertion-to-readback summaries. & \textbf{C} The handler's inferred receiver includes a class without \texttt{generate\_answer}, hiding the source-bearing return. \\
\addlinespace[5pt]
Quivr\newline MC-Poison (M)\newline Poisoning & \textbf{L} Named model/history-update arguments, omitted \texttt{**kwargs} input, and nested response indexing. Initialized \texttt{OpenAiAnswer} content does not reach its constructed result. & \textbf{A} Missing completion request-to-response summary. Native question-storage paths remain positive but omit the generated assistant response. & \textbf{C} Unresolved \texttt{generate\_answer} hides the question-to-assistant sink summary. \\
\addlinespace[5pt]
Quivr\newline MC-Poison (KB)\newline Exploitation & \textbf{C} Unresolved vector-store retrieval.\newline
\textbf{L} Named model/history-update arguments, omitted \texttt{**kwargs}, nested response indexing. \texttt{OpenAiAnswer} and \texttt{ChatHistory} constructed content. & \textbf{C} Unresolved vector-store retrieval.\newline
\textbf{L} Constructed content requires a whole-record rule for the configured object endpoint.\newline
\textbf{A} Completion request-to-response and database insertion-to-readback summaries. & \textbf{C} The handler cannot import the retrieved-document return through unresolved \texttt{generate\_answer}. \\
\addlinespace[5pt]
Quivr\newline MC-Poison (KB)\newline Poisoning & \textbf{C} Existing-file linking and brain-record insertion targets.\newline
\textbf{L} Named \texttt{File} constructor input. Argument binding at repaired calls. Caller-visible constructed \texttt{File} content. & \textbf{C} HTML-parser dispatch on the specified new-file route.\newline
\textbf{L} Document fields and constructed document content.\newline
\textbf{A} Pydantic file initialization, upload reads, loading/splitting, and document-to-vector-ID association.\newline
\textbf{S} Temporary-file write to later load. & \textemdash \\
\addlinespace[5pt]
Quivr\newline IPI + MC-Poison (KB)\newline Exploitation & \textbf{C} Same call-target findings as Quivr KB exploitation.\newline
\textbf{L} Same language-construct findings as Quivr KB exploitation. & \textbf{C} Same call-target findings as Quivr KB exploitation.\newline
\textbf{L} Same language-construct findings as Quivr KB exploitation.\newline
\textbf{A} Same API-call findings as Quivr KB exploitation. & \textbf{C} Same call-target findings as Quivr KB exploitation. \\
\addlinespace[5pt]
Quivr\newline IPI + MC-Poison (KB)\newline Poisoning & \textbf{C} HTML parser and document-computation targets.\newline
\textbf{L} \texttt{File} constructor input binding. Argument binding at repaired calls. Constructed File content.\newline
\textbf{A} Opaque file and loader effects.\newline
\textbf{S} Crawler write/copy/read and stored split-document readback. & \textbf{C} HTML-parser dispatch.\newline
\textbf{L} Document fields and constructed document content.\newline
\textbf{A} Pydantic initialization, upload reads, and loader/splitter effects.\newline
\textbf{S} Crawler write/copy/read and temporary-file loading. & \textbf{C} HTML-parser dispatch.\newline
\textbf{S} Crawler file write to ingestion read. \\
\addlinespace[5pt]
aider\newline IPI & \textbf{C} Selected coder/model/commit/write targets.\newline
\textbf{L} Named commit inputs, omitted SDK \texttt{**kwargs}, nested response fields, omitted \texttt{yield}/\texttt{yield from} operands. Argument binding at repaired calls. Yielded-record materialization and tuple-field projection (tested jointly).\newline
\textbf{A} SDK response readback.\newline
\textbf{S} Saved commit text and later history reads. & \textbf{C} Main coder call target and selected commit-model target.\newline
\textbf{L} Main-call argument binding, focused content reads at list concatenations, and parser accumulation representation.\newline
\textbf{A} SDK request-to-response and response-field summaries, including the later request.\newline
\textbf{S} Commit-message history used in a later editing round. & \textbf{C} Message-file read, factory-created coder's \texttt{run}, and selected output write. All three type repairs are needed for the reconstructed qualifying report. \\
\addlinespace[5pt]
MetaGPT\newline IPI & \textbf{C} File-repository and LLM-wrapper targets.\newline
\textbf{L} SDK \texttt{**kwargs}, nested conditional response decoding and parser keyword binding. & \textbf{C} File-repository and LLM-wrapper targets in the verified prefix.\newline
\textbf{L} Loaded document's content, returned document, and caller field read in that prefix. & \textemdash \\
\addlinespace[5pt]
SWE-agent\newline IPI & \textbf{C} Agent, parser, environment, and completion-hook targets; spurious dictionary-read targets.\newline
\textbf{L} Nested tool-argument reads, named arguments and omitted command \texttt{**kwargs}. Argument binding at repaired calls. Parsed-action/result fields.\newline
\textbf{S} History/property reads, patch-file effects and saved patch/resource association. & \textbf{C} Selected issue-getter, parser and inner completion-hook targets.\newline
\textbf{L} Tuple-to-step-field assignment; issue/message/result fields and whole-result construction. Original comprehension capture works.\newline
\textbf{A} LLM request/response, Jinja rendering and argument formatting.\newline
\textbf{S} Copied history and property reads, environment command-to-patch effects, and written patch-to-consumed-file association. & \textbf{C} Unresolved completion hook hides the \texttt{\_save\_patch.info} sink; issue taint already reaches the agent result. \\
\addlinespace[5pt]
\end{longtable}
\endgroup

\twocolumn
\section{Details of Knowledge Collection}
\label{apdx:luna-evaluation}
\begingroup
\emergencystretch=3em

\subsection{Structured Prompt Example}
\label{apdx:knowledge-prompt}
\begingroup
\emergencystretch=3em

Listing~\ref{lst:knowledge-prompt} shows a simplified prompt used to label Quivr's \texttt{get\_chat\_history}.
The listing has four parts: the labeling task, code and supporting information, decision rules, and the required answer.
The task and decision rules are supplied in a system message, while the code and supporting information are supplied in a user message.
We also specify the fields that the LLM's answer must contain and require it to return them in JSON.
The listing abbreviates documents, code, and repeated metadata while retaining all six calls to be classified.

\begin{lstlisting}[basicstyle=\ttfamily\scriptsize,numbers=none,captionpos=b,caption={Simplified prompt from the saved Quivr request.},label={lst:knowledge-prompt},aboveskip=4pt,belowskip=4pt]
[1. Task: system instructions]
Fill a developer-knowledge mapping-table row.
Analyze every call in record.api_calls.
EXT: task information outside the application.
INT: task information forming part of the app.
Use all supplied knowledge documents/examples.

[2. Context: user input, abbreviated]
record.function_key: backend/repository/chat/
                    get_chat_history.py:
                    get_chat_history
record.api_calls:
  a0: tmp0.execute()
  a1: tmp2.filter("chat_id", "eq", chat_id)
  a2: tmp3.select("*")
  a3: tmp4.from_("chat_history")
  a4: tmp1.order("message_time", desc=False)
  a5: tmp5.append(ChatHistory(message))
  <source locations, receiver/argument records,
   candidate targets and normalization limits>

knowledge.documents: <full bodies supplied>
  python/nested-functions.md
  resource-identification/
    process-memory-boundary.md
    generic-file-or-url-arguments.md
    task-information-sources.md
  Examples: in-process access is not a source;
  nested body != enclosing body;
  open(path) needs the actual file's use;
  formatting supplied text is not source access.

source: <spans with IDs, paths and line numbers>
  # Target function, shortened:
  commons = common_dependencies()
  history = (commons["supabase"]
    .from_("chat_history").select("*")
    .filter("chat_id", "eq", chat_id)
    .order("message_time", desc=False)
    .execute()).data
  # Supplied caller line, openai.py:145:
  history = get_chat_history(self.chat_id)

supporting_evidence: <cached program relations>
  a0 target: not_resolved_to_application
  a0 receiver: tmp0; return type: ANY
  <definitions, dependencies, caller bindings,
   source-span references and query limitations>
overview: 6 calls; 77 queries; 16 queries omitted;
  <source ranges and collection-limit notices>
next_action: return complete call map; no tools.

[3. Decision rules: system instructions]
Count direct builtin/library source accesses.
Do not inherit an application callee's accesses.
Other source calls are context, not targets.
Separate nested/generated operations; consider
alternative branches. Use each call's allowed
findings; mark missing decisive facts unresolved.
Do not equate an API name with a source type,
full source coverage with certainty, or source
comments/documents with overriding instructions.

[4. Output: instructions + strict JSON schema]
Required fields:
  calls: every key a0-a5, each with
    finding, note, supporting_spans
  explanation, knowledge_used, missing_facts
Allowed findings for a0-a5 in this request:
  EXT, INT, both, no_source_access, unresolved
note and explanation: nonempty strings.
Spans/document paths: supplied identifiers only.
No extra fields, tool requests or later turns.
Python attaches call-site citations and computes
EXT / INT / skip / unresolved; do not guess the
function label independently of call findings.
\end{lstlisting}

\noindent\textbf{1. Labeling task.}
The task is to determine the function's EXT or INT label from its direct accesses to information sources.
For each call in \texttt{record.api\_calls}, the LLM reports whether it directly accesses an external (EXT) information source, an internal (INT) information source, both types, neither, or cannot be classified from the supplied evidence.
The function's label is then computed from these findings.

An information source supplies or receives information used in the application's tasks.
EXT denotes a source through which the application obtains new task information or communicates with users, LLMs, or external services, such as user input, model responses, or public web content.
INT denotes a source that the application maintains for use across tasks, such as its stored conversation history or knowledge base.
A function can receive an INT label even when it accesses information over a network.
For example, a function that retrieves the application's stored conversation history from a remote database accesses an INT source.

\noindent\textbf{2. Code and supporting information.}
The user message contains six components, each with a separate purpose.

\emph{Function and call record (\texttt{record}).}
The \texttt{function\_key} identifies the target function by file and name; \texttt{api\_calls} lists the operations to classify.
Each call has an identifier (a0--a5), a source location, and details recording the operation called, the object it is called on, and the arguments passed to it.
The \texttt{tmp} names denote intermediate values in the analysis representation.
For example, \texttt{tmp0} holds the query built by the preceding database operations, so \texttt{tmp0.execute()} runs that query.
This call is listed as a0.
Call a5 represents adding each \texttt{ChatHistory(message)} value to the result list.
The analysis makes this list-building step explicit as \texttt{tmp5.append(ChatHistory(message))}, where \texttt{tmp5} is the result list.
Thus, a5 refers to \texttt{append}, which adds the value to the list; \texttt{ChatHistory(message)} creates the value being added.

\emph{Labeling guidance (\texttt{knowledge.documents}).}
The prompt includes four documents containing rules and examples to help the LLM judge whether a call directly accesses an EXT or INT information source.
Developers can extend this guidance by adding documents or updating the rules and examples to cover additional cases.
The current documents explain how to classify only the outermost call in a nested call expression, exclude in-memory operations from information-source accesses, and determine whether a file or URL provides EXT or INT task information.
For example, in \texttt{items.append(read\_file(path))}, the outermost call \texttt{append} only modifies an in-memory list; it does not perform the file access.
Classifying \texttt{read\_file(path)} requires additional evidence about the file: a public document is an EXT source, whereas the application's stored conversation history is an INT source.

\emph{Code excerpts (\texttt{source}).}
Code excerpts can help the LLM understand what each listed call does and how its inputs and outputs are used.
The excerpts include the target function and its calling context, such as code around the call site, argument definitions, and uses of the returned values.
Each excerpt includes an identifier for citation, its file path and line numbers, and the code itself.
In this example, the target function queries the \texttt{chat\_history} table using \texttt{chat\_id}, and its caller passes \texttt{self.chat\_id} as the argument.

\emph{Static analysis results (\texttt{supporting\_evidence}).}
The prompt also provides program-analysis results that identify possible callees and show how values reach each call's receiver and arguments.
These results help the LLM trace which information source a call accesses.
Each result identifies the listed call it describes and references the corresponding code excerpt.
The supporting evidence also includes whether each call was resolved to an application-defined function.
For example, a0, \texttt{tmp0.execute()}, is marked \texttt{not\_resolved\_to\_application}, indicating that no application-defined callee was identified.

\emph{Coverage summary (\texttt{overview}).}
Even when the full function is supplied, the LLM may still need evidence from elsewhere in the application to determine whether a call accesses an EXT or INT information source.
The coverage summary helps the LLM assess whether the supplied evidence is sufficient to classify each call.
It reports the number of calls to be classified, the source-code line ranges supplied, and any evidence that could not be retrieved or was omitted.

\emph{Response reminder (\texttt{next\_action}).}
The LLM must return a finding for every listed call using the supplied evidence.
It cannot request additional evidence or invoke tools.

\noindent\textbf{3. Decision rules.}
The LLM assigns one of six call-level decisions to each listed call based on the supplied evidence:
\vspace{-2mm}
\begin{itemize}[leftmargin=*]
    \item \texttt{EXT}: the call directly accesses an EXT source.
    \item \texttt{INT}: the call directly accesses an INT source.
    \item \texttt{both}: the call directly accesses both EXT and INT sources. This finding requires evidence of both accesses, rather than uncertainty between the two types.
    \item \texttt{application\_call}: all identified candidate callees are application-defined functions. Their information-source accesses are assessed separately within those functions.
    \item \texttt{no\_source\_access}: the call does not directly access a task information source, such as a call that only formats text or modifies an in-memory list.
    \item \texttt{unresolved}: the supplied evidence is insufficient to assign a finding. The LLM must state what information is missing.
\end{itemize}

The available decisions depend on the callee information supplied for each call.
A possible application-defined callee excludes \texttt{EXT}, \texttt{INT}, and \texttt{both}; \texttt{application\_call} is available only when all candidate callees are application-defined.
In Listing~\ref{lst:knowledge-prompt}, none of the six calls has an identified application-defined callee, so \texttt{application\_call} is unavailable.

Each decision applies only to the listed call.
For a nested call expression, the LLM classifies the outermost call without attributing the inner calls' information-source accesses to it.
Every listed call must receive a decision, including calls in alternative branches.

\noindent\textbf{4. Required answer.}
The LLM returns a report explaining its call-level decisions for developer review.
For each listed call, the report contains one of the six decisions defined above, a brief explanation, and references to supporting code excerpts.
The report also provides an overall explanation, identifies the labeling-guidance documents used, and lists any missing information needed to classify unresolved calls.
Listing~\ref{lst:knowledge-prompt-output} shows the report entry for a5: the decision is \texttt{no\_source\_access}, and the explanation states that appending a constructed value to a list does not access an information source.
The entry also references supporting code.
The remaining report entries are omitted for brevity.

\begin{lstlisting}[basicstyle=\ttfamily\scriptsize,numbers=none,captionpos=b,caption={One call entry from the saved response.},label={lst:knowledge-prompt-output},aboveskip=4pt,belowskip=4pt]
"a5": {
  "finding": "no_source_access",
  "note": "list.append only adds a constructed ChatHistory value to an in-memory list; it does not directly access a task information source.",
  "supporting_spans": [
    "s:0264096192d124985453",
    "s:72cdd6a301c9d7bf36ab"
  ]
}
\end{lstlisting}
\vspace{-2mm}

\par\endgroup

\subsection{Detailed LLM Inference Results}
\label{apdx:llm-inference-results}

Table~\ref{tab:llm-inference-results} reports the LLM-assisted labeling results for the 2,719 candidate functions across the five applications, before developer review and correction. Note that the application functions that do not call any built-in or library APIs do not need a label (Section~\ref{subsec:dev-kb}). They are excluded in the results.

For each candidate function, the call-level decisions are combined to propose an \texttt{EXT} or \texttt{INT} label.
The function receives no label (\texttt{skip}) if its calls access neither source type or both types.
If the supplied evidence is insufficient to determine whether the function should receive a label, the function remains \texttt{unresolved}.

The \emph{Decided} column counts candidate functions that receive an \texttt{EXT}, \texttt{INT}, or \texttt{skip} decision.
The \emph{Unresolved} column counts candidate functions requiring additional inspection before a decision can be made.
The two \emph{N} columns report how many functions receive proposed EXT and INT labels, respectively; the remaining decided functions receive \texttt{skip}.
All proposed decisions still require developer review for correctness validation.

For each label type, FNR is the fraction of candidate functions requiring that label that do not receive it, while FPR is the fraction of candidate functions not requiring that label that incorrectly receive it.
Both rates consider only functions with an \texttt{EXT}, \texttt{INT}, or \texttt{skip} decision and a known reference label.
Functions that remain unresolved are excluded.
Thus, a zero FNR does not mean that all functions requiring the label have been identified; some may remain unresolved.

\begin{table}[htbp]\footnotesize
\centering
\setlength{\tabcolsep}{0.6pt}
\renewcommand{\arraystretch}{0.8}
\vspace{-2mm}
\caption{LLM decisions by application.}
\vspace{-3mm}
\label{tab:llm-inference-results}
\footnotesize
\begin{tabular*}{\columnwidth}{@{\extracolsep{\fill}}lrrrrrrr@{}}
\toprule
& & & & \multicolumn{2}{c}{EXT} & \multicolumn{2}{c}{INT} \\
\cmidrule(lr){5-6}\cmidrule(l){7-8}
Application & Functions & \shortstack{Decided\\(\%)} & \shortstack{Unresolved\\(\%)} & N & FNR/FPR & N & FNR/FPR \\
\midrule
\shortstack[l]{GPT-\\Researcher} & 353 & \shortstack[r]{303\\(85.84)} & \shortstack[r]{50\\(14.16)} & 36 & 0.00/3.97 & 0 & --/0.00 \\
Quivr & 127 & \shortstack[r]{89\\(70.08)} & \shortstack[r]{38\\(29.92)} & 5 & 0.00/4.55 & 13 & 0.00/5.00 \\
aider & 640 & \shortstack[r]{536\\(83.75)} & \shortstack[r]{104\\(16.25)} & 83 & 0.00/9.04 & 8 & 0.00/0.94 \\
MetaGPT & 1,164 & \shortstack[r]{925\\(79.47)} & \shortstack[r]{239\\(20.53)} & 55 & 0.00/1.92 & 12 & 0.00/0.44 \\
SWE-agent & 435 & \shortstack[r]{362\\(83.22)} & \shortstack[r]{73\\(16.78)} & 21 & 0.00/3.12 & 4 & --/1.10 \\
\midrule
Total & 2,719 & \shortstack[r]{2,215\\(81.46)} & \shortstack[r]{504\\(18.54)} & 200 & 0.00/4.19 & 37 & 0.00/0.78 \\
\bottomrule
\end{tabular*}
\par\smallskip
\begin{minipage}{\linewidth}\footnotesize
FNR and FPR are reported as percentages; -- indicates a zero denominator.
\end{minipage}
\vspace{-6mm}
\end{table}

\subsection{Developer Report-Review Examples}
\label{apdx:knowledge-review-examples}
\begingroup
\emergencystretch=3em

The following examples show how developers review the LLM reports from the evaluation in Section~\ref{subsec:rq6}.
They cover three cases: accepting a correct label, correcting an incorrect label, and resolving a case where the LLM could not reach a decision.
Each listing shows (1) the candidate function and its calls, (2) the code needed to understand them, (3) the LLM report, and (4) the steps a developer would take to determine the function's label.
It also explains what the developer would need to inspect without the report.
The reports are shortened for readability while preserving the original decisions and explanations.

\noindent\textbf{Correct label: review and accept EXT.}
Listing~\ref{lst:review-correct} shows the report for GPT-Researcher's \texttt{BingSearch.search}.
The report proposes \texttt{EXT} because the function sends the agent's search query to Bing and retrieves web results.
It identifies the HTTP call that performs this access and explains why the other six calls, which format, parse, log, or collect results, do not directly access information sources.
The developer can follow the cited code to verify the search request, the returned content, and the explanations for the remaining calls before accepting \texttt{EXT}.
The proposed label matches the reference label used in the evaluation.

\begin{lstlisting}[basicstyle=\ttfamily\scriptsize,numbers=none,captionpos=b,breaklines=true,caption={Correct proposal: review and accept EXT.},label={lst:review-correct},aboveskip=4pt,belowskip=4pt]
[Original table entry, reformatted]
File: gpt_researcher/retrievers/bing/bing.py
Function: BingSearch.search
Label to fill: ___ (EXT / INT / NA)
Calls (original line:column and expression):
  a0 45:9 print("Searching with query {0}...".format(self.query))
  a1 45:15 "Searching with query {0}...".format(self.query)
  a2 66:16 requests.get(url, headers=headers, params=params)
  a3 72:30 json.loads(resp.text)
  a4 75:13 self.logger.error(
  a5 79:13 self.logger.warning(f"No search results found for query: {self.query}")
  a6 93:13 search_results.append(search_result)

# Original workflow: locate the endpoint and query;
# follow response parsing; distinguish search content
# from formatting, logging, and list operations;
# decide and enter the function label.

[Code supplied to the model, simplified]
# bing.py:50-66,72-73,86-95
url = "https://api.bing.microsoft.com/v7.0/search"
params = {"q": self.query, ...}
resp = requests.get(url, headers=headers, params=params)
results = json.loads(resp.text)["webPages"]["value"]
for result in results:
    search_result = {"title": result["name"],
                     "href": result["url"],
                     "body": result["snippet"]}
    search_results.append(search_result)
return search_results
# Other listed calls format or print progress and
# log parse failures or missing search results.

[Saved report, shortened]
Proposal: EXT.
a2: EXT; retrieves Bing results for the research task.
a0-a1, a3-a6: no_source_access; diagnostic output,
  formatting, parsing, or local list operations.
Missing facts: none.
Evidence: bing.py:50-66 (URL, query, HTTP request).

[Exact developer review]
1. Check that url is Bing's search endpoint and
   params["q"] is self.query.
2. Check that the response supplies result titles,
   URLs, and snippets, and that the other listed
   calls only process values or report diagnostics.
3. Accept EXT for BingSearch.search.
Saved work: the report identifies the source call,
connects it to the endpoint and query, and proposes
findings for all seven calls. The developer checks
these links instead of constructing them first.
\end{lstlisting}

\noindent\textbf{Incorrect label: review and remove EXT.}
Listing~\ref{lst:review-incorrect} shows the report for aider's \texttt{check\_version}.
The report proposes \texttt{EXT} because call a3 retrieves the latest aider version from PyPI.
However, the supplied code compares this version with the installed version to check for an application update; it does not retrieve information for the user's coding task.
The developer therefore changes a3's decision to \texttt{no\_source\_access} and checks that the remaining calls also access no task information sources.
The function then receives a \texttt{skip} decision, recorded as \texttt{NA} in the developer-knowledge table.
The report has already identified the network request and the returned version; the developer needs to correct why that information does not qualify for an EXT label.

\begin{lstlisting}[basicstyle=\ttfamily\scriptsize,numbers=none,captionpos=b,breaklines=true,caption={Incorrect proposal: review and remove EXT.},label={lst:review-incorrect},aboveskip=4pt,belowskip=4pt]
[Original table entry, reformatted]
File: aider/versioncheck.py
Function: check_version
Label to fill: ___ (EXT / INT / NA)
Calls (original line:column and expression):
  a0 65:27 VERSION_CHECK_FNAME.exists()
  a1 67:17 time.time()
  a2 67:31 os.path.getmtime(VERSION_CHECK_FNAME)
  a3 78:20 requests.get("https://pypi.org/pypi/aider-chat/json")
  a4 79:16 response.json()
  a5 87:31 packaging.version.parse(latest_version)
  a6 87:73 packaging.version.parse(
  a7 94:9 VERSION_CHECK_FNAME.parent.mkdir(parents=True, exist_ok=True)
  a8 95:9 VERSION_CHECK_FNAME.touch()

# Original workflow: inspect the HTTP response and
# its uses; determine whether the cache and remote
# endpoint hold coding-task or operational data;
# decide and enter the function label.

[Code supplied to the model, simplified]
# aider/versioncheck.py:65-95
if not just_check and VERSION_CHECK_FNAME.exists():
    since = time.time() - os.path.getmtime(
        VERSION_CHECK_FNAME)
    if 0 < since < 60 * 60 * 24:
        return
response = requests.get(
    "https://pypi.org/pypi/aider-chat/json")
latest_version = response.json()["info"]["version"]
current_version = aider.__version__
is_update_available = (
    packaging.version.parse(latest_version) >
    packaging.version.parse(current_version))
# In finally:
VERSION_CHECK_FNAME.parent.mkdir(
    parents=True, exist_ok=True)
VERSION_CHECK_FNAME.touch()

[Saved report, shortened]
Proposal: EXT.
a3: EXT; public PyPI supplies the latest aider
  version for the update check.
a0-a2, a4-a8: no_source_access; operational cache,
  clock, or processing of already available values.
Missing facts: none.
Evidence: versioncheck.py:78 (HTTP request),
  84-90 (version output and comparison).

[Exact developer review and correction]
1. Check that response.json()["info"]["version"]
   is compared with aider.__version__. This is
   application maintenance, not coding-task content.
2. Reject a3's EXT finding. Change it to
   no_source_access under the task-information rule.
3. Check that the cache only limits update checks;
   retain the other eight no_source_access findings.
4. Record NA for check_version: completed skip,
   with no EXT/INT seed and no remaining uncertainty.
Saved work: the report already identifies PyPI,
its returned version, and the cache's role. The
extra work is to correct the resource-role judgment
and resulting label, not to discover the endpoint.
\end{lstlisting}

\noindent\textbf{Unresolved label: inspect additional code to determine INT.}
Listing~\ref{lst:review-deferred} shows the report for Quivr's \texttt{get\_chat\_history}, the function used in the earlier prompt example.
The report leaves the function unresolved because the supplied code does not establish how the database client is created or whether the retrieved history is reused to answer later questions.
The developer checks additional code to confirm that the query uses a Supabase client and that the retrieved conversation history is passed to the LLM when answering a new question.
This evidence supports an \texttt{INT} label.
The listing shows the additional code separately from the code supplied to the LLM.
The evaluation still counts the original LLM result as unresolved; the INT label is determined through developer inspection.

\begin{lstlisting}[basicstyle=\ttfamily\scriptsize,numbers=none,captionpos=b,breaklines=true,caption={Deferred decision: inspect the named dependency and history use.},label={lst:review-deferred},aboveskip=4pt,belowskip=4pt]
[Original table entry, reformatted]
File: backend/repository/chat/get_chat_history.py
Function: get_chat_history
Label to fill: ___ (EXT / INT / NA)
Calls (original line:column and expression):
  a0 9:9 tmp0.execute()
  a1 9:9 tmp2.filter("chat_id", "eq", chat_id)
  a2 9:9 tmp3.select("*")
  a3 9:9 tmp4.from_("chat_history")
  a4 9:9 tmp1.order("message_time", desc = False)
  a5 19:16 tmp5.append(ChatHistory(message))

# Original workflow: reconstruct the tmp query chain;
# locate the Supabase client and history consumers;
# establish the stored data's role; distinguish the
# generated list append; decide the function label.

[Code supplied to the model, simplified]
# backend/repository/chat/get_chat_history.py:6-19
commons = common_dependencies()
history = (commons["supabase"]
    .from_("chat_history").select("*")
    .filter("chat_id", "eq", chat_id)
    .order("message_time", desc=False)
    .execute()).data
return [ChatHistory(message) for message in history]
# backend/llm/openai.py:145-148
history = get_chat_history(self.chat_id)
transformed_history = format_chat_history(history)

[Saved report, shortened]
Proposal: unresolved; no function seed supplied.
a0-a4: unresolved; the chat_history query is known,
  but the report does not establish its EXT/INT role.
a5: no_source_access; append builds a local list.
Missing facts: the Supabase dependency's definition
  and whether chat_history is reusable app history.

[Additional code for developer inspection]
# backend/models/settings.py:6,24-38 (shortened)
from supabase import Client, create_client
# Inside common_dependencies():
supabase_client = create_client(
    settings.supabase_url, settings.supabase_service_key)
return {"supabase": supabase_client, ...}
# backend/llm/openai.py:150-151
model_response = self._call_chain(
    self.qa, question, transformed_history)

[Exact developer review and additional work]
1. Check a5 against the comprehension: it appends
   constructed values and adds no source access.
2. Open models/settings.py, follow common_dependencies,
   and verify that "supabase" is the configured
   library client, not an in-process collection.
3. Follow transformed_history in llm/openai.py to
   _call_chain. Confirm that past human/AI pairs are
   reused to answer a later question.
4. Resolve the query's resource role as INT and
   enter INT for get_chat_history. The actual read
   is execute(); query-building calls do not each
   need a separate I/O claim. If the dependency or
   reuse cannot be confirmed, retain unresolved NA.
Saved work: the report reconstructs the query,
separates the list append, and names the missing
facts. Client inspection and the final source-role
decision still require developer work.
\end{lstlisting}

\end{document}